\documentstyle[prd,epsf,aps]{revtex}
\begin{document}

\title{STATIC AXIALLY SYMMETRIC EINSTEIN-YANG-MILLS-DILATON SOLUTIONS:
I.REGULAR SOLUTIONS}

\vspace{1.5truecm}
\author{
{\bf Burkhard Kleihaus, and Jutta Kunz}\\
Fachbereich Physik, Universit\"at Oldenburg, Postfach 2503\\
D-26111 Oldenburg, Germany}

\vspace{1.5truecm}

%\date{June 1, 1997}

\maketitle
\vspace{1.0truecm}

\begin{abstract}
We discuss the static axially symmetric regular solutions,
obtained recently in Einstein-Yang-Mills and 
Einstein-Yang-Mills-dilaton theory [1].
These asymptotically flat solutions are characterized by
the winding number $n>1$ and the node number $k$ 
of the purely magnetic gauge field.
The well-known spherically symmetric solutions have 
winding number $n=1$.
The axially symmetric solutions satisfy the same relations
between the metric and the dilaton field
as their spherically symmetric counterparts.
Exhibiting a strong peak along the $\rho$-axis,
the energy density of the matter fields of 
the axially symmetric solutions has a torus-like shape.
For fixed winding number $n$ with increasing node number $k$
the solutions form sequences.
The sequences of magnetically neutral non-abelian
axially symmetric regular solutions with winding number $n$
tend to magnetically charged abelian spherically symmetric
limiting solutions, corresponding to
``extremal'' Einstein-Maxwell-dilaton solutions
for finite values of $\gamma$ and to extremal Reissner-Nordstr\o m
solutions for $\gamma=0$, with $n$ units of magnetic charge.
\end{abstract}
\vfill
\noindent {Preprint gr-qc/9707045} \hfill\break
\vfill\eject

\section{Introduction}

SU(2) Einstein-Yang-Mills (EYM) 
and Einstein-Yang-Mills-dilaton (EYMD) theory
possess a rich variety of solutions.
Recently, we have constructed static axially symmetric
regular and black hole solutions
in these theories \cite{kk2,kk3}.
These solutions represent generalizations
of the spherically symmetric regular and black hole solutions,
known since several years \cite{bm,su2,eymd}.
All these solutions are characterized by two integers,
the node number $k$ of the gauge field function(s)
and their winding number $n$ 
with respect to the azimuthal angle $\phi$,
since the fields wind $n$ times around
while $\phi$ covers the full trigonometric circle once.
The spherically symmetric solutions have winding number $n=1$,
a winding number $n>1$ leads to axially symmetric solutions.

All these EYM and EYMD solutions are asymptotically flat
and have non-trivial magnetic gauge field configurations,
but no global charge.
To every regular spherically symmetric solution
there exists a corresponding family of black hole solutions
with regular event horizon $x_{\rm H}>0$ \cite{su2,eymd}.
The same holds true for the axially symmetric solutions \cite{kk3}.
Outside their event horizon
these black hole solutions possess non-abelian hair.
Therefore they represent counterexamples to the
``no-hair'' conjecture.
The axially symmetric black hole solutions additionally
contradict the expectation,
that static black hole solutions are spherically symmetric.
The spherically symmetric EYM and EYMD solutions 
are unstable \cite{stab,eymd},
and there is all reason to believe, that
the axially symmetric solutions are unstable, too.

For fixed $n$ and increasing $k$ the solutions form
sequences, tending to limiting solutions.
Although all solutions of a given sequence are neutral,
the limiting solution is charged,
and the winding number $n$ of the solutions
represents the charge of the limiting solution.
The limiting solutions are spherically symmetric
and abelian, representing Einstein-Maxwell-dilaton (EMD) \cite{emd}
or Reissner-Nordstr\o m (RN) solutions for EYMD and EYM theory,
respectively. In particular,
the limiting solutions of the sequences of regular solutions
correspond to extremal EMD and RN solutions.

The regular axially symmetric solutions have a torus-like shape.
With the $z$-axis as symmetry axis,
the energy density has a strong peak along the $\rho$-axis
and decreases monotonically along the $z$-axis \cite{kk2}.
It has the same shape known from axially symmetric
multimonopoles \cite{rr,mm}, from axially symmetric multiskyrmions
\cite{ms} and from multisphalerons \cite{kk,kk1}.
The energy density of the axially symmetric black hole solutions
looks torus-like only for small horizon radii,
for larger horizon radii it changes shape \cite{kk3}.

In EYMD theory
the dilaton coupling constant $\gamma$ represents a parameter.
In the limit $\gamma \rightarrow 0$ 
the dilaton decouples and EYM theory is obtained,
for $\gamma = 1$ contact with the low energy effective action
of string theory is made,
and in the limit $\gamma \rightarrow \infty$ gravity decouples and
Yang-Mills-dilaton (YMD) theory is obtained.
The YMD limit is of particular interest, since YMD theory
also possesses static spherically symmetric and 
axially symmetric solutions \cite{ymd,kk1}.
In fact,
the axially symmetric YMD solutions serve as starting configurations
in our numerical construction of the regular EYMD and EYM solutions.

This paper is the first of two papers on static axially
symmetric solutions in EYM and EYMD theory and
presents a detailed account of the regular solutions \cite{kk2}.
The second paper will give a detailed account of the
axially symmetric black hole solutions \cite{kk3,kknew}.
In section II of this paper we present the action, the axially
symmetric ansatz in isotropic spherical coordinates
and the boundary conditions. 
Further we derive relations between the metric and the dilaton field
and expressions for the mass.
In section III we recall the spherically symmetric solutions,
presenting them in isotropic coordinates.
In section IV we discuss
the properties of the axially symmetric regular EYM and EYMD solutions
and the convergence of the sequences of regular non-abelian solutions
to limiting abelian solutions.
We present our conclusions in section V.
In Appendix A we formulate the ansatz
in cylindrical coordinates. We present the equations
of motion in Appendix B and the expansion of the functions
at the origin and at infinity in Appendix C.
Finally, in Appendix D we discuss some
technical aspects concerning
the construction of the axially symmetric solutions.

\section{\bf Einstein-Yang-Mills-Dilaton Equations of Motion}

\subsection{\bf Einstein-Yang-Mills-dilaton action}

We consider the SU(2) Einstein-Yang-Mills-dilaton action
\begin{equation}
S=\int \left ( \frac{R}{16\pi G} + L_M \right ) \sqrt{-g} d^4x
\ \label{action} \end{equation}
with the matter Lagrangian
\begin{equation}
L_M=-\frac{1}{2}\partial_\mu \Phi \partial^\mu \Phi
 -e^{2 \kappa \Phi }\frac{1}{2} {\rm Tr} (F_{\mu\nu} F^{\mu\nu})
\ , \label{lagm} \end{equation}
the field strength tensor
\begin{equation}
F_{\mu \nu} = 
\partial_\mu A_\nu -\partial_\nu A_\mu + i e \left[A_\mu , A_\nu \right] 
\ , \label{fmn} \end{equation}
the gauge field
\begin{equation}
A_{\mu} = \frac{1}{2} \tau^a A_\mu^a
\ , \label{amu} \end{equation}
the dilaton field $\Phi$,
and the Yang-Mills and dilaton coupling constants
$e$ and $\kappa$, respectively.

Variation of the action (\ref{action}) with respect to the metric
$g^{\mu\nu}$ leads to the Einstein equations
\begin{equation}
G_{\mu\nu}= R_{\mu\nu}-\frac{1}{2}g_{\mu\nu}R = 8\pi G T_{\mu\nu}
\  \label{ee} \end{equation}
with stress-energy tensor
\begin{eqnarray}
T_{\mu\nu} &=& g_{\mu\nu}L_M -2 \frac{\partial L_M}{\partial g^{\mu\nu}} 
 \nonumber \\
  &=& \partial_\mu \Phi \partial_\nu \Phi 
    -\frac{1}{2} g_{\mu\nu} \partial_\alpha \Phi \partial^\alpha \Phi
    +2e^{2 \kappa \Phi}  {\rm Tr} 
    ( F_{\mu\alpha} F_{\nu\beta} g^{\alpha\beta}
   -\frac{1}{4} g_{\mu\nu} F_{\alpha\beta} F^{\alpha\beta})
\ . \end{eqnarray}
Variation with respect to the gauge field $A_\mu$ 
and the dilaton field $\Phi$ 
leads to the matter field equations.

\subsection{\bf Static axially symmetric ansatz}

We here formulate the ansatz in terms of spherical coordinates
\cite{kk3}. The equivalent formulation 
in terms of cylindrical coordinates \cite{kk2} is given in Appendix A.

Instead of the Schwarzschild-like coordinates, used for the
spherically symmetric EYM and EYMD solutions \cite{bm,su2,eymd}
(see section III),
we adopt isotropic coordinates
to obtain the static axially symmetric solutions.
In terms of the spherical coordinates $r$, $\theta$ and $\phi$
the isotropic metric reads \cite{foot2}
\begin{equation}
ds^2=
  - f dt^2 +  \frac{m}{f} d r^2 + \frac{m r^2}{f} d \theta^2 
           +  \frac{l r^2 \sin^2 \theta}{f} d\phi^2
\ , \label{metric2} \end{equation}
where the metric functions
$f$, $m$ and $l$ are only functions of 
the coordinates $r$ and $\theta$.
The $z$-axis ($\theta=0$) represents the symmetry axis.
Regularity on this axis requires \cite{book}
\begin{equation}
m|_{\theta=0}=l|_{\theta=0}
\ . \label{lm} \end{equation}

We take a purely magnetic gauge field, $A_0=0$,
and choose for the gauge field the ansatz \cite{rr,kk,kk1,kk2,kk3}
\begin{equation}
A_\mu dx^\mu =
\frac{1}{2er} \left[ \tau^n_\phi 
 \left( H_1 dr + \left(1-H_2\right) r d\theta \right)
 -n \left( \tau^n_r H_3 + \tau^n_\theta \left(1-H_4\right) \right)
  r \sin \theta d\phi \right]
\ . \label{gf1} \end{equation}
Here the symbols $\tau^n_r$, $\tau^n_\theta$ and $\tau^n_\phi$
denote the dot products of the cartesian vector
of Pauli matrices, $\vec \tau = ( \tau_x, \tau_y, \tau_z) $,
with the spatial unit vectors
\begin{eqnarray}
\vec e_r^{\, n}      &=& 
(\sin \theta \cos n \phi, \sin \theta \sin n \phi, \cos \theta)
\ , \nonumber \\
\vec e_\theta^{\, n} &=& 
(\cos \theta \cos n \phi, \cos \theta \sin n \phi,-\sin \theta)
\ , \nonumber \\
\vec e_\phi^{\, n}   &=& (-\sin n \phi, \cos n \phi,0) 
\ , \label{rtp} \end{eqnarray}
respectively.
Since the fields wind $n$ times around, while the
azimuthal angle $\phi$ covers the full trigonometric circle once,
we refer to the integer $n$ as the winding number of the solutions.
The four gauge field functions $H_i$ 
and the dilaton function $\Phi$ depend only on 
the coordinates $r$ and $\theta$.
For $n=1$ and $H_1=H_3=0$, $H_2=H_4=w(r)$ and $\Phi=\Phi(r)$,
the spherically symmetric ansatz of ref.~\cite{eymd} 
is recovered.

The ansatz (\ref{gf1}) is axially symmetric in the sense,
that a rotation around the $z$-axis can be compensated
by a gauge rotation.
The most general axially symmetric ansatz would
contain nine gauge field functions (when $A_0=0$).
By requiring additional discrete symmetries
this number is reduced.
The ansatz (\ref{gf1}) respects the discrete mirror
symmetry $M_{xz}\otimes C$,
where the first factor represents reflection through the
$xz$-plane and the second factor denotes charge conjugation
\cite{kkb,kk,bk}.

The ansatz is form-invariant under the abelian gauge transformation
\cite{kkb,kk,kk1}
\begin{equation}
 U= \exp \left({\frac{i}{2} \tau^n_\phi \Gamma(r,\theta)} \right)
\ .\label{gauge} \end{equation}
The functions $H_1$ and $H_2$ transform inhomogeneously
under this gauge transformation,
\begin{eqnarray}
  H_1 & \rightarrow & H_1 -  r \partial_r \Gamma \ , \nonumber \\
  H_2 & \rightarrow & H_2 +   \partial_\theta \Gamma
\ , \label{gt1} \end{eqnarray}
like a 2-dimensional gauge field.
The functions $H_3$ and $H_4$ combine to form a scalar doublet,
$(H_3+{\rm ctg} \theta, H_4)$.
The choice of an appropriate gauge
leading to regular solutions
presents a non-trivial problem \cite{kkb,bkk}.
We therefore fix the gauge by choosing the same gauge condition
as previously \cite{kkb,kk,kk1,kk2}.
In terms of the functions $H_i$ it reads
\begin{equation}
 r \partial_r H_1 - \partial_\theta H_2 = 0 
\ . \label{gc1} \end{equation}

With the ansatz (\ref{metric2})-(\ref{gf1})
and the gauge condition (\ref{gc1}) 
we obtain the set of EYMD field equations.
These are given in Appendix B.
Here we only present the energy density 
of the matter fields $\epsilon =-T_0^0=-L_M$ 
\begin{eqnarray}
-T_0^0 &= & \frac{f}{2m} \left[
 (\partial_r \Phi )^2 + \frac{1}{r^2} (\partial_\theta \Phi )^2 \right]
       + e^{2 \kappa \Phi} \frac{f^2}{2 e^2 r^4 m} \left\{
 \frac{1}{m} \left(r \partial_r H_2 + \partial_\theta H_1\right)^2 
 \right.
\nonumber \\
      & +&  \left.
   \frac{n^2}{l} \left [
  \left(  r \partial_r H_3 - H_1 H_4 \right)^2
+ \left(r \partial_r H_4 + H_1 \left( H_3 + {\rm ctg} \theta \right)
    \right)^2 \right. \right.
\nonumber \\
      & + & \left. \left.
  \left(\partial_\theta H_3 - 1 + {\rm ctg} \theta H_3 + H_2 H_4
     \right)^2 +
  \left(\partial_\theta H_4 + {\rm ctg} \theta \left( H_4-H_2 \right) 
   - H_2 H_3 \right)^2 \right] \right\}
\ , \label{edens} \end{eqnarray}
where the first gauge field term derives from $F_{r\theta}$, 
the second and third derive from $F_{r\phi}$ 
and the fourth and fifth from $F_{\theta\phi}$.

\subsection{Boundary conditions}

To obtain asymptotically flat solutions
which are globally regular
and possess the proper symmetries,
we must impose certain boundary conditions \cite{kk2,kk3}.
Here we are looking for solutions
with parity reflection symmetry.
Therefore we need to consider the solutions only
in the region $0 \le \theta \le \pi/2$,
imposing boundary conditions along the 
$\rho$- and $z$-axis (i.e.~for
$\theta=\pi/2$ and $\theta=0$).
The complete solutions then follow from the symmetries.
In the following we discuss the boundary conditions
for the regular solutions, which
possess axial and parity reflection symmetry.
The expansions of the functions at the origin
(in powers of $r$) and at infinity (in powers of $1/r$)
are given in Appendix C.

{\sl Boundary conditions at infinity}

Asymptotic flatness imposes
for the metric functions of the regular solutions 
at infinity ($r=\infty$) the boundary conditions
\begin{equation}
f|_{r=\infty}= m|_{r=\infty}= l|_{r=\infty}=1
\ . \label{bc1a} \end{equation}
For the dilaton function we require
\begin{equation}
\Phi|_{r=\infty}=0 
\ , \label{bc1b} \end{equation}
since any finite value of the dilaton field at infinity
can always be transformed to zero via
$\Phi \rightarrow \Phi - \Phi(\infty)$, 
$r \rightarrow r e^{-\kappa \Phi(\infty)} $.
For magnetically neutral solutions,
the gauge field functions $H_i$ must satisfy
\begin{equation}
H_2|_{r=\infty}=H_4|_{r=\infty}=\pm 1, \ \ \ 
H_1|_{r=\infty}=H_3|_{r=\infty}=0
\ . \label{bc1c} \end{equation}

{\sl Boundary conditions at the origin}

Requiring the solutions to be regular at the origin
($r=0$) leads 
to the boundary conditions for the metric functions
\begin{equation}
\partial_r f|_{r=0}= \partial_r m|_{r=0}= \partial_r l|_{r=0}= 0
\ . \label{bc2a} \end{equation}
Analogously, the dilaton function must satisfy
\begin{equation}
\partial_r \Phi|_{r=0} =0 
\ , \label{bc2b} \end{equation}
and the set of gauge field functions $H_i$ satisfies
\begin{equation}
H_2|_{r=0}=H_4|_{r=0}= 1, \ \ \ H_1|_{r=0}=H_3|_{r=0}=0
\ . \label{bc2c} \end{equation}
As for the spherically symmetric solutions \cite{bm,su2,eymd},
this choice of boundary conditions for the gauge field functions
is always possible, because of the symmetry
with respect to $H_i \rightarrow -H_i$.
Solutions with an even number of nodes then
have $H_2(\infty)=H_4(\infty)=1$, whereas solutions
with an odd number of nodes 
have $H_2(\infty)=H_4(\infty)=-1$.

{\sl Boundary conditions along the axes}

The boundary conditions along the $\rho$- and $z$-axis
($\theta=\pi/2$ and $\theta=0$) are determined by the 
symmetries.
The metric functions satisfy along the axes
\begin{equation}
\begin{array}{lllllll}
\partial_\theta f|_{\theta=0} &=& \partial_\theta m|_{\theta=0} &=&
\partial_\theta l|_{\theta=0} &=&0 
\ , \\
\partial_\theta f|_{\theta=\pi/2} &=&
\partial_\theta m|_{\theta=\pi/2} &=&
\partial_\theta l|_{\theta=\pi/2} &=&0 \ ,
\end{array}
\   \label{bc4a} \end{equation}
in addition to the condition (\ref{lm}) along the $z$-axis.
Likewise the dilaton function satisfies
\begin{equation}
\begin{array}{lll}
\partial_\theta \Phi|_{\theta=0} &=& 0
\ , \\
\partial_\theta \Phi|_{\theta=\pi/2} &=& 0
\end{array}
\   \label{bc4b} \end{equation}
along the axes.
For the gauge field functions $H_i$ symmetry considerations
lead to the boundary conditions
\begin{equation}
\begin{array}{lllllllllll}
H_1|_{\theta=0}&=&H_3|_{\theta=0}&=&0 &\ , \ \ \ &
\partial_\theta H_2|_{\theta=0} &=& \partial_\theta H_4|_{\theta=0} 
 &=& 0 \ ,
\\
H_1|_{\theta=\pi/2}&=&H_3|_{\theta=\pi/2}&=&0 &\ , \ \ \ &
\partial_\theta H_2|_{\theta=\pi/2} &=& 
\partial_\theta H_4|_{\theta=\pi/2} &=& 0
\end{array}
\   \label{bc4c} \end{equation}
along the axes.

\subsection{Dimensionless Quantities}

It is convenient to introduce dimensionless quantities.
We introduce the dimensionless coordinate $x$,
\begin{equation}
x=\frac{e}{\sqrt{4\pi G}} r
\ , \label{dimx} \end{equation}
the dimensionless dilaton function $\varphi$,
\begin{equation}
\varphi = \sqrt{4\pi G} \Phi
\ , \label{dimp} \end{equation}
and the dimensionless dilaton coupling constant $\gamma$,
\begin{equation}
\gamma =\frac{1}{\sqrt{4\pi G}} \kappa
\ . \label{dimg} \end{equation}
The value $\gamma=1$ corresponds to string theory.

The dimensionless mass $\mu$ is related to 
the mass $M$ via
\begin{equation}
\mu = \frac{e G}{\sqrt{4 \pi G }} M
\ . \label{ms0} \end{equation}

\subsection{Relations between metric and dilaton field}

We now derive two relations between metric and dilaton field
for static axially symmetric solutions
of EYMD theory with purely magnetic 
gauge fields and a general gauge group.
These relations are known to hold for
the spherically symmetric solutions \cite{kks3}.

Let us return to eq.~(\ref{lagm}) and
introduce the gauge field Lagrangian $L_A$
\begin{equation}
L_A=-\frac{1}{2} {\rm Tr} (F_{\mu\nu} F^{\mu\nu})
\ . \label{lagaa} \end{equation}
 From the action (\ref{action}) we obtain
the equation of motion for the dilaton field
\begin{equation}
\partial_\mu \left( \sqrt{-g} \partial^\mu \Phi \right) =
  -2  \kappa \sqrt{-g} e^{2 \kappa \Phi} L_A
\ . \label{eqdil1} \end{equation}
Our aim now is to replace the matter field functions
on the r.h.s.~of this equation by metric functions alone.
Therefore we consider the matter lagrangian
$$
L_M=-\frac{1}{2}\partial_\mu \Phi \partial^\mu \Phi
 +e^{2 \kappa \Phi } L_A \ .
$$
We replace $L_M$ on the l.h.s.~with help of the
($00$) component of the Einstein equations
$$
G_{00}= 8 \pi G T_{00} = 8 \pi G g_{00} L_M \ ,
$$
and we replace the dilaton term on the r.h.s.~with help of the
contracted Einstein equations,
\begin{equation}
 R = -8 \pi G T_\mu^{\ \mu} 
   = 8 \pi G \partial_\mu \Phi \partial^\mu \Phi
\ . \label{R} \end{equation}
(For purely magnetic gauge fields
the contracted gauge field stress energy tensor vanishes.)
This gives for the matter part on the r.h.s.~of eq.~(\ref{eqdil1})
\begin{equation}
   2 e^{2 \kappa \Phi} L_A =
   2 g^{00} T_{00} - g^{\mu \nu} T_{\mu \nu} 
 = \frac{1}{8 \pi G} \left(2 g^{00} G_{00} + R \right)
 = \frac{1}{4 \pi G} g^{00} R_{00} 
\ , \label{r00} \end{equation}
and the dilaton field equation becomes
\begin{equation}
\partial_\mu \left( \sqrt{-g} \partial^\mu \Phi \right) =
  - \frac{\kappa}{4 \pi G}  \sqrt{-g} g^{00} R_{00}
\ . \label{eqdil2} \end{equation}
By evaluating $R_{00}$ explicitly for the metric (\ref{metric2}), 
we find 
\begin{equation}
- \sqrt{-g} g^{00} R_{00} = \frac{1}{2} \partial_\mu
 \left( \sqrt{-g} \partial^\mu \ln f \right)
\ . \end{equation}
The dilaton equation then becomes
\begin{equation}
 \partial_\mu \left( \sqrt{-g} \partial^\mu 
 \left[ \Phi - \frac{\kappa}{8 \pi G} \ln f \right] \right) = 0
\ . \label{eqdil3} \end{equation}
The derivation of the relations between metric and dilaton field
is based on this equation.

Abbreviating the expression 
in square brackets in eq.~(\ref{eqdil3}) by $B$,
\begin{equation}
B= \Phi - \frac{\kappa}{8 \pi G} \ln f
\ , \label{eqdilB} \end{equation}
we now show that the regular static axially symmetric
solutions satisfy $B = 0$.
Therefore we multiply eq.~(\ref{eqdil3}) by $B$
and integrate over $r$ and $\theta$.
When integrating by parts, the surface terms vanish,
(the boundary conditions eq.~(\ref{bc1a})
and (\ref{bc1b}) give $B(\infty)=0$,)
and the integral becomes
\begin{equation}
\int_0^\infty \int_0^\pi \left( r^2 (\partial_r B)^2 + 
 (\partial_\theta B)^2 \right)
\sin \theta \sqrt{l} dr d\theta = 0
\ . \label{pr1} \end{equation}
Since the integrand is positive this implies that $B$ is constant.
With $B(\infty)=0$ this implies $B\equiv 0$,
yielding the first relation between metric and dilaton field,
\begin{equation}
\Phi = \frac{\kappa}{8 \pi G} \ln f
\ . \label{rel1dim} \end{equation}
Using dimensionless quantities, this relation reads
\begin{equation}
\varphi = \frac{\gamma}{2} \ln f
= \frac{\gamma}{2} \ln (- g_{00}) 
\ , \label{rel1} \end{equation}
representing the same relation obtained in the case of
regular spherically symmetric solutions \cite{kks3}.

The second relation is either obtained by
integrating eq.~(\ref{eqdil3}) over $r$ and $\theta$,
or directly from eq.~(\ref{rel1dim}), which yields
\begin{equation}
 \lim_{r \rightarrow \infty} r^2 \partial_r  \Phi
 = \frac{\kappa}{8 \pi G} \lim_{r \rightarrow \infty} r^2 \partial_r  f
\ , \label{der3} \end{equation}
since $f(\infty)=1$.
Changing to dimensionless quantities we find
\begin{equation}
 \lim_{x \rightarrow \infty} x^2 \partial_x  \varphi
 = \frac{\gamma}{2} \lim_{x \rightarrow \infty} x^2 \partial_x  f
\ . \label{der4} \end{equation}
The l.h.s.~of eq.~(\ref{der4}) defines the dilaton charge
\begin{equation}
 D = \lim_{x \rightarrow \infty} x^2 \partial_x  \varphi
\ . \label{der5} \end{equation}
The r.h.s.~of eq.~(\ref{der4}) is proportional to the
dimensionless mass $\mu$
\begin{equation}
  \mu =
  \frac{1}{2} \lim_{x \rightarrow \infty} x^2 \partial_x  f
\ , \label{der6} \end{equation}
as shown below.
Together this yields the second relation
\begin{equation}
D = \gamma \mu
\ , \label{rel2} \end{equation}
which also holds for regular spherically symmetric solutions
\cite{kks3}.

\subsection{\bf Mass}

The mass $M$ of the static axially symmetric
solutions can be obtained directly from
the total energy-momentum ``tensor'' $\tau^{\mu\nu}$
of matter and gravitation \cite{wein}
\begin{equation}
M=\int \tau^{00} d^3r
\ . \end{equation}
Adopting the procedure of \cite{wein},
we start from the metric $g_{\mu \nu}$ in isotropic coordinates,
eq.~(\ref{metric2}).
We then change to quasi-Minkowskian coordinates,
in the sense that the metric $\tilde g_{\mu \nu}$
approaches the Minkowski metric $\eta_{\mu \nu}$
at great distances,
i.e.~we introduce $h_{\mu \nu}$
\begin{equation}
\tilde g_{\mu \nu} = \eta_{\mu \nu} + h_{\mu \nu}
\   \label{h1} \end{equation}
such that $h_{\mu \nu}$ vanishes at infinity.
Identifying $h_{\mu \nu}$
from the metric equation, eq.~(\ref{metric2}),
and employing the coordinates
\begin{equation}
x^1= r \sin \theta \cos \phi \ , \ \ 
x^2= r \sin \theta \sin \phi \ , \ \ 
x^3= r \cos \theta 
\ , \label{xyz} \end{equation}
we obtain
\begin{equation}
h_{\mu \nu} = \left( \begin{array}{cccc}
-f+1 & 0 & 0 & 0 \\
 0 & \frac{m}{f}\cos^2\phi + \frac{l}{f}\sin^2\phi -1
   & \left( \frac{m}{f}-\frac{l}{f} \right) \sin\phi \cos\phi & 0 \\
 0 & \left( \frac{m}{f}-\frac{l}{f} \right) \sin\phi \cos\phi 
   & \frac{m}{f}\sin^2\phi + \frac{l}{f}\cos^2\phi -1 & 0 \\
 0 & 0 & 0 & \frac{m}{f} -1
\end{array} \right)
\ . \label{h2} \end{equation}
Indeed, $h_{\mu \nu}$ vanishes at infinity.

The exact Einstein equations can be written as
\begin{equation}
R^{(1)}_{\ \ \mu \nu} - \frac{1}{2} \eta_{\mu \nu}
R^{(1)\lambda}_{\ \ \ \lambda}
= - 8 \pi G \left( T_{\mu \nu} + t_{\mu \nu} \right)
\ , \label{einlin} \end{equation}
where
$R^{(1)}_{\ \ \mu \nu}$ is the part of the Ricci tensor linear in 
$h_{\mu \nu}$.
Interpreting the quantity
\begin{equation}
\tau^{\alpha \beta} = \eta^{\alpha \mu} \eta^{\beta \nu}
\left( T_{\mu \nu} + t_{\mu \nu} \right)
\   \label{einlin1} \end{equation}
as the total energy-momentum ``tensor'' of matter and gravitation,
the mass is given as the volume integral over $\tau^{00}$.
Transforming this integral to a surface integral
yields for the mass \cite{wein}
\begin{equation}
M = - \frac{1}{16 \pi G} \int \left( 
 \frac{\partial h_{jj}}{\partial x^i}
 - \frac{\partial h_{ij}}{\partial x^j} \right) 
 \frac{x_i}{r} r^2 \sin\theta d\theta d\phi
\ , \label{h3} \end{equation}
where $x^i$ are the coordinates, eq.~(\ref{xyz}),
and the integral is taken over a large sphere of radius 
$r \rightarrow \infty$.
%With
%\begin{equation}
%h_{jj}= \frac{2m+l}{f}-3
%\ , \end{equation}
%\begin{equation}
% \frac{\partial h_{jj}}{\partial x^i} \frac{x_i}{r}
% =\frac{\partial h_{jj}}{\partial r}
% =\frac{\partial}{\partial r} \left( \frac{2m+l}{f} \right)
%\ , \label{part1} \end{equation}
%and
%\begin{equation}
% \frac{\partial h_{ij}}{\partial x^j} \frac{x_i}{r} 
% =\frac{\partial}{\partial r} \left( \frac{m}{f} \right)
% +\frac{m-l}{rf}
%\   \label{part2} \end{equation}
Evaluating the integrand, we obtain
 \begin{equation}
 M = - \frac{1}{16 \pi G} \int \left( 
 \frac{\partial}{\partial r} \left( \frac{m+l}{f} \right)
 +\frac{m-l}{rf} \right)
  r^2 \sin\theta d\theta d\phi
\ . \label{h4} \end{equation}
Since the term in square brackets is independent of the angle
for $r \rightarrow \infty$,
the integration is trivial, yielding
\begin{equation}
M = - \frac{1}{4 G} \lim_{r \rightarrow \infty}  \left(
 \frac{\partial}{\partial r} \left( \frac{m+l}{f} \right)
 +\frac{m-l}{rf} \right) r^2
 = \frac{1}{2G} \lim_{r \rightarrow \infty} r^2 \partial_r  f
\ , \label{der7} \end{equation}
where the last step makes use of
the asymptotic behaviour of the metric functions
(see Appendix C)
\begin{equation}
f = 1 +\frac{\bar f_{1}}{r} + O\left(\frac{1}{r^2}\right)\ , \ \ 
m = 1 + O\left(\frac{1}{r^2}\right)\ , \ \ 
l = 1 + O\left(\frac{1}{r^2}\right)
\ . \label{bcinf} \end{equation}
In dimensionless quantities, eq.~(\ref{der7})
yields the desired relation for the mass, eq.~(\ref{der6}).  

We next show, that the mass of the regular axially symmetric solutions 
can also be obtained from 
\begin{equation} 
 M = - \int_0^\infty \int_0^\pi \int_0^{2 \pi}
  \sqrt{-g} \left( 2 T_0^{\ 0} - T_\mu^{\ \mu} \right)
 dr d\theta d\phi
\ .  \label{ms} \end{equation}
We start from the dilaton field equation (\ref{eqdil2})
\begin{equation}
\partial_\mu \left( \sqrt{-g} \partial^\mu \Phi \right) =
  \frac{\kappa}{4 \pi G}  \sqrt{-g} \frac{R_{00}}{f}
\   \nonumber \end{equation}
and replace the r.h.s~with help of eq.~(\ref{r00}) 
\begin{equation}
\partial_\mu \left( \sqrt{-g} \partial^\mu \Phi \right) =
 - \kappa \sqrt{-g} \left( 2 T_0^{\ 0} - T_\mu^{\ \mu} \right)
\ . \label{ms1} \end{equation}
Integrating both sides over $r$, $\theta$ and $\phi$,
we obtain
\begin{equation}
4 \pi \lim_{r \rightarrow \infty} r^2 \partial_r  \Phi
 = - \kappa \int_0^\infty \int_0^\pi \int_0^{2 \pi}
  \sqrt{-g} \left( 2 T_0^{\ 0} - T_\mu^{\ \mu} \right)
 dr d\theta d\phi = \kappa M
\ , \label{ms2} \end{equation}
where the last equality makes use of relation (\ref{der3})
together with eq.~(\ref{der7}).

\section{\bf Spherically Symmetric Solutions}

The spherically symmetric EYM and EYMD solutions 
were obtained previously in Schwarzschild-like coordinates
with metric \cite{bm,eymd,kks3}
\begin{equation}
ds^2=
  - A^2 N dt^2 +  \frac{1}{N} d \tilde r^2 
  + \tilde r^2 \left( d \theta^2 + \sin^2 \theta d\phi^2 \right)
\ , \label{metric3} \end{equation}
where the metric functions $A$ and $N$
are only functions of the radial coordinate $\tilde r$, and
\begin{equation}
N(\tilde r) = 1 - \frac{2 \tilde m(\tilde r)}{\tilde r}
\ . \label{N} \end{equation}
Since we here construct the static axially symmetric solutions
in isotropic coordinates, we briefly review the spherical solutions
in isotropic coordinates, derive the coordinate transformation between
both sets of coordinates and discuss the limiting solutions.

\subsection{Coordinate transformation}

By requiring $l=m$ and the metric functions  $f$ and $m$
to be only functions of the coordinate $r$,
the axially symmetric isotropic metric (\ref{metric2})
reduces to the spherically symmetric isotropic metric
\begin{equation}
ds^2=
  - f dt^2 +  \frac{m}{f} \left ( d r^2 + r^2 \left( d \theta^2 
           +  \sin^2 \theta d\phi^2 \right) \right)
\ . \label{metric4} \end{equation}
Comparison with the metric (\ref{metric3}) yields
\begin{equation}
\frac{m(r)}{f(r)} = \frac{\tilde r^2}{r^2}
\   \label{co1} \end{equation}
and, with (\ref{co1}),
\begin{equation}
\frac{dr}{r} = \frac{1}{\sqrt{N(\tilde r)}} \frac{d \tilde r}{\tilde r}
\ . \label{co2} \end{equation}

The function $N(\tilde r)$ (or equivalently the mass function
$\tilde m(\tilde r)$) is only known numerically.
Therefore the coordinate function $r(\tilde r)$ can only be
obtained numerically from eq.~(\ref{co2}).
To obtain the coordinate transformation,
we switch to dimensionless coordinates $x$ and $\tilde x$
and introduce the function $\beta$,
\begin{equation}
\beta = \frac{x}{\tilde x}
\ . \label{co3} \end{equation}
Rewriting eq.~(\ref{co2}) in terms of $\beta$,
\begin{equation}
\frac{d {\rm ln} \beta}{ d \tilde x} = \frac{1}{\tilde x}
  \left( \frac{1}{\sqrt{N}}-1 \right)
\ , \label{co4} \end{equation}
we find
\begin{equation}
\beta(\tilde x) = \exp \left[ { -\int_{\tilde x}^\infty
 \frac{1}{\tilde x'} \left( \frac{1}{\sqrt{N}}-1 \right) d \tilde x' }
 \right]
\ , \label{co5} \end{equation}
with
\begin{equation}
N(\tilde x) = 1 - \frac{2\mu(\tilde x)}{\tilde x}
\ . \label{co6} \end{equation}
The integrand is well behaved at the origin, since
$\mu(\tilde x) \sim \tilde x^3$,
and the integration constant is adjusted such that at infinity
$\beta =1$, i.e.~$x= \tilde x$.

Fig.~1 demonstrates the coordinate transformation
for the spherically symmetric EYMD solutions, 
for $\gamma=1$ and $k=1-4$.
Figs.~2a,b show the metric functions $f$ and $m$,
and Figs.~3a,b show the gauge field function $w$
and the dilaton function $\varphi$.
These solutions satisfy relation (\ref{rel1}),
$\varphi = (\gamma/2) {\rm ln}{f}$
and relation (\ref{rel2}), $D=\gamma \mu$,
where $\mu = \mu(\infty)$ represents the dimensionless mass.

Fig.~4 demonstrates the coordinate transformation
for the spherically symmetric EYM solutions, for $k=1-4$.
Figs.~5a,b show the metric functions $f$ and $m$.

\subsection{Limiting solutions}

For fixed dilaton coupling constant $\gamma$,
the sequence of neutral regular spherically symmetric EYMD solutions 
converges to a limiting solution \cite{eymd,kks3}.
This limiting solution is the ``extremal'' EMD solution \cite{emd}
with horizon $\tilde x=0$,
charge $P=1$, mass 
\begin{equation}
\mu = P/\sqrt{1+\gamma^2}
\   \label{mp} \end{equation}
and dilaton charge $D=\gamma\mu$ (\ref{rel1}).
For $\gamma=1$ the coordinate transformation
for the limiting solution
\begin{equation}
x= \frac{1}{\sqrt{2}}
 \left(\sqrt{ P^2+ 2 \tilde x^2} - P \right)
\   \label{emd1} \end{equation}
is also shown in Fig.~1,
the metric functions $f$ and $m$ of the limiting solution
\begin{equation}
ds^2= -\left( \frac{x}{\sqrt{2}P+x} \right) dt^2
 + \left( \frac{\sqrt{2}P+x}{x} \right)
   \left ( d x^2 + x^2 \left( d \theta^2 
           +  \sin^2 \theta d\phi^2 \right) \right)
\   \label{emd2} \end{equation}
are also shown in Figs.~2, and
the dilaton function $\varphi$ of the limiting solution,
\begin{equation}
e^{2 \varphi_\infty} = \frac{x}{\sqrt{2}P+x}
\ , \label{emd3} \end{equation}
is also shown in Fig.~3.
The gauge field function of the limiting solution is trivial, 
$w_\infty=0$.
The limiting functions for $f$ and $\varphi$ are reached uniformly.
In contrast, the function $m$ 
approaches the limiting function $m_\infty=1$ non-uniformly, since it
tends to a constant $m(0) \ne 1$ at $x=0$.
Likewise the function $w$ approaches the limiting function $w_\infty=0$
non-uniformly, since the boundary conditions
require $w(0) \ne 0$ as well as $w(\infty) \ne 0$.

The sequence of spherically symmetric regular EYM solutions
also approaches a limiting solution \cite{lim}.
In Schwarzschild-like coordinates this limiting solution
consists of two parts. A non-trivial part for $\tilde x<1$,
representing an oscillating solution, and a simple part
for $\tilde x>1$, representing the exterior of an extremal 
RN solution with mass $\mu=1$, 
horizon radius $\tilde x_{\rm H}=1$ and charge $P=1$.

 From Fig.~4 we observe, that with increasing $k$
the interval $0 < \tilde x < 1$ is mapped to an ever smaller
interval $0 < x < x_m$, with $x_m$ tending to zero
for $k \rightarrow \infty$.
Therefore the limiting solution becomes rather simple in isotropic
coordinates. The non-trivial part 
of the solution in Schwarzschild-like
coordinates is mapped to zero, and the limiting solution
corresponds to the exterior of an
extremal RN solution with 
mass $\mu=1$, horizon $x_{\rm H}=0$ and charge $P=1$.
The coordinate transformation for this limiting 
RN solution,
\begin{equation}
x = \tilde x-P
\ , \label{rn2} \end{equation}
is also shown in Fig.~4, and the metric functions $f_\infty$ and $m_\infty$,
given by the metric
\begin{equation}
ds^2= -\left( \frac{x}{P+x} \right)^2 dt^2
 + \left( \frac{P+x}{x} \right)^2
   \left ( d x^2 + x^2 \left( d \theta^2 
           +  \sin^2 \theta d\phi^2 \right) \right)
\   \label{rn1} \end{equation}
(i.e.~$m_\infty=1$), are also shown in Figs.~5.
Again, the gauge field function of the limiting solution is trivial,
$w_\infty=0$, and the functions $m$ and $w$ approach the
limiting functions non-uniformly.

\section{\bf Axially Symmetric Solutions}

Beginning with a short discription of the numerical technique,
we here give a detailed description of the numerically constructed 
axially symmetric regular solutions.
We discuss the properties of these
solutions and the convergence of the various sequences
of solutions to limiting solutions.
Some technical aspects concerning
the construction of the axially symmetric solutions
are discussed in Appendix D.

\subsection{\bf Numerical technique}

Subject to the above boundary conditions,
we solve the system of coupled non-linear partial
differential equations numerically.
We employ the radial coordinate 
\begin{equation}
\bar{x} = \frac{x}{1+x}
\   \label{barx} \end{equation}
instead of $x$,
to map spatial infinity to the finite value $\bar{x}=1$ 
(see also Appendix B).
The numerical calculations are performed with help of the program
FIDISOL, which is extensively documented in \cite{schoen}.
The equations are discretized on a non-equidistant
grid in $\bar{x}$ and  $\theta$.
Typical grids used have sizes $150 \times 30$, 
covering the integration region 
$0\leq\bar{x}\leq 1$ and $0\leq\theta\leq\pi/2$.

The numerical method is based on the Newton-Raphson
method, an iterative procedure to find a good approximation to
the exact solution. 
Let us put the partial differential equations into the form 
$P(u)=0$, where $u$ denotes the unknown functions 
(and their derivatives). For an approximate solution $u^{(1)}$,
$P(u^{(1)})$ does not vanish. If we could find a small correction 
$\Delta u$, such that $u^{(1)}+\Delta u$ is the exact solution, 
$P(u^{(1)}+\Delta u)=0$ should hold. Approximately the expansion in 
$\Delta u$ gives 
$$
0=P(u^{(1)}+\Delta u) \approx 
P(u^{(1)})+\frac{\partial P}{\partial u }(u^{(1)}) \Delta u \ .
$$
The equation 
$P(u^{(1)})= -\frac{\partial P}{\partial u }(u^{(1)}) \Delta u $
can be solved to determine the correction  $\Delta u^{(1)}= \Delta u$.
$u^{(2)}=u^{(1)}+\Delta u^{(1)}$ will not be the exact solution
but an improved approximate solution. Repeating the calculations 
iteratively, the approximate solutions will converge to the exact solutuion,
provided the initial guess solution is close enough to the exact solution. 
The iteration stops after $i$ steps if the Newton residual $P(u^{(i)})$ 
is smaller than a prescribed tolerance.
Therefore it is essential to have a good
first guess, to start the iteration procedure.
Our strategy therefore is to use a known solution as guess
and then vary some parameter to produce the next solution.
To construct regular axially symmetric EYMD and EYM solutions, 
we have used the known regular axially symmetric 
YMD solutions \cite{kk1} as starting solutions, 
corresponding to the limit $\gamma \rightarrow \infty$.
Beginning with large values of the parameter $\gamma$, $\gamma$
is slowly decreased, until EYMD solutions with $\gamma=1$
and EYM solutions ($\gamma=0$) are obtained.
A second procedure, also employed, is to start
from the spherically symmetric regular solutions with $n=1$,
and then increase the `parameter' $n$ slowly, to obtain
the desired EYMD and EYM solutions at integer values of $n$.

For a numerical solution it is important to have information 
about its quality, i.~e.~to have an error estimate. The error originates 
from the discretization of the system of partial differential equations.
It depends on the number of gridpoints and on the order of consistency of 
the differential formulae for the derivatives. FIDISOL provides
an error estimate for each unknown function, corresponding to 
the maximum of the discretization error divided by the 
maximum of the function. For the solutions presented here  
the estimations of the relative error for the functions are 
on the order of $10^{-3}$ and $10^{-2}$ for $k<4$ and $k=4$,
respectively. 

\subsection{\bf Properties of the solutions}

For the static axially symmetric solutions constructed here,
the functions depend on two variables, the radial coordinate
$x$ and the angle $\theta$. 
In the following we either present the functions
in three-dimensional plots, employing the compactified
coordinates $\bar \rho=\bar x \sin \theta$ 
and $\bar z=\bar x \cos \theta$ 
(with $\bar x $ given in eq.~(\ref{barx})),
or we present the functions in two-dimensional plots
exhibiting the $x$-dependence for three fixed angles,
$\theta=0$, $\theta=\pi/4$ and $\theta=\pi/2$.
While the three-dimensional plots give a good qualitative
picture of the solutions, the two-dimensional plots
are better for a quantitative analysis of the solutions and
in particular for a comparison of solutions with differing
values of $n$, $k$ or $\gamma$.

We begin with a description of the lowest
axially symmetric regular solution of EYMD and EYM theory.
This solution has winding number $n=2$ and node number $k=1$.
The EYMD solution for $\gamma=1$ is shown in Figs.~6-9.
Figs.~6a-c show the metric functions $f$, $m$ and $l$.
The functions $m$ and $l$ have a rather similar shape.
Figs.~7a-d show the gauge field functions $H_1$-$H_4$.
The single (non-trivial) node of the functions $H_2$, $H_3$ and $H_4$ is
seen in Fig.~7e. The lines representing the position of the node
of $H_2$ and $H_4$ are close together.
The non-trivial node of the function $H_3$ is 
located farther outwards.
Interestingly, the nodal lines of the functions 
$H_2$-$H_4$ are almost spherical.
The function $H_1$ does not possess a non-trivial node.
Fig.~8 shows the dilaton function $\varphi$,
which is related to the metric function $f$ via (\ref{rel1}).
Fig.~9 presents the energy density of the matter fields $\epsilon$,
showing a pronounced peak along the $\rho$-axis
and decreasing monotonically along the $z$-axis.
Thus equal density contours reveal a torus-like shape
of the solutions.
The maximum of the energy density $\epsilon_{\rm max}$
and the location $\rho_{\rm max}$ of the maximum of the energy density
are shown in Table~1 along with the mass of the solution.

The corresponding EYM solution looks very similar
to the EYMD solution. We therefore exhibit only the energy density
of the matter fields for this solution in Fig.~10.
The mass of the solution is shown in Table~2 along with
the maximum of the energy density $\epsilon_{\rm max}$
and the location $\rho_{\rm max}$ of the maximum of the energy density.

Let us now consider the lowest solutions for higher 
winding number.
To see the change of the functions for 
node number $k=1$ with increasing winding number $n$,
we exhibit in Figs.~11-14 the EYMD solutions for $\gamma=1$
with $n=1$, $2$ and $4$.
Figs.~11a-c show the metric functions,
Figs.~12a-d the gauge field functions,
Fig.~13 the dilaton function
and Fig.~14 the energy density of the matter fields.
Obviously, the angle-dependence of the metric and matter functions
increases strongly with $n$.
The location of the biggest angular splitting
of the metric and the matter field functions 
moves further outward with increasing $n$ 
along with the location of the nodes of the gauge field functions
$H_2$-$H_4$.
At the same time the peak
of the energy density along the $\rho$-axis 
shifts outward with increasing $n$ and decreases in height.
At the origin the values of the metric functions increase with $n$,
while the central energy density decreases.
The nodal line of the functions $H_2$-$H_4$ remains almost spherical
with increasing $n$, and $H_1$ remains nodeless.
(Note, that $H_1$ and $H_3$ are zero on the axes in Figs.~12a,c.)

By decreasing $\gamma$ from one to zero, we obtain the
corresponding EYM solutions, shown in Figs.~15-17.
Qualitatively the EYM solutions look like their EYMD counterparts.
But there are a number of quantitative differences.
The metric functions of the EYM solutions are
considerably smaller at the origin, the gauge field functions
have their peaks and nodes shifted inwards,
and the energy density of the matter fields
has slightly higher peaks, which are also shifted inwards,
as compared to the EYMD solutions.
The masses increase with decreasing $\gamma$, as seen in
Tables~1-3.

We now consider the radially excited solutions.
To see the change of the functions for fixed winding number $n$
with increasing node number $k$,
we exhibit in Figs.~18-21 the EYMD solutions for $\gamma=1$
winding number $n=2$ and node number $k=1-4$.
Figs.~18a-c show the metric functions,
Figs.~19a-d the gauge field functions,
Fig.~20 the dilaton function
and Fig.~21 the energy density of the matter fields.
The magnitude of the angular dependence of the metric and matter functions
stays roughly the same with increasing $k$ in these figures.
The peak of the energy density of the matter fields 
moves inward with increasing $k$ and at the same time
increases strongly in height, along with the central density.
The gauge field functions $H_2$ and $H_4$ have precisely $k$ nodes,
the gauge field function $H_1$ has $k-1$ nodes,
and the gauge field function $H_3$ has one node.

Again, the corresponding EYM solutions are similar to these EYMD
solutions. We exhibit only the metric functions in Figs.~22
and the energy density of the matter fields in Fig.~23.
Again, the metric functions of the EYM solutions are
considerably smaller at the origin, and the gauge field functions
have their peaks and nodes shifted inwards,
as compared to the EYMD solutions.
But the energy density of the matter fields
increases only moderately with $k$, indicating a finite limit
for the energy density of the EYM solutions. In contrast, 
the dramatic increase of the energy density
of the matter fields of the EYMD solutions
with increasing $k$ indicates a diverging limit.

The masses of the EYMD solutions with $\gamma=1$ are shown in Table~1,
and the masses of the EYM solutions are shown in Table~2.
The EYMD solutions for other values of $\gamma$ are similar
to these solutions.
For comparison, Table~3 presents the masses of the solutions
with $n=3$ and $k=1-4$ for several values of $\gamma$.
In the limit $\gamma \rightarrow 0$ the EYM solutions are obtained,
and in the limit $\gamma \rightarrow \infty$ 
the corresponding YMD solutions \cite{kk2,kk1}.

\subsection{\bf Limiting solutions}

Let us now consider the limiting solutions of the sequences
of axially symmetric solutions
with fixed winding number $n$, dilaton coupling constant $\gamma$
and increasing node number $k$.
For $n>1$ the pattern of convergence is similar to
the one observed in the spherically symmetric case, $n=1$, considered
in section III.

For fixed winding number $n$ and dilaton coupling constant $\gamma$,
we consider the solutions with $k=1-4$, shown in
Figs.~18-23 and Tables~1-3, and extrapolate to
the limit $k \rightarrow \infty$, to obtain the limiting solution,
also shown.
For finite $\gamma$ this limiting solution is
an ``extremal'' EMD solution with mass 
$\mu_\infty = n/\sqrt{1+\gamma^2}$ (\ref{mp}),
charge $P=n$ and metric (\ref{emd2}).
For $\gamma=0$ it represents the exterior solution of 
an extremal RN solution 
with mass $\mu_\infty = n$, horizon $x_{\rm H}=0$,
charge $P=n$ and metric (\ref{rn2}).
So we find,
that this limiting solution is spherically symmetric,
although the solutions of the sequence are axially symmetric,
and that this limiting solution carries charge $n$,
although the solutions of the sequence are magnetically neutral.

The limiting values for the mass
represent upper bounds for the sequences.
The convergence of the mass $\mu_k$ to the mass of the 
corresponding limiting solution $\mu_\infty$ is exponential.
This is demonstrated in Figs.~24,
where we show the deviation of the mass of the
$k$-th solution from the mass of the limiting solution, $\Delta \mu_k$,
\begin{equation}
\Delta \mu_k = \mu_\infty (\infty)-\mu_k (\infty)
\ , \label{muinf} \end{equation}
for fixed winding number $n$, $n=1-4$, and $\gamma=1$ 
(Fig.~24a) as well as $\gamma=0$ (Fig.~24b).
As required for an exponential convergence, 
the logarithm of the deviation $\Delta \mu_k$ is a straight line
as a function of $k$. The slope of this straight line
decreases with increasing $n$,
thus the larger $n$ the slower is the 
convergence to the limiting solution.

The convergence of the functions of the axially symmetric solutions
is less rapid than the convergence of the mass.
The metric function $f$, the gauge field functions
$H_1$ and $H_3$ and the dilaton function $\varphi$
converge uniformly to the corresponding functions
$f_\infty$, $H_{1,\infty}$, $H_{3,\infty}$
and $\varphi_\infty$ of the limiting solution.
For instance we observe from Fig.~20, that the dilaton function $\varphi$ 
deviates from $\varphi_\infty$
only in an inner region, which decreases exponentially with $k$.
The metric functions $m$ and $l$ also deviate from the limiting
functions $m_\infty=1$ and $l_\infty=1$ 
only in an exponentially decreasing inner region.
But the metric functions $m$ and $l$
as well as the gauge field functions $H_2$ and $H_4$
converge non-uniformly to the corresponding limiting functions.
The metric functions $m$ and $l$ tend to constants
$m(0) \ne 1$ and $l(0) \ne 1$ at the origin,
while the gauge field functions
$H_2$ and $H_4$ differ from the value zero of the limiting solution
both at the origin and at infinity because of their boundary conditions.

\section{Conclusions}

We have given strong numerical evidence, that
both EYMD and EYM theory possess
sequences of regular static axially symmetric solutions,
in addition to the known static spherically symmetric solutions.
These sequences are characterized by the winding number
$n>1$, and the solutions within each sequence by the node number $k$.
(For $n=1$ the spherically symmetric solutions are recovered.)
To rigorously establish the existence of these solutions
a formal existence proof is desirable,
analogous to the proof given for the spherically
symmetric solutions \cite{smo}.

For fixed $n$ and $\gamma>0$ ($\gamma=0$)
the sequences of regular axially symmetric
solutions tend to limiting solutions.
These are the ``extremal'' EMD solutions \cite{emd} 
(the exterior solutions of the extremal RN solutions)
with magnetic charge $n$ and the same $\gamma$.
This generalizes the corresponding
observation for the sequences of spherically symmetric solutions
\cite{eymd,kks3}.
For the YMD solutions the limiting solutions are magnetic
monopoles with $n$ units of charge \cite{kk1}.

The regular axially symmetric solutions have a torus-like shape. 
Otherwise,
many properties of the axially symmetric solutions are similar
to those of their spherically symmetric counterparts.
In particular,
there is all reason to believe, that these static regular
axially symmetric EYMD and EYM solutions are unstable.
Since we can also associate the Chern-Simons number 
$N_{\rm CS}=n/2$ \cite{kk,bk} (for odd $k$ \cite{gv})
with these solutions, we interpret them 
as {\sl gravitating multisphalerons}.

Having constructed the axially symmetric solutions
in EYMD and EYM theory, it appears straightforward
to construct analogous solutions in other non-abelian theories.
For instance, we expect to find
gravitating axially symmetric multimonopoles
in SU(2) Einstein-Yang-Mills-Higgs (EYMH) theory 
with a Higgs triplet,
gravitating axially symmetric multisphalerons
in SU(2) EYMH theory with a Higgs doublet,
or gravitating axially symmetric multiskyrmions
in SU(2) Einstein-Skyrme (ES) theory.
In particular we expect, that the lowest gravitating
axially symmetric $n=2$ multimonopole and multiskyrmion
solutions are stable.

We expect, that the axially symmetric solutions 
represent only the simplest type of non-spherical
gravitating regular solutions 
in EYMD and EYM theory as well as in other non-abelian theories.
We conjecture, that there are gravitating regular solutions
with much more complex shapes and only discrete symmetries left.
Such solutions arise in flat space
for instance in SU(2) Skyrme theory \cite{bc,bs}
and in SU(2) Yang-Mills-Higgs theory \cite{mon},
representing multiskyrmions with higher baryon number
and multimonopoles with higher charge, respectively.

But EYMD and EYM theory also possess black hole solutions.
The non-abelian spherically symmetric black hole solutions
may be regarded as black holes inside sphale\-rons \cite{gv}.
The axially symmetric black hole solutions in EYM and EYMD
theory \cite{kk3}, corresponding to black holes inside
multisphalerons, will be discussed in detail in \cite{kknew}.
We expect analogous black hole solutions in EYMH and ES theory.
Indeed, a perturbative calculation has revealed
not only axially symmetric EYMH black holes,
but also black holes without rotational symmetry \cite{ewein}.

We would like to thank the RRZN in Hannover for computing time.

\vfill\eject
\section{Appendix A}

We here present the ansatz, the gauge condition
and the boundary conditions
in cylindrical coordinates, as employed in \cite{kk2}.
We also show the relation of the functions $H_i$,
used in this paper, to the functions $F_i$, employed in \cite{kk2}.

\subsection{Ansatz in cylindrical coordinates}

In terms of the cylindrical coordinates $\rho$ and $z$ 
the metric in isotropic coordinates reads
\begin{equation}
ds^2=
  - f dt^2 +  \frac{m}{f} \left( d \rho^2+ dz^2 \right)
           +  \frac{l}{f} \rho^2 d\phi^2
\ , \label{metric1} \end{equation}
where the metric functions $f$, $m$ and $l$ are
only functions of $\rho$ and $z$.

The gauge field is parametrized by \cite{rr,kk,kk1,kk2}
\begin{equation}
A_\mu dx^\mu =
\frac{1}{2} \left[ \tau^n_\phi 
 \left( w_1^3 d\rho +  w_2^3 dz \right)
 + \left( \tau^n_\rho w_3^1 + \tau_z w_3^2 \right)
  \rho d\phi \right]
\ , \label{gf} \end{equation}
where $\tau^n_\phi$, $\tau^n_\rho$ and $\tau_z$
denote the dot products of the cartesian vector
of Pauli matrices with the spatial unit vectors
\begin{equation}
\vec e_\phi^{\, n} = (-\sin n \phi, \cos n \phi,0) \ , \ \ \
\vec e_\rho^{\, n} = ( \cos n \phi, \sin n \phi,0) \ , \ \ \
\vec e_z=(0,0,1)
\ , \end{equation}
respectively.
The four functions $w^i_j$ and the dilaton function $\Phi$
depend only on the coordinates $\rho$ and $z$.

With respect to the gauge transformation (\ref{gauge}),
\begin{equation}
 U= \exp\left(\frac{i}{2} \tau^n_\phi \Gamma(\rho,z) \right)
\ ,\label{gaugerz} \end{equation}
the functions
$(\rho w_3^1,\rho w_3^2-n/e)$ transform like a scalar doublet,
and the functions $(w_1^3,w_2^3)$ transform
like a 2-dimensional gauge field.
The gauge condition reads \cite{kkb,kk,kk1,kk2}
\begin{equation}
\partial_\rho w_1^3 + \partial_z w_2^3 =0 
\ . \label{gc} \end{equation}

With the ansatz (\ref{metric1})-(\ref{gf})
and the gauge condition (\ref{gc}) 
the set of EYMD field equations is obtained.
The energy density of the matter fields
$\epsilon =-T_0^0=-L_M$ reads
\begin{eqnarray}
-T_0^0& = & \frac{f}{2m} \left[
 (\partial_\rho \Phi )^2 + (\partial_z \Phi )^2 \right]
       + e^{2 \kappa \Phi} \frac{f^2}{2 m^2}
 (\partial_\rho w_2^3 - \partial_z w_1^3 )^2 
\nonumber \\
      & + &e^{2 \kappa \Phi} \frac{f^2}{2 m l} \left [
  \left( \partial_\rho w_3^1 + \frac{( n w_1^3 + w_3^1 )}{\rho}
	 - e w_1^3 w_3^2 \right)^2 
+ \left (\partial_\rho w_3^2 + \frac{            w_3^2  }{\rho}
	 + e w_1^3 w_3^1 \right)^2  \right.
\nonumber \\
      & + & \left.
  \left (\partial_z    w_3^1 + \frac{ n w_2^3           }{\rho}
	 - e w_2^3 w_3^2 \right)^2 
+ \left (\partial_z    w_3^2 
	 + e w_2^3 w_3^1 \right)^2
      \right]
\ . \end{eqnarray}

Transforming to spherical coordinates
$r$ and $\theta$ ($\rho=r \sin \theta$, $z=r \cos \theta$),
the gauge field functions $F_i(r,\theta)$ are introduced via
\cite{kk,kk1,kk2}
\begin{eqnarray}
w_1^3 \  & 
= &  \ \ {1 \over{er}}(1 - F_1) \cos \theta \ , \ \ \ \ 
w_2^3 \    
= - {1 \over{er}} (1 - F_2)\sin \theta    \ ,     
\nonumber \\
w_3^1 \  & 
= & - {{ n}\over{er}}(1 - F_3)\cos \theta    \ , \ \ \ \
w_3^2 \    
=  \ \ {{ n}\over{er}}(1 - F_4)\sin \theta 
\ . \end{eqnarray}
The spherically symmetric ansatz of ref.~\cite{eymd} is recovered
for $F_1=F_2=F_3=F_4=w(r)$, $\Phi=\Phi(r)$ and $n=1$.

\subsection{Boundary conditions}

At infinity
the boundary conditions for the metric functions and the dilaton
functions are given by eqs.~(\ref{bc1a})-(\ref{bc1b}),
whereas the gauge field functions $F_i$ satisfy
\begin{equation}
F_1|_{r=\infty}=F_2|_{r=\infty}=F_3|_{r=\infty}=F_4|_{r=\infty}=\pm 1
\ . \label{bc1d} \end{equation}
At the origin 
the boundary conditions for the metric functions and the dilaton
functions are given by eqs.~(\ref{bc2a})-(\ref{bc2b}),
and the set of gauge field functions $F_i$ satisfies
\begin{equation}
F_1|_{r=0}=F_2|_{r=0}=F_3|_{r=0}=F_4|_{r=0}=1
\ . \label{bc2d} \end{equation}
Along the $\rho$- and $z$-axis,
the boundary conditions for the metric functions
and the dilaton function are given by
eqs.~(\ref{bc4a})-(\ref{bc4b}),
and for the gauge field functions $F_i$ by
\begin{equation}
\begin{array}{lllllllll}
\partial_\theta F_1|_{\theta=0} &=& \partial_\theta F_2|_{\theta=0} &=&
\partial_\theta F_3|_{\theta=0} &=& \partial_\theta F_4|_{\theta=0} 
 &=& 0
\\
\partial_\theta F_1|_{\theta=\pi/2} &=& 
\partial_\theta F_2|_{\theta=\pi/2} &=& 
\partial_\theta F_3|_{\theta=\pi/2} &=& 
\partial_\theta F_4|_{\theta=\pi/2} &=& 0
\end{array}
\   \label{bc4d} \end{equation}
with the exception of $F_2(r,\theta)$ for $n=2$,
which has 
\begin{equation}
\partial_\theta F_2(r,\theta)|_{\theta=0} \ne 0 \ \ \
({\rm for}  \ n=2)
\ . \label{bc4e} \end{equation}

\boldmath
\subsection{Relation between $H_i$ and $F_i$}
\unboldmath

The functions $F_i$ are related to
the functions $H_i(r,\theta)$ via
\begin{eqnarray}
H_1&=&\sin \theta \cos \theta (F_2-F_1) \ , \nonumber \\
H_3&=&\sin \theta \cos \theta (F_4-F_3) \ , \nonumber \\
H_2&=&\cos^2\theta F_1 + \sin^2\theta F_2 \ , \nonumber \\
H_4&=&\cos^2\theta F_3 + \sin^2\theta F_4
\ . \label{foot} \end{eqnarray}
Note, that the boundary conditions on the $z$-axis for the functions 
$F_i$, eqs.~(\ref{bc4d}), imply the boundary conditions 
for the functions $H_i$, eqs.~(\ref{bc4c}), but not vice versa.
The boundary conditions (\ref{bc4c}) allow for 
$\partial_\theta F_2(r,\theta)|_{\theta=0} \ne 0 $
i.~e.~they properly incorporate the case $n=2$ with
condition (\ref{bc4e}).
 
\section{Appendix B}

We here derive the equations of motion in isotropic spherical
coordinates.

\boldmath
\subsection{Tensors $F_{\mu\nu}$, $T_{\mu\nu}$, $G_{\mu\nu}$}
\unboldmath

We expand the field strength tensor with respect to the Pauli matrices 
$\tau^{n}_{\lambda}$, ($\lambda=r$, $\theta$, $\phi$,
see eq.~(\ref{rtp}))
\begin{eqnarray*}
F_{\mu \nu} & = & F^{(\lambda)}_{\mu \nu} \ \frac{\tau^{n}_{\lambda}}{2}
 \ .\\
\end{eqnarray*}
Inserting ansatz (\ref{gf1}) for the gauge field,
we obtain the non-vanishing coefficients $F_{\mu\nu}^{(\lambda)}$,
\begin{eqnarray}
F_{r\theta }^{(\phi )}
&  = & 
 - \frac{1}{r}\left( H_{1,\theta} + r H_{2,r} \right) 
\ , \nonumber \\
F_{r\phi }^{(r)} 
& = & 
-n \frac{ \sin \theta}{r}\left( r H_{3,r} - H_1  H_4  \right) 
\ , \nonumber \\
F_{r\phi }^{(\theta )} 
& = & 
n \frac{ \sin \theta}{r}\left( r H_{4,r} + H_1 H_3   
+ \cot \theta H_1 \right) 
\ , \nonumber \\
F_{\theta\phi }^{(r)} 
& = & 
-n \sin \theta  \left( H_{3,\theta} - 1 +  H_2 H_4  
+ \cot \theta H_3  \right) 
\ , \nonumber \\
F_{\theta\phi }^{(\theta )} 
& = &  
n \sin \theta \left( H_{4,\theta} - H_2 H_3   
 - \cot \theta \left( H_2 - H_4 \right)  \right) 
\ , \nonumber \\
\end{eqnarray}
and $F^{(\lambda)}_{\mu \nu} = -F^{(\lambda)}_{\nu \mu}$.
It is convenient to define
\begin{eqnarray}
F^2_{r \theta} 
& = & 
\left( F_{r\theta }^{(\phi )}\right)^2 
+\frac{1}{r^2} \left( r H_{1,r}-H_{2,\theta}\right)^2 
\ , \nonumber \\
F^2_{r \phi} 
& = & 
\left( F_{r\phi }^{(r)}\right)^2 
+\left( F_{r\phi }^{(\theta )}\right)^2 
\ , \nonumber \\ 
F^2_{\theta \phi} 
& = & 
\left( F_{\theta\phi }^{(r)}\right)^2 
+\left(F_{\theta\phi }^{(\theta )}\right)^2 
\ , \nonumber \\ 
\end{eqnarray}
where the second term in the definition of $F^2_{r \theta}$ 
represents the gauge fixing term.

With the ansatz for the metric (\ref{metric2}) we obtain the
Lagrange densities
\begin{eqnarray}
L_F      
& = & 
-\frac{f}{2 m} \left( \frac{f}{r^2 m} F^2_{r \theta} 
 + \frac{f}{r^2 \sin^2 \theta l} (F^2_{r \phi}
+\frac{1}{r^2}F^2_{\theta \phi}) \right) 
\ , \nonumber \\
L_{\Phi} 
& = & 
-\frac{f}{2 \ m} \left(\Phi_{,r}^2 
+ \frac{1}{r^2}\Phi_{,\theta}^2 \right)
\ , \nonumber \\
L_M 
& = & 
L_{\Phi}+e^{2 \kappa \Phi} L_{F}
\ , \nonumber \\   
\end{eqnarray}
and the non-vanishing components of the stress energy tensor, 
\begin{eqnarray}
T_{00} 
& = & 
\frac{f^2}{2 m} \left( \Phi_{,r}^2 +\frac{1}{r^2} \Phi_{,\theta}^2 
    +e^{2 \kappa \Phi}
     \left[ \frac{f}{r^2 m} F^2_{r \theta} 
     + \frac{f}{r^2 \sin^2 \theta l}(F^2_{r \phi} 
     + \frac{1}{r^2} F^2_{\theta \phi})\right] \right) 
\ , \nonumber \\
T_{rr} 
& = & 
\frac{1}{2} \left( \Phi_{,r}^2 -\frac{1}{r^2} \Phi_{,\theta}^2 
    +e^{2 \kappa \Phi}
     \left[ \frac{f}{r^2 m} F^2_{r \theta} 
     + \frac{f}{r^2 \sin^2 \theta l}(F^2_{r \phi} 
     - \frac{1}{r^2} F^2_{\theta \phi})\right] \right) 
\ , \nonumber \\
T_{\theta \theta} 
& = & 
\frac{r^2}{2} \left( -\Phi_{,r}^2 +\frac{1}{r^2} \Phi_{,\theta}^2 
    +e^{2 \kappa \Phi}
     \left[ \frac{f}{r^2 m} F^2_{r \theta} 
     + \frac{f}{r^2 \sin^2 \theta l}(-F^2_{r \phi} 
     + \frac{1}{r^2} F^2_{\theta \phi})\right] \right)
\ , \nonumber \\
T_{\phi \phi} 
& = & 
\frac{r^2 \sin \theta l }{2 m} \left( 
-\Phi_{,r}^2 -\frac{1}{r^2} \Phi_{,\theta}^2 
    +e^{2 \kappa \Phi}
     \left[ -\frac{f}{r^2 m} F^2_{r \theta} 
     + \frac{f}{r^2 \sin^2 \theta l}(F^2_{r \phi} 
     + \frac{1}{r^2} F^2_{\theta \phi})\right] \right)
\ , \nonumber \\
T_{r \theta} 
& = &
\Phi_{,r} \Phi_{,\theta} +e^{2 \kappa \Phi}\frac{f}{r^2 \sin^2 \theta l}
               (F_{r \phi}^{(r)} F_{\theta \phi}^{(r)} 
               +F_{r \phi}^{(\theta)} F_{\theta \phi}^{(\theta)})
\ . \nonumber \\
\end{eqnarray}
The Einstein tensor has the non-vanishing components
\begin{eqnarray}
G_{00} 
& = & 
-\frac{f^2}{4 r^2 m }
\nonumber \\ & &
\left( 2 \left( \frac{m_{,\theta ,\theta} + r^2 m_{,r,r} + r m_{,r}}{m} 
-\left( \frac{m_{,\theta}}{m}\right)^2 
- \left( \frac{r m_{,r}}{m}\right)^2 \right) 
  +\left( 2 \frac{l_{,\theta ,\theta} + r^2 l_{,r,r} + 3 r l_{,r}}{l} 
-\left( \frac{l_{,\theta}}{l}\right)^2 
- \left( \frac{r l_{,r}}{l}\right)^2 \right) \right.
\nonumber \\ & &
- 4 \frac{ f_{,\theta ,\theta} + r^2 f_{,r,r}}{f} 
+ 5 \left( \left( \frac{f_{,\theta}}{f}\right)^2 
+ \left( \frac{r f_{,r}}{f}\right)^2 \right)  
\nonumber \\ & & 
\left.
- 2 \left( \left( 4 + \frac{r l_{,r}}{l}\right) \frac{r f_{,r}}{f} 
+ \frac{l_{,\theta}}{l} \frac{f_{,\theta}}{f} \right) 
+ 4\left( \frac{l_{,\theta}}{l} 
- \frac{f_{,\theta}}{f} \right) \cot \theta \right) 
%\nonumber \\  & &
\ , \nonumber \\
G_{rr} 
& = & 
\frac{1}{ 4 r^2 } \left( 
 2 \frac{l_{,\theta ,\theta}}{l} - \left( \frac{l_{,\theta}}{l}\right)^2 
+ 2 \left( \frac{ r l_{,r}}{l} + \frac{ r m_{,r}}{m} \right)   
+\frac{ r m_{,r}}{m} \frac{ r l{,r}}{l}
 -\frac{ m_{,\theta}}{m} \frac{ l_{,\theta}}{l}
+\left( \frac{f_{,\theta}}{f}\right)^2 
- \left( \frac{r f_{,r}}{f}\right)^2 \right.
\nonumber \\  & & 
\left.
-2 \left( \frac{m_{,\theta}}{m} 
- 2  \frac{l_{,\theta}}{l} \right) \cot \theta  \right)
\ , \nonumber \\ 
G_{\theta \theta} 
& = & 
\frac{1}{ 4 } \left(                    
2 \frac{ r^2 l_{,r,r}}{l} -\left( \frac{ r l_{,r}}{l}\right)^2 
 + 4 \frac{r l_{,r}}{l} - 2 \frac{r m_{,r}}{m}  
 - \frac{r m_{,r}}{m} \frac{r l_{,r}}{l} 
 + \frac{ m_{,\theta}}{m} \frac{ l_{,\theta}}{l} 
+ \left( \frac{r f_{,r}}{f}\right)^2 
-\left( \frac{f_{,\theta}}{f}\right)^2 
+ 2 \frac{m_{,\theta}}{m} \cot \theta \right)                 
\ , \nonumber \\ 
G_{\phi \phi} 
& = & 
\frac{l \sin^2 \theta}{4 m} \left( 2 \left(
\frac{r^2 m_{,r,r} + r m_{,r} + m_{,\theta ,\theta}}{m} 
-\left( \frac{m_{,\theta}}{m}\right)^2 
- \left( \frac{r m_{,r}}{m}\right)^2 \right)
+\left( \frac{f_{,\theta}}{f}\right)^2 
+ \left( \frac{r f_{,r}}{f}\right)^2 \right)                           
\ , \nonumber \\ 
G_{r \theta} 
& = & 
\frac{1}{4 r} \left(
 -2 \frac{r l_{,r,\theta}}{l} + \frac{r l_{,r}}{l} \frac{l_{,\theta}}{l}
+\frac{r m_{,r}}{m} \frac{l_{,\theta}}{l}
+\frac{r l_{,r}}{l} \frac{m_{,\theta}}{m} 
+2 \frac{m_{,\theta}}{m} - 2 \frac{r f_{,r}}{f} \frac{f_{,\theta}}{f}
+ 2 \left( \frac{r m_{,r}}{m} 
-\frac{r l_{,r}}{l}\right) \cot \theta \right)
\ . \nonumber \\ 
\end{eqnarray}
With these results at hand we now derive the equations of motion.

\subsection{Matter equations}

The equations of motion for the gauge field functions and for 
the dilaton function are obtained
from the variation of the matter part of the action (\ref{action}),
\begin{eqnarray}
0 & = & 
\left( r^2 H_{1,r,r} + H_{1,\theta ,\theta} + H_{2,\theta} 
- n^2 \frac{m}{l} \left( r H_{4,r} H_3 - r H_{3,r} H_4 
                   +\left( H_3^2 + H_4^2 - 1 \right) 
   H_1 \right)  \right)
 \sin^2 \theta
\nonumber \\  & &
+\left( r H_{2,r} + H_{1, \theta} 
- n^2 \frac{m}{l}\left( 2 H_1 H_3 + r H_{4,r}\right) \right)
 \sin \theta \cos \theta 
 - n^2 \frac{m}{l} H_1                    
\nonumber \\  & &
+\sin^2 \theta \left[H_{1, \theta} + r H_{2,r}  \right]  
 \ln \left( \Sigma \frac{l}{m} \right)_{,\theta} 
+\sin^2 \theta  \left[ r H_{1,r} - H_{2,\theta} \right] r 
\ln \left( \Sigma \frac{l}{m} \right)_{, r}
\nonumber \\  & &
\label{eqh1} \\
0 & = & 
\left( -(r^2 H_{2,r,r} + H_{2,\theta ,\theta} - H_{1,\theta}) 
+n^2 \frac{m}{l} \left( H_4 H_{3,\theta} -  H_3 H_{4, \theta}
                      +\left( H_3^2 +H_4^2 - 1 \right) H_2 \right)
\right) \sin^2 \theta               
\nonumber \\  & &
+\left( r H_{1,r} - H_{2,\theta} 
+ n^2 \frac{m}{l}\left( 2 H_2 H_3 - H_{4,\theta} \right) \right) 
\sin \theta \cos \theta                                                   
+ n^2  \frac{m}{l}\left( H_2 - H_4 \right) 
\nonumber \\  & &
+   \sin^2 \theta \left[ r H_{1,r} - H_{2,\theta}) \right] 
\ln \left( \Sigma \frac{l}{m} \right)_{,\theta}
        -\sin^2 \theta  \left[ H_{1, \theta} + r H_{2,r}  \right] r 
        \ln \left( \Sigma \frac{l}{m} \right)_{, r}
\nonumber \\  & &
\label{eqh2} \\
0 & = & 
\left( r^2 H_{3,r,r} +H_{3,\theta ,\theta} 
- H_3 \left( H_1^2 + H_2^2 \right) 
+ H_1 \left( H_4 - 2 r H_{4,r} \right) 
- H_4 \left( r H_{1,r} - H_{2,\theta} \right) + 2 H_2 H_{4,\theta} 
\right)  \sin^2 \theta
\nonumber \\  & &
+\left( H_{3,\theta} + H_2 H_4 - H_1^2 - H_2^2 
\right) \sin \theta \cos \theta 
-H_3
\nonumber \\  & &
+  
 \sin^2 \theta \left[ H_{3,\theta} -1 + H_2 H_4 
 + \cot \theta H_3  \right] 
   \ln \left( \Sigma \right)_{,\theta}
 + \sin^2 \theta \left[ r H_{3,r} - H_1 H_4 \right] 
 r \ln \left( \Sigma \right)_{, r}          
\nonumber \\  & &
\label{eqh3} \\
0 & = & 
\left( -r^2 H_{4,r,r} - H_{4,\theta ,\theta} 
+H_4 \left( H_1^2 +H_2^2 \right) 
+ H_1 H_3 
- H_2 \left( 1 - 2 H_{3,\theta} \right)
- 2 r H_{3,r} H_1 
-H_3 \left( r H_{1,r} - H_{2,\theta} \right) 
 \right) \sin^2 \theta
\nonumber \\  & &
+\left( H_2 H_3 + H_1 - r H_{1,r} + H_{2,\theta} - H_{4,\theta}  
 \right) \sin \theta \cos \theta 
- \left( H_2 - H_4 \right)
\nonumber \\  & &
-\sin^2 \theta \left[ H_{4,\theta} - H_3 H_2 
- \left( H_2 - H_4 \right) \cot \theta \right]
   \ln \left( \Sigma \right)_{,\theta}
- \sin^2 \theta \left[  r H_{4,r} + H_1 H_3 +  \cot \theta H_1 \right] 
r \ln \left( \Sigma \right)_{, r}             
\nonumber \\  & &
\label{eqh4} \\
0 & = & 
\Phi_{,r,r} + \frac{1}{r^2} \Phi_{,\theta ,\theta} 
- \kappa e^{2 \kappa \Phi} \left( \frac{f}{r^2 m} F^2_{r \theta} 
+ \frac{f}{r^2 \sin^2 \theta l} (F^2_{r \phi}
+\frac{1}{r^2}F^2_{\theta \phi}) \right)                                
\nonumber \\  & &
+\ln ( \sin \theta \sqrt{l} r^2 )_{,r} \Phi_{,r} 
+\frac{1}{r^2} \ln ( \sin \theta \sqrt{l} r^2 )_{,\theta} \Phi_{,\theta}
\nonumber \\ 
\label{eqph} 
\end{eqnarray}
with $\Sigma = e^{2 \kappa \Phi} \frac{f}{\sqrt{l}}$.
The differential equations for the gauge field functions and 
the dilaton function are 
diagonal in the second derivatives with respect to the variable $r$. 

\subsection{Metric equations}

The equations for the metric functions are the Einstein equations
\begin{eqnarray}
8 \pi G \ T_{\mu \nu} & = & G_{\mu \nu} 
\ . \nonumber \\
\end{eqnarray} 
They represent five differential equations for 
three metric functions.
Two of these equations are redundant and can be used to achieve a 
convenient form of the three equations to be solved numerically
(see Appendix D).

We diagonalize the Einstein equations with respect to the
set of derivatives 
$(f_{,r,r}, m_{,r,r}, l_{,r,r}, l_{,\theta ,\theta}, l_{,r,\theta})$.
Except for the ($00$) equation,
these equations are already diagonal. 
The resulting equation for the ($00$) component is
\begin{eqnarray}
  8 \pi G \frac{m}{f} \left( T_\mu^\mu - 2 T_0^0 \right) 
  & = &   
\frac{1}{r^2} \left( \frac{r^2 f_{,r,r}+f_{,\theta ,\theta}+2 r f_{,r}}{f} 
     -  \left( \left( \frac{f_{,\theta}}{f} \right)^2
               + \left( \frac{r f_{,r}}{f}\right)^2 \right)
               + \cot \theta \frac{f_{,\theta}}{f} 
 +\frac{1}{2} \left( \frac{r f_{,r}}{f} \frac{r l_{,r}}{l}
  + \frac{f_{,\theta}}{f} \frac{l_{,\theta}}{l} \right) \right) 
\ . \nonumber \\ \label{eqm1} 
\end{eqnarray}
Then we solve the ($rr$) component for $(\frac{r f_{,r}}{f})^2$ 
and substitute this 
expression into the ($\phi \phi$) and ($\theta \theta$) components,
yielding
\begin{eqnarray}
8 \pi G \frac{m}{f} \left( T_r^r +  T_{\phi}^{\phi} \right)
& = &
\frac{1}{ 4 r^2} \left(
2 \left( 
\frac{ r^2 m_{,r,r} + r m_{,r} + m_{,\theta ,\theta}}{m}
-\left( \frac{m_{,\theta }}{m} \right)^2 
-\left( \frac{ r m_{,r}}{m} \right)^2 
\right) 
+2 \frac{l_{,\theta ,\theta}}{l} 
+ 2 \left( \frac{f_{,\theta}}{f} \right)^2 
\right.
\nonumber \\ & & 
\left.                                                           
- \left( \frac{l_{,\theta}}{l} \right)^2 
+ 2 \left( \frac{r l_{,r}}{l} + \frac{ r m_{,r}}{m} \right)
+ \frac{r m_{,r}}{m} \frac{r l_{,r}}{l}
- \frac{ m_{,\theta}}{m} \frac{ l_{,\theta}}{l}
-2 \left( \frac{m_{,\theta}}{m} -2 \frac{l_{,\theta}}{l} 
\right) \cot \theta 
\right) 
\ , \nonumber \\  & &
\label{eqm2} \\
8 \pi G \frac{m}{f} \left( T_r^r 
+  T_{\theta}^{\theta} \right)
& = &
\frac{1}{4 r^2} \left(
 2  \frac{r^2 l_{,r,r}+l_{,\theta ,\theta}}{l} 
 -\left( \frac{l_{,\theta}}{l} \right)^2
 -\left( \frac{r l_{,r}}{l} \right)^2 
 + 6 \frac{r l_{,r}}{l} + 4 \frac{l_{,\theta}}{l} \cot \theta \right)
\ . \nonumber \\ 
\label{eqm3} \end{eqnarray}
The differential equations (\ref{eqh1})-(\ref{eqph}) 
and (\ref{eqm1})-(\ref{eqm3})
are then diagonal in the second derivatives 
with respect to the radial variable $r$.

Finally we change to dimensionless quantities $x$, $\varphi$
and $\gamma$,
(eqs.~(\ref{dimx})-(\ref{dimg}))
and furthermore to the radial variable $\bar x$,
(eq.~(\ref{barx})),
thus mapping the infinite interval of the variable $x$ onto 
the finite interval $[0,1]$ of the variable $\bar x$.
For the derivatives this leads to the substitutions 
\begin{eqnarray}
r F_{,r}     & \longrightarrow & \bar x (1-\bar x) F_{,\bar x} \ \ , 
\nonumber \\ 
r^2 F_{,r,r} & \longrightarrow & 
\bar x^2 \left( (1-\bar x)^2 F_{,\bar x,\bar x} 
  - 2 (1-\bar x) F_{,\bar x} \right) \\ 
\end{eqnarray}
for any function $F$ in the differential equations. 
In this form we have solved the system of differential equations 
numerically.  

\section{Appendix C}

Here we present the expansion of the functions 
at the origin in powers of $x$
and at infinity in powers of $1/x$.

\subsection{Expansion at the origin}
\noindent
Expanding the gauge field functions, the dilaton function 
and the metric functions at the origin 
up to the second order in $x$ we obtain
\begin{eqnarray}
H_1 & = & \ \ \ \ \ x^2 \sin \theta \cos \theta H_{11}
\ \ \ \ \ \ \ \ \  + O(x^3) \ ,
                                                                 \\ \nonumber
H_2 & = & 1 - x^2 \frac{H_{11}}{2} \left(1 -  2 \sin^2 \theta \right)
\ \ \ + O(x^3)\ ,
                                                                 \\ \nonumber
H_3 & = & \ \ \ \ \ x^2 \sin \theta \cos \theta H_{31}
\ \ \ \ \ \ \ \ \  + O(x^3)\ ,
                                                                 \\ \nonumber
H_4 & = & 1 - x^2 \left( \frac{H_{11}}{2}  -  H_{31} \sin^2 \theta \right)
 + O(x^3)\ ,
                                                                 \\ \nonumber
\varphi & = & \varphi_0 -x^2 \left( 
\left( 1-3 \cos^2 \theta \right) \frac{\varphi_1}{2} 
- n^2 \gamma \sin^2 \theta \frac{f_0}{4 l_0} 
    e^{2 \gamma \varphi_0} \left( H_{11} - 2H_{31} \right)^2 \right)
\ \ \  + O(x^3)\ ,
                                                                 \\ \nonumber
f & = & f_0 \left( 1 
-x^2 \left( \frac{f_1}{2} \left( 1- 3 \cos^2 \theta \right)
- n^2  \sin^2 \theta \frac{f_0}{2 l_0} 
    e^{2 \gamma \varphi_0} \left( H_{11} - 2H_{31} \right)^2 \right)
 \right)    
 + O(x^3)\ ,
                                                                 \\ \nonumber
m & = & l_0 \left( 1 
-x^2 \left( l_1 \left( 1- 2 \cos^2 \theta \right)
- n^2  \sin^2 \theta \frac{f_0}{l_0} 
    e^{2 \gamma \varphi_0} \left( H_{11} - 2H_{31} \right)^2 \right)
 \right)    
\ \   + O(x^3)\ ,
                                                                 \\ \nonumber
l & = & l_0 \left( 1 
-x^2 \frac{l_1}{3} \left( 1- 4 \cos^2 \theta \right) \right)
 + O(x^3)\ ,
                                                                 \\ \nonumber
\end{eqnarray}
where $H_{11}$, $H_{31}$, $\varphi_0$, $\varphi_1$, $f_0$, $f_1$, 
$l_0$ and $l_1$ are constants.

\subsection{Expansion at infinity}

Expanding the gauge field functions, the dilaton function 
and the metric functions at infinity in powers of
$\frac{1}{x}$ we obtain
\begin{eqnarray}
H_1 & = & 
\frac{1}{x^2} \bar{H}_{12} \sin \theta \cos \theta
 + O\left(\frac{1}{x^3}\right)\ ,
                                                                 \\ \nonumber
H_2 & = & 
\pm 1 + \frac{1}{2 x^2} \bar{H}_{12} 
\left( \cos^2 \theta - \sin^2 \theta \right)
 + O\left(\frac{1}{x^3}\right)\ ,
                                                                 \\ \nonumber
H_3 & = & 
\frac{1}{x} \sin \theta \cos \theta \bar{H}_{31}
+\frac{1}{4 x^2} \sin \theta \cos \theta \left( \pm 2 \bar{H}_{12} 
+ \bar{H}_{31} \left( \bar{f}_1 + 2 \gamma \bar{\varphi}_1 \right) \right)
 + O\left(\frac{1}{x^3}\right)\ ,
                                                                 \\ \nonumber
H_4 & = & 
\pm \left( 1+ \frac{1}{x} \bar{H}_{31} \sin^2 \theta 
         -\frac{1}{4 x^2} \sin^2 \theta 
         \left( 
         \pm 2 \bar{H}_{12} 
         + \bar{H}_{31} \left( \bar{f}_1 + 2 \gamma \bar{\varphi}_1 \right)
         \right)
         \right) 
 + O\left(\frac{1}{x^3}\right)\ ,
                                                                 \\ \nonumber
\varphi & = & \frac{1}{x} \bar{\varphi}_1 
 + O\left(\frac{1}{x^3}\right)\ ,
                                                                 \\ \nonumber
f & = & 
1+ \frac{1}{x} \bar{f}_1 + \frac{1}{2 x^2} \bar{f}_1^2
 + O\left(\frac{1}{x^3}\right)\ ,
                                                                 \\ \nonumber
m & = &
1+ \frac{1}{x^2} \bar{l}_2 + \frac{1}{x^2} \sin^2 \theta \bar{m}_2
 + O\left(\frac{1}{x^3}\right)\ ,
                                                                 \\ \nonumber
l & = &
1 + \frac{1}{x^2} \bar{l}_2
 + O\left(\frac{1}{x^3}\right)\ ,
                                                                 \\ \nonumber
\end{eqnarray}
with constants $\bar{H}_{12}$, $\bar{H}_{31}$, $\bar{f}_1$, $\bar{\varphi}_1$,
$\bar{m}_2$ and $\bar{l}_2$.

\section{Appendix D}

Here we discuss some technical aspects concerning
the construction of the axially symmetric solutions.

\subsection{Metric}

Prior to the metric (\ref{metric2})
we have employed the isotropic metric in the form \cite{foot3}
\begin{equation}
ds^2=
  - f dt^2 +  e^{\bar \mu} \left( d r^2 + r^2 d \theta^2 \right)
           +  \bar l r^2 \sin^2 \theta d\phi^2
\ , \label{metrics} \end{equation}
where the metric functions
$f$, $\bar \mu$ and $\bar l$ are functions of 
the coordinates $r$ and $\theta$.
At infinity, the functions satisfy
\begin{eqnarray}
f & = & 
1+ \frac{1}{x} \bar{f}_1 + O\left(\frac{1}{x^2}\right)\ ,
\nonumber \\
\bar \mu & = &
 - \frac{1}{x} \bar{f}_1 + O\left(\frac{1}{x^2}\right)\ ,
\nonumber \\
l & = &
1- \frac{1}{x} \bar{f}_1 + O\left(\frac{1}{x^2}\right)\ ,
\\ 
\end{eqnarray}
with constant $\bar{f}_1$.
We have also replaced
the function $\bar \mu$ by the new function $\bar m$,
\begin{equation}
\bar m = e^{\bar \mu}
\ . \end{equation}
Our final choice of metric parametrization with
\begin{equation}
\bar m = \frac{m}{f}  \ \ \ {\rm and}  \ \ \
\bar l = \frac{l}{f} \ ,
\end{equation}
has the advantage that it
does lead to a faster convergence of the metric functions
$m$ and $l$ to their limiting values
(see Appendix C).

\subsection{Einstein equations}

We have diagonalized the Einstein equations 
with respect to the second derivatives in $r$ of the metric functions.
But this diagonalization is ambiguous, 
since only 3 of 5 equations are needed.
In the earliest calculations we followed the diagonalization 
scheme employed in ref.~\cite{schunck}.
However, these calculations did not properly produce the
asymptotic behaviour of the derivative $\partial_x \bar \mu$.
Using another linear combination of the equations,
the problem was shifted to
the asymptotic behaviour of the derivative $\partial_x f$.
The problem persisted even for the spherical solutions,
indicating that its origin resided in the program package
FIDISOL.
Therefore we looked for a combination of the equations
which FIDISOL would deal with properly.
We realized, that eliminating the square of the first derivatives of
$f$, $(\partial_r f)^2$, from the metric equations,
would precisely do this. 
Eqs. (\ref{eqm2}) and (\ref{eqm3}) then become identical 
in the spherically symmetric case, $n=1$. 
In this form the computations with FIDISOL give the correct
behaviour of the functions for $n=1$ and also for $n > 1$.
This motivates our final choice of the equations,
(\ref{eqm1})-\ref{eqm3}).

\subsection{Regularity condition}

Regularity on the $z$-axis requires the condition (\ref{lm})
to be satisfied.
When the numerical calculations do not impose this condition,
it is not exactly satisfied.
Typical values for the ratio $m/l$ are 0.995 instead of 
precisely one.
To strictly satisfy the regularity condition (\ref{lm})
we have introduced a new function
\begin{equation}
g=\frac{m}{l}
\ , \end{equation}
imposing condition (\ref{lm}) as boundary condition for this
function.
The resulting solutions respect the regularity condition
and are otherwise in excellent agreement
with the solutions obtained before.

\newpage
\begin{table}[p!]
\begin{center}
\begin{tabular}{|c|cccc|} 
 \hline 
\multicolumn{1}{|c|} { $ $ }&
\multicolumn{4}{ c|}  {$\mu$ } \\
 \hline 
$k/n$ &  $1$  & $2$ & $3$ & $4$ \\
 \hline 
$1$ &  $0.577$   & $0.961$  & $1.297$  & $1.607$  \\ 
$2$ &  $0.685$   & $1.262$  & $1.770$  & $2.239$  \\ 
$3$ &  $0.703$   & $1.365$  & $1.976$  & $2.549$  \\  
$4$ &  $0.707$   & $1.399$  & $2.063$  & $2.698$  \\ 
\hline  
$-$ &  $0.707$   & $1.414$  & $2.121$  & $2.828$  \\
\hline
\multicolumn{1}{|c|} { $ $ }&
\multicolumn{4}{ c|}  {$\epsilon_{max}(\rho_{max})$ } \\
 \hline 
$k/n$ &  $1$  & $2$ & $3$ & $4$ \\
\hline
$1$ &  $1.075 \ (0.)$   & $0.177 \ (0.90)$  & $0.098 \ (1.59)$ 
 & $0.072 \ (2.37)$  \\ 
$2$ &  $11.63 \ (0.)$   & $0.910 \ (0.30)$  & $0.380 \ (0.66)$  
& $0.235 \ (1.10)$  \\ 
$3$ &  $79.70 \ (0.)$   & $3.443 \ (0.09)$  & $1.124 \ (0.28)$  
& $0.601 \ (0.51)$  \\  
$4$ &  $498.2 \ (0.)$   & $12.01 \ (0.03)$  & $3.064 \ (0.11)$  
& $1.435 \ (0.24)$  \\ 
\hline
\end{tabular}
\end{center} 
{\bf Table 1}\\
The dimensionless mass $\mu$, 
 the maximum of the energy density $\epsilon_{\rm max}$
and the location $\rho_{max}$ of the maximum of the energy density 
of the EYMD solutions of the sequences $n=1-4$
with node numbers $k=1-4$ and $\gamma=1$.
For comparison, the mass of the limiting
solutions is also shown.
Note, that $\mu/n$ decreases with $n$ for fixed finite $k$.
\end{table}

\newpage
\begin{table}[p!]
\begin{center}
\begin{tabular}{|c|cccc|} 
 \hline 
\multicolumn{1}{|c|} { $ $ }&
\multicolumn{4}{ c|}  {$\mu$ } \\
 \hline 
$k/n$ &  $1$  & $2$ & $3$ & $4$ \\
 \hline 
$1$ &  $.829$   & $1.385$  & $1.870$  & $2.319$  \\ 
$2$ &  $.971$   & $1.796$  & $2.527$  & $3.197$  \\ 
$3$ &  $.995$   & $1.935$  & $2.805$  & $3.621$  \\  
$4$ &  $.999$   & $1.980$  & $2.921$  & $3.824$  \\ 
\hline  
$-$ &  $1.$   & $2.$  & $3.$  & $4.$  \\
\hline
\multicolumn{1}{|c|} { $ $ }&
\multicolumn{4}{ c|}  {$\epsilon_{max}(\rho_{max})$ } \\
 \hline 
$k/n$ &  $1$  & $2$ & $3$ & $4$ \\
\hline
$1$ &  $1.23 \ (0.)$   & $0.26 \ (0.7 )$  & $0.16 \ (1.3)$  
& $0.12 \ (1.8)$  \\ 
$2$ &  $2.55 \ (0.)$   & $0.49 \ (0.2 )$  & $0.26 \ (0.5)$  
& $0.19 \ (0.8)$  \\ 
$3$ &  $2.92 \ (0.)$   & $0.62 \ (0.07)$  & $0.35 \ (0.2)$  
& $0.25 \ (0.4)$  \\  
$4$ &  $2.98 \ (0.)$   & $0.68 \ (0.02)$  & $0.40 \ (0.1)$  
& $0.28 \ (0.2)$  \\ 
\hline
\end{tabular}
\end{center} 
{\bf Table 2}\\
The dimensionless mass $\mu$, 
 the maximum of the energy density $\epsilon_{\rm max}$
and the location $\rho_{max}$ of the maximum of the energy density 
of the EYM solutions of the sequences $n=1-4$
with node numbers $k=1-4$.
For comparison, the mass of the limiting
solutions is also shown.
Note, that $\mu/n$ decreases with $n$ for fixed finite $k$.
\end{table}

\newpage
\begin{table}[p!]
\begin{center}
\begin{tabular}{|c|c|ccc|c|} \hline
\multicolumn{1}{|c|} { $ $ }&
\multicolumn{1}{ c|} { $EYM $ }&
\multicolumn{3}{ c|}  {$EYMD$ } & 
\multicolumn{1}{ c|} { $YMD$ }\\   
 \hline
 $k/\gamma$ &  $0$     &    $0.5$  &  $1.0$   & $2.0$ &   $\infty$ \\
 \hline
 $1$        &  $1.870$ &   $1.659$ &  $1.297$ & $0.811$ & $1.800$ \\ 
 $2$        &  $2.524$ &   $2.250$ &  $1.770$ & $1.114$ & $2.482$ \\  
 $3$        &  $2.805$ &   $2.505$ &  $1.976$ & $1.247$ & $2.785$ \\  
 $4$        &  $2.922$ &   $2.611$ &  $2.063$ & $1.304$ & $2.913$ \\   
 \hline  
 $-$        &  $3    $ &   $2.683$ &  $2.121$ & $1.342$ & $3    $ \\ 
\hline
\end{tabular}
\end{center} 
\vspace{1.cm} 
{\bf Table 3}\\
The dimensionless mass $\mu$ of the EYMD solutions
with winding number $n=3$ and up to 4 nodes
for several values of the dilaton coupling constant $\gamma$.
For comparison, the last row gives the mass of the limiting
solutions, the first column gives the mass
of the EYM solutions, and the last column the
scaled mass of the corresponding YMD solutions \cite{kk1}.
\end{table}

\newpage
\begin{figure}
\centering
\epsfysize=12cm
\mbox{\epsffile{
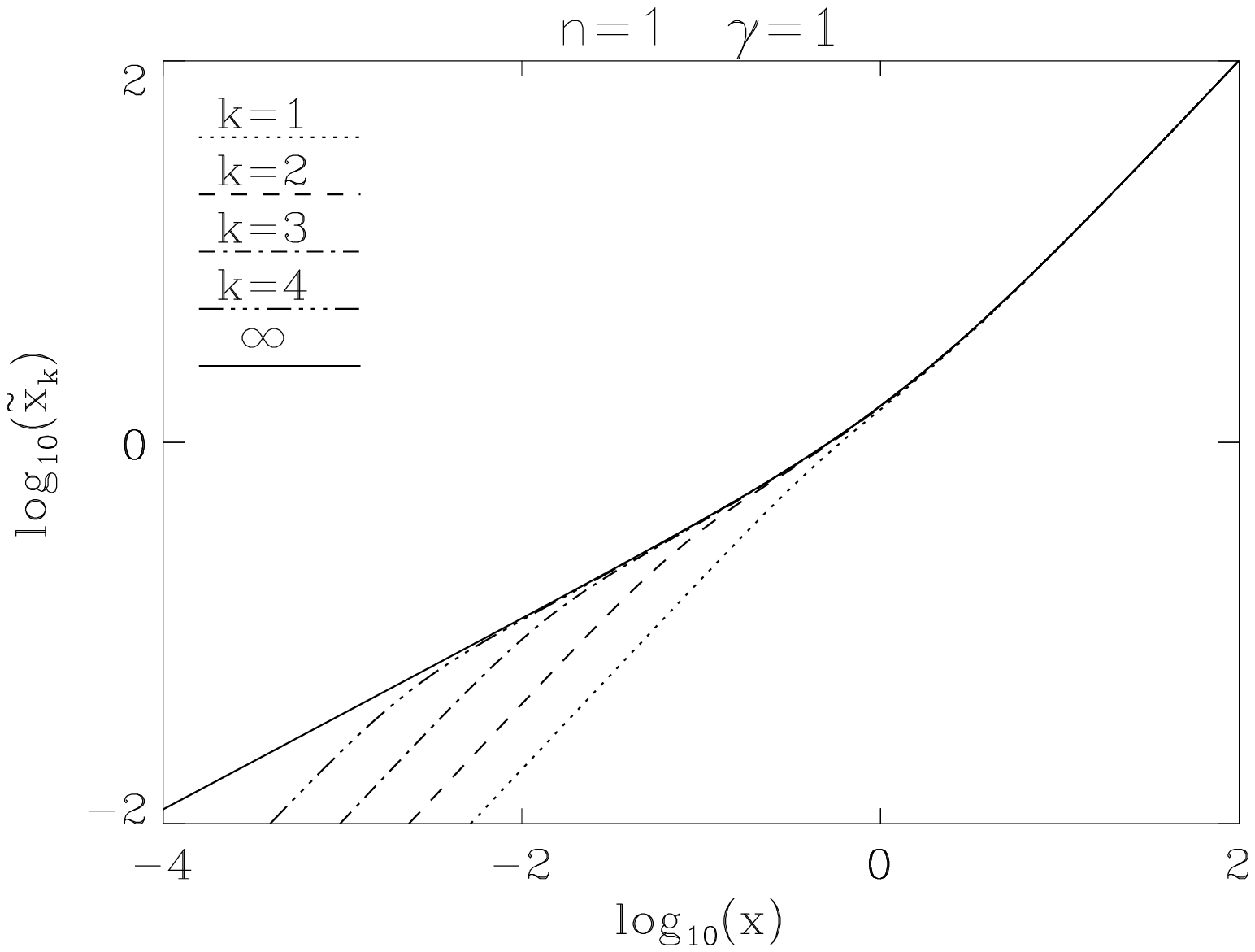
}}\\
Fig.~1\\
The coordinate transformation between the isotropic coordinate $x$
and the Schwarzschild-like coordinate $\tilde x$
is shown for the spherically symmetric solutions ($n=1$)
of EYMD theory with $\gamma=1$ and $k=1-4$.
Also shown is the coordinate transformation for the limiting EMD solution.
\end{figure}

\newpage
\begin{figure}
\centering
\epsfysize=12cm
\mbox{\epsffile{
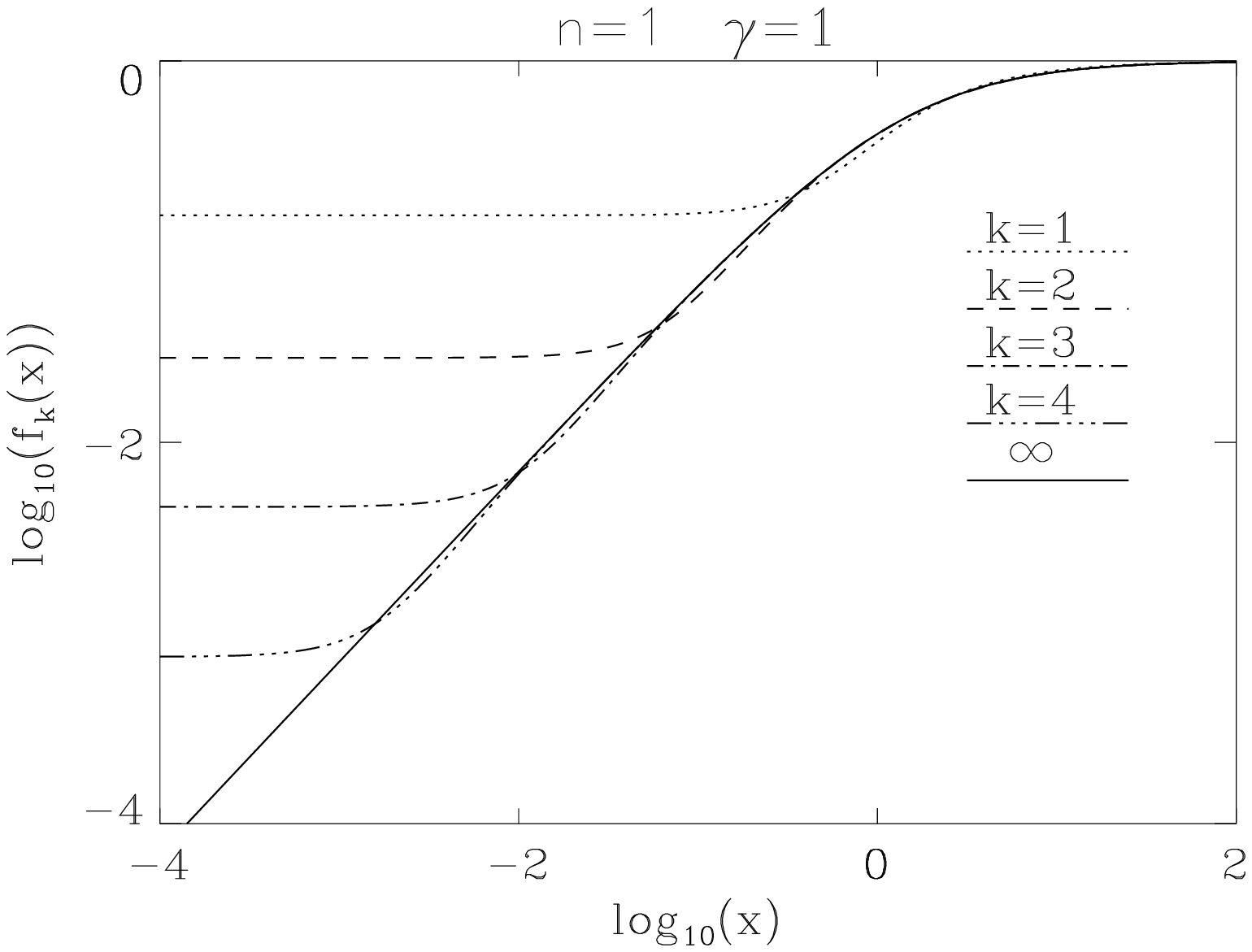
}}\\
Fig.~2a\\
The metric function $f$ is shown 
for the spherically symmetric solutions ($n=1$)
of EYMD theory with $\gamma=1$ and $k=1-4$.
Also shown is the metric function of the limiting EMD solution.
\end{figure} 

\newpage
\begin{figure}
\centering
\epsfysize=12cm
\mbox{\epsffile{
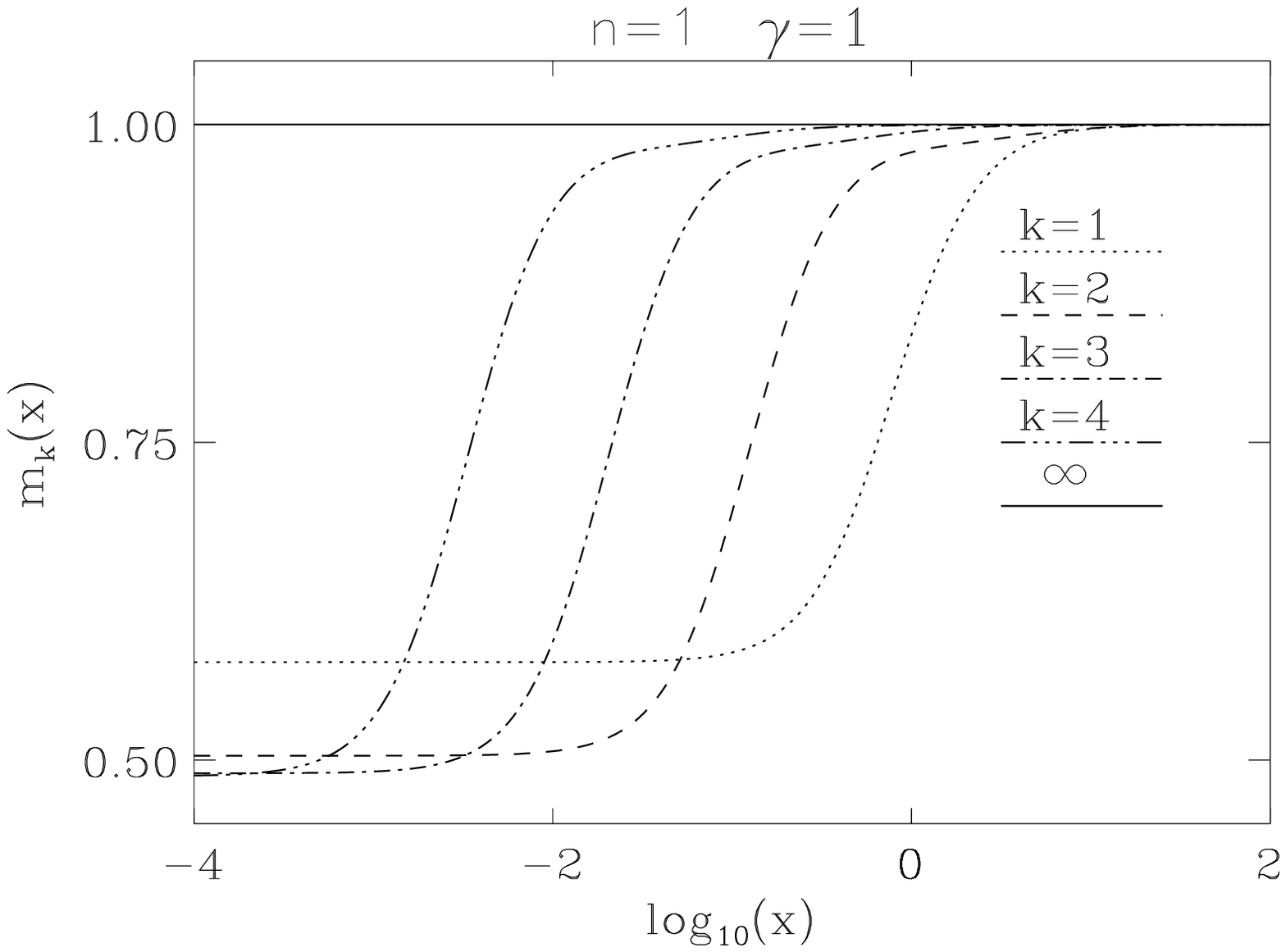
}}\\
Fig.~2b\\
Same as Fig.~2a for the metric function $m$.
\end{figure}

\newpage
\begin{figure}
\centering
\epsfysize=12cm
\mbox{\epsffile{
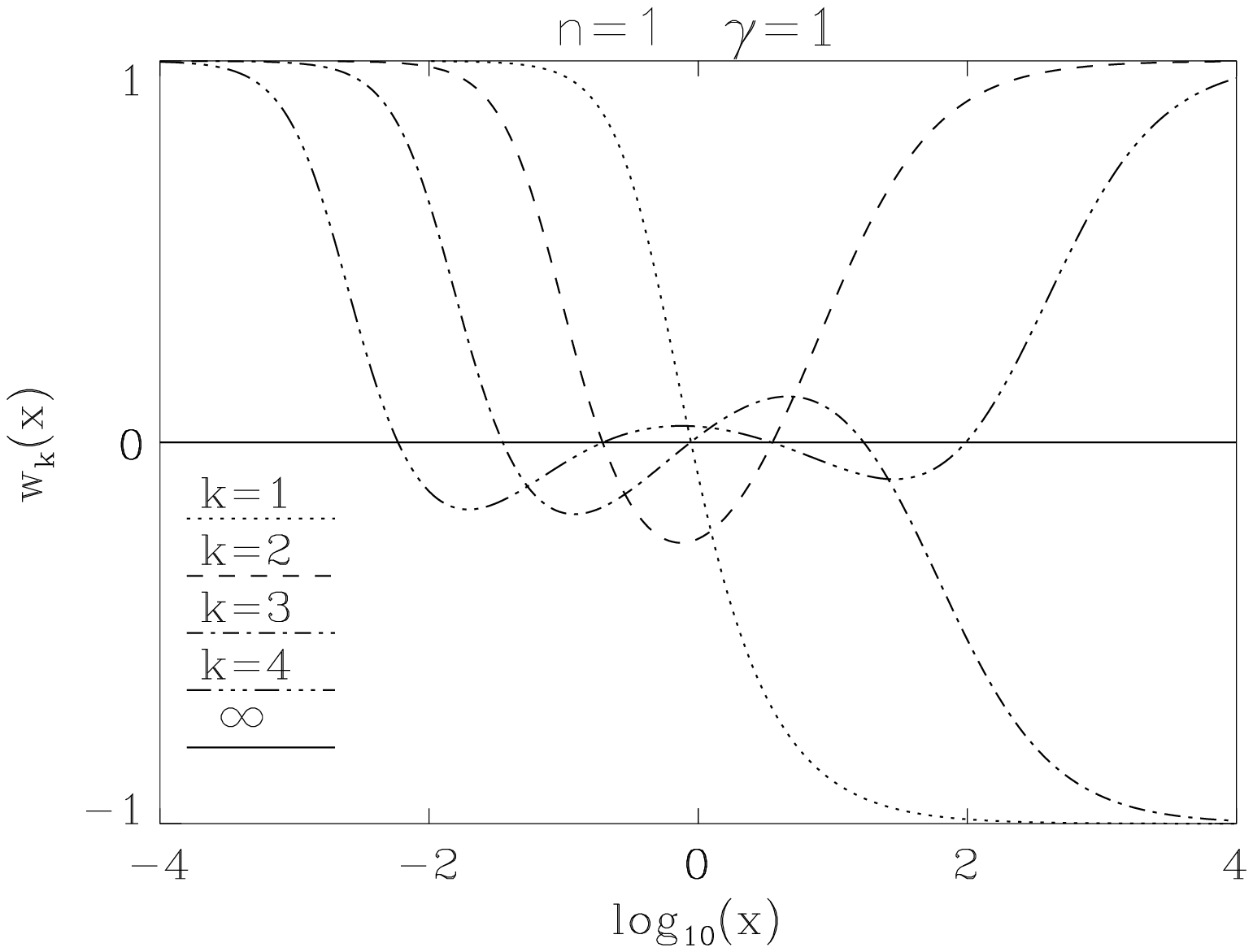
}}\\
Fig.~3a\\
Same as Fig.~2a for the gauge field function $w$.
\end{figure}
 
\newpage
\begin{figure}
\centering
\epsfysize=12cm
\mbox{\epsffile{
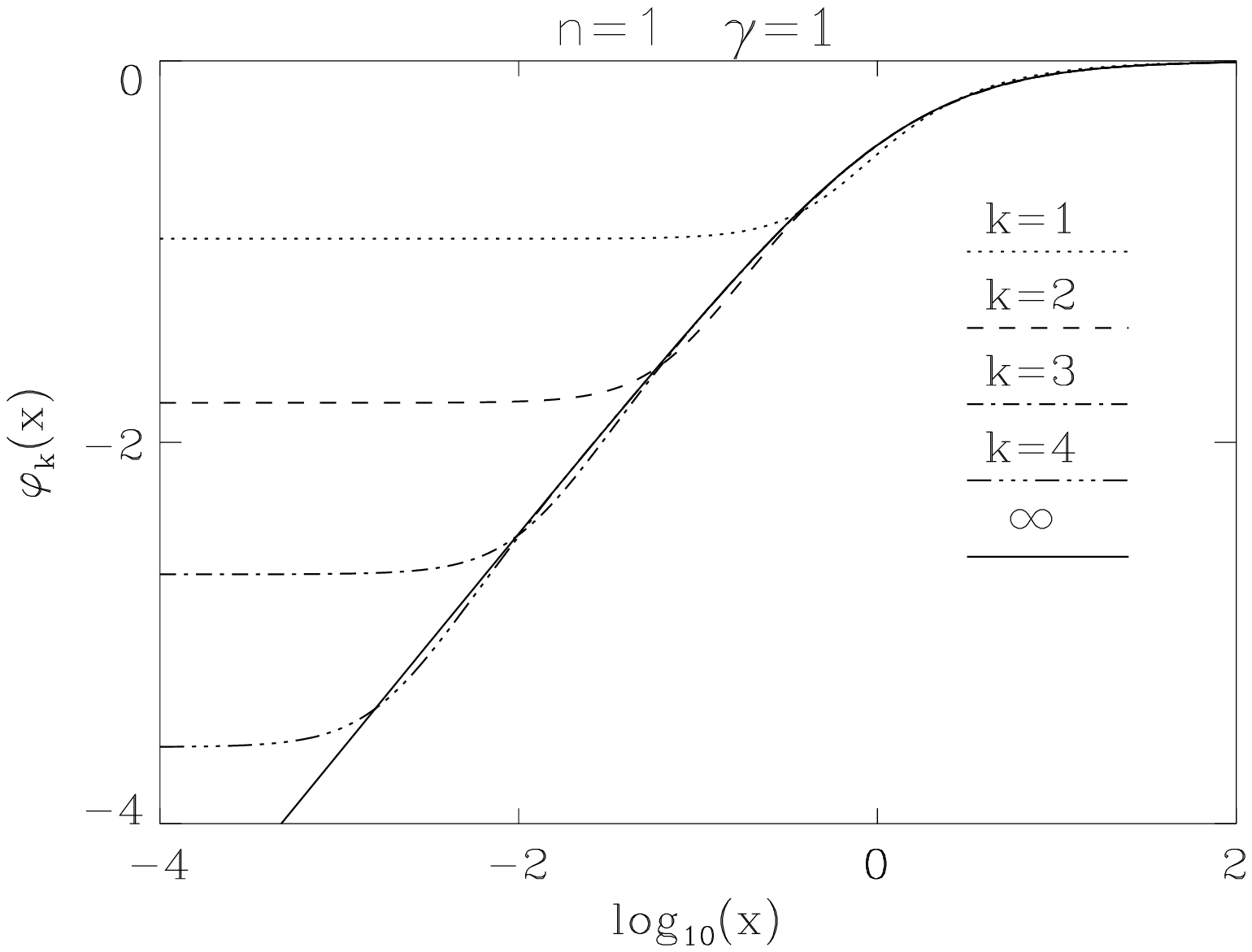
}}\\
Fig.~3b\\
Same as Fig.~2a for the dilaton function $\varphi$.
\end{figure}
 
\newpage
\begin{figure}
\centering
\epsfysize=12cm
\mbox{\epsffile{
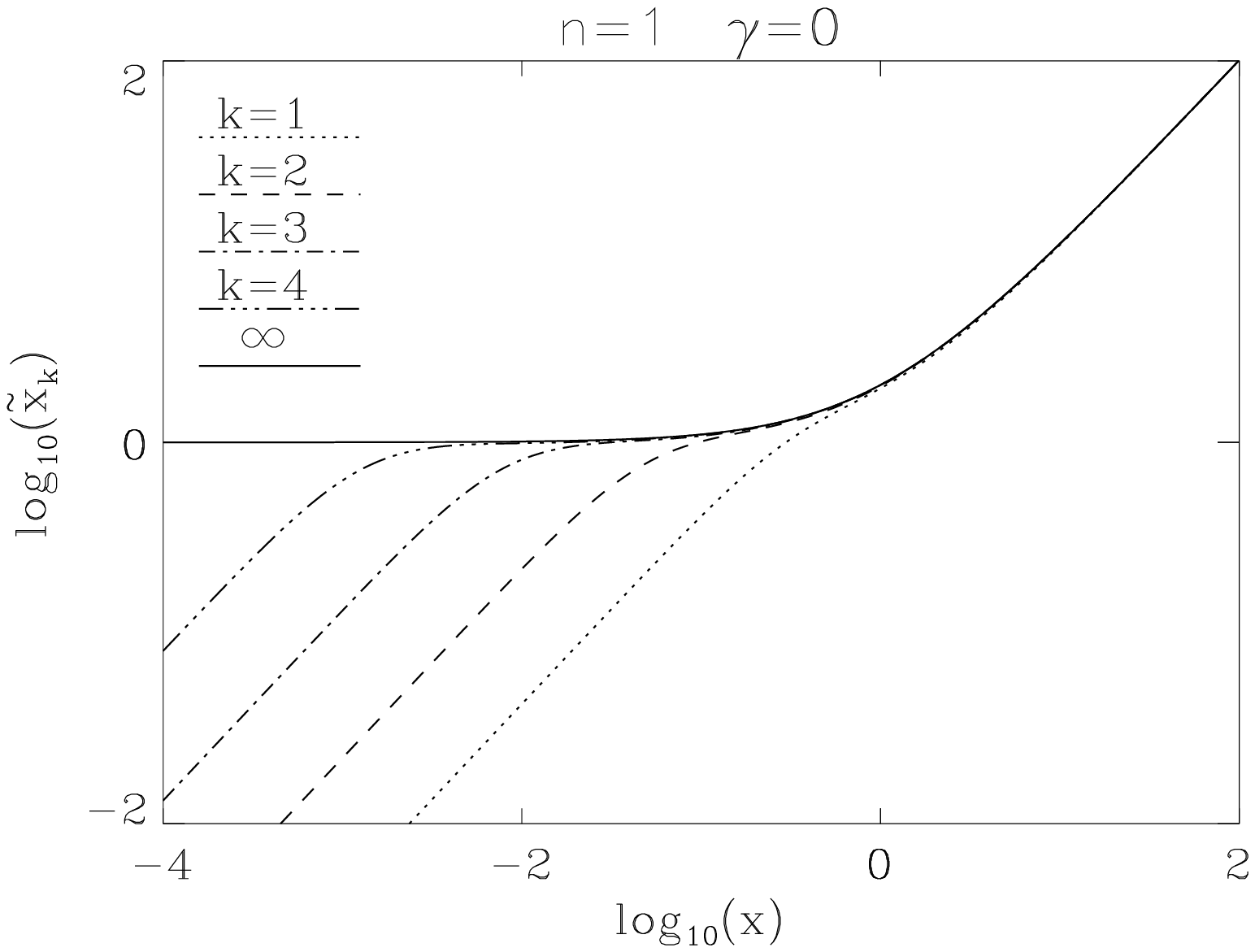
}}
Fig.~4\\
The coordinate transformation between the isotropic coordinate $x$
and the Schwarzschild-like coordinate $\tilde x$
is shown for the spherically symmetric solutions ($n=1$)
of EYM theory with $k=1-4$.
Also shown is the coordinate transformation for the limiting RN solution.
\end{figure}
 
\newpage
\begin{figure}
\centering
\epsfysize=12cm
\mbox{\epsffile{
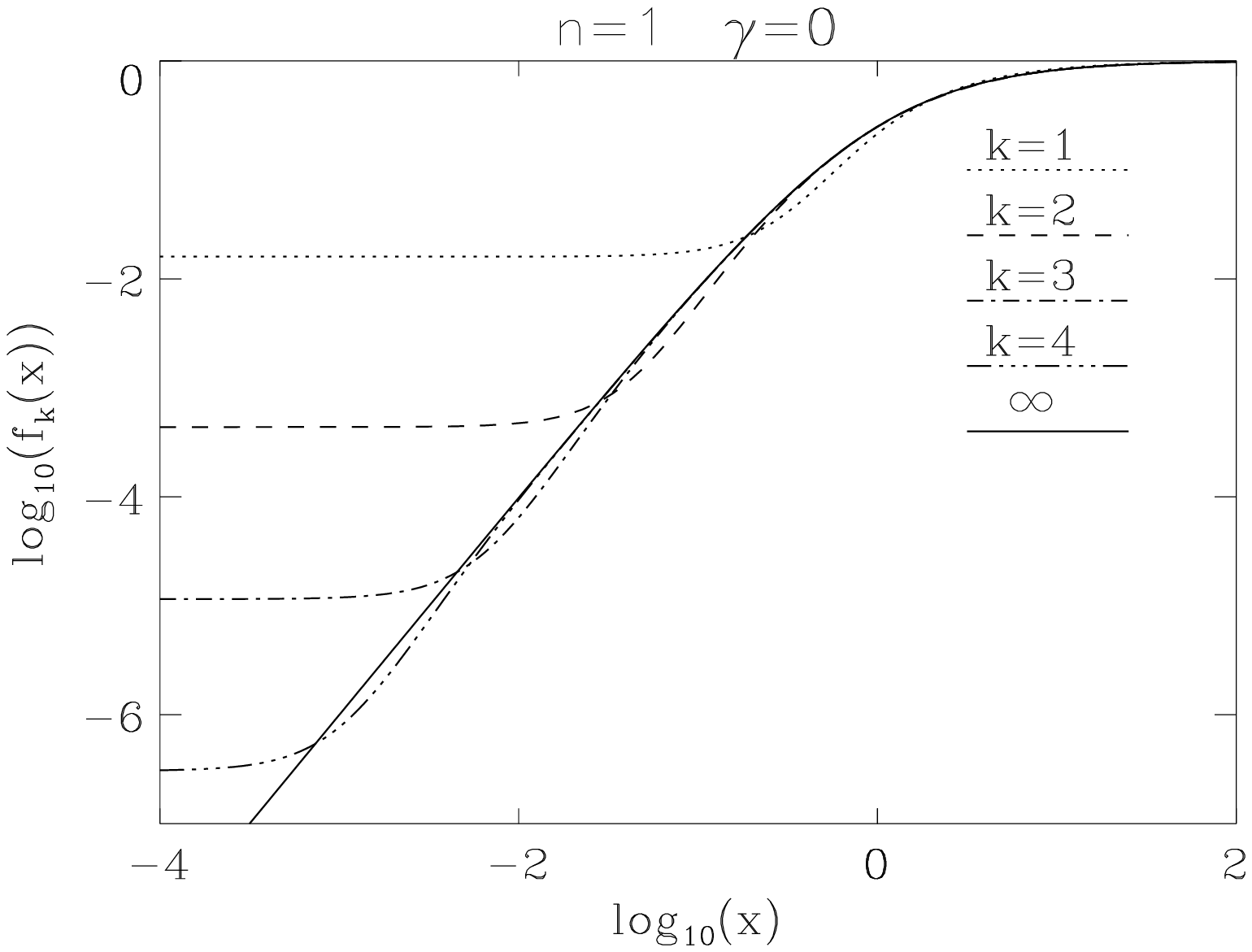
}}\\
Fig.~5a\\
The metric function $f$ is shown 
for the spherically symmetric solutions ($n=1$)
of EYM theory with $k=1-4$.
Also shown is the metric function of the limiting RN solution.
\end{figure} 

\newpage
\begin{figure}
\centering
\epsfysize=12cm
\mbox{\epsffile{
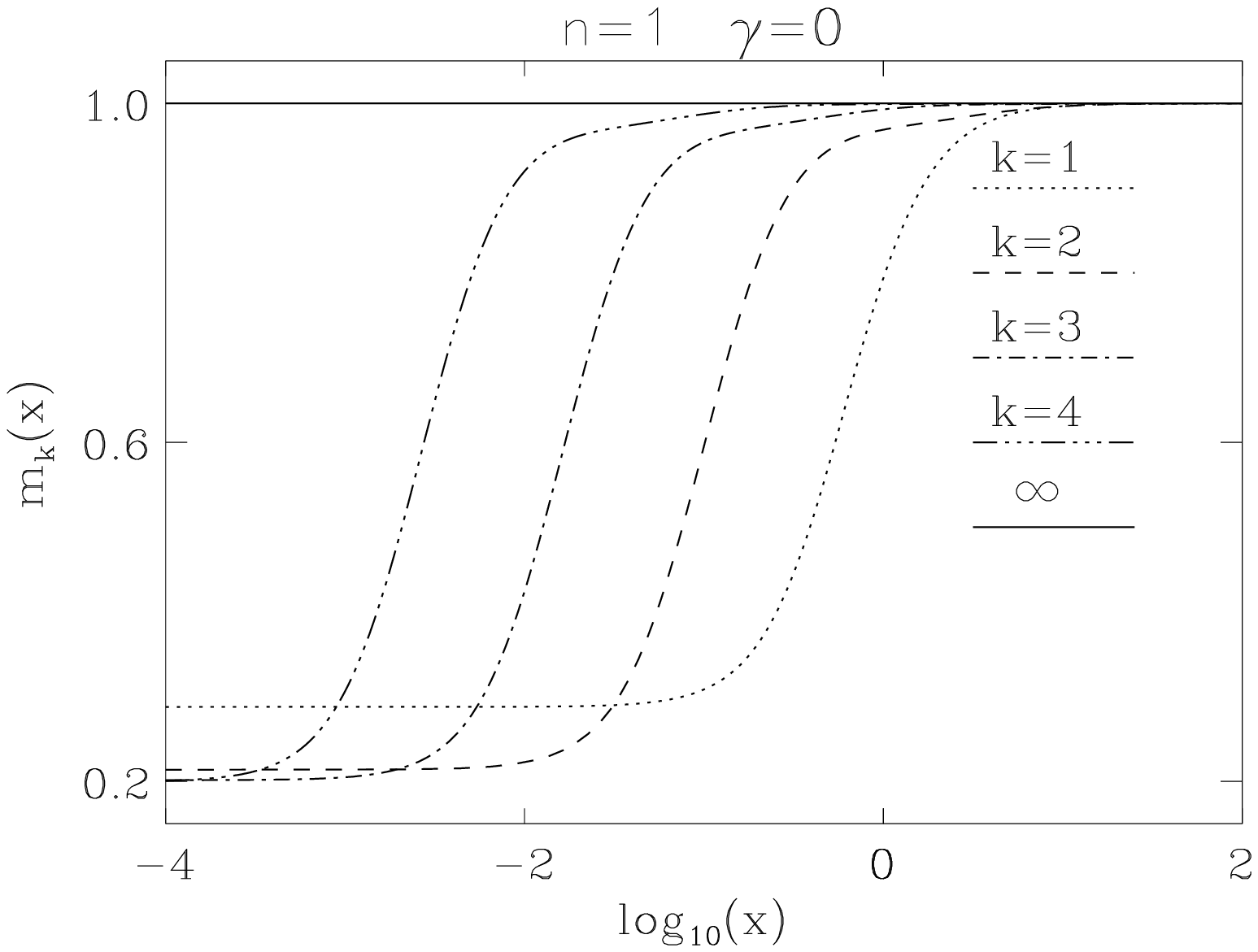
}}\\
Fig.~5b\\
Same as Fig.~5a for the metric function $m$.
\end{figure}
\clearpage

\newpage
\begin{figure}
\centering
\epsfysize=12cm
\mbox{\epsffile{
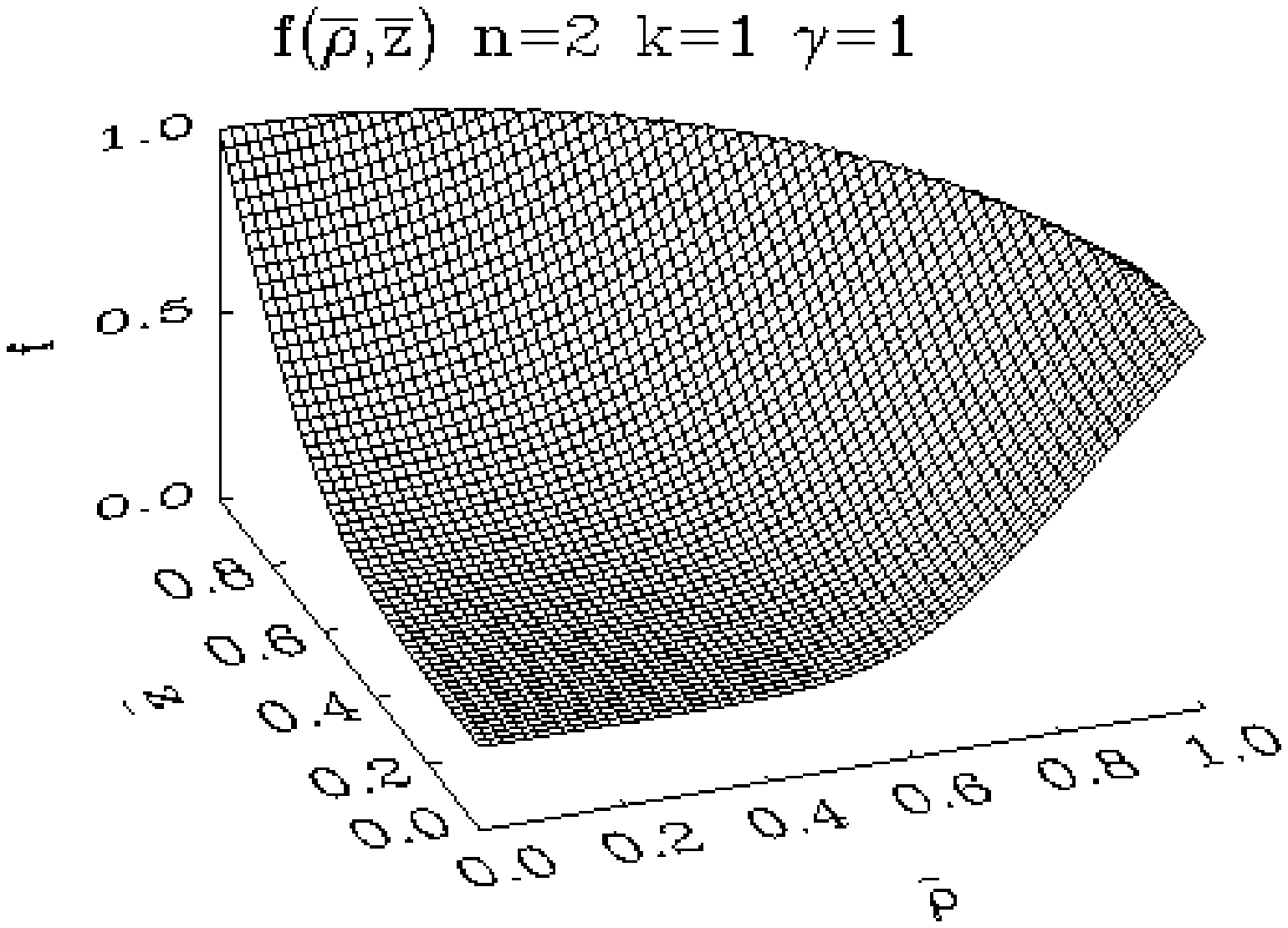
}}\\
Fig.~6a\\
The metric function $f$ of the EYMD solution with $\gamma=1$, $n=2$,
$k=1$ is shown as a function of the compactified coordinates
$\bar \rho$ and $\bar z$.
\end{figure} 

\newpage
\begin{figure}
\centering
\epsfysize=12cm
\mbox{\epsffile{
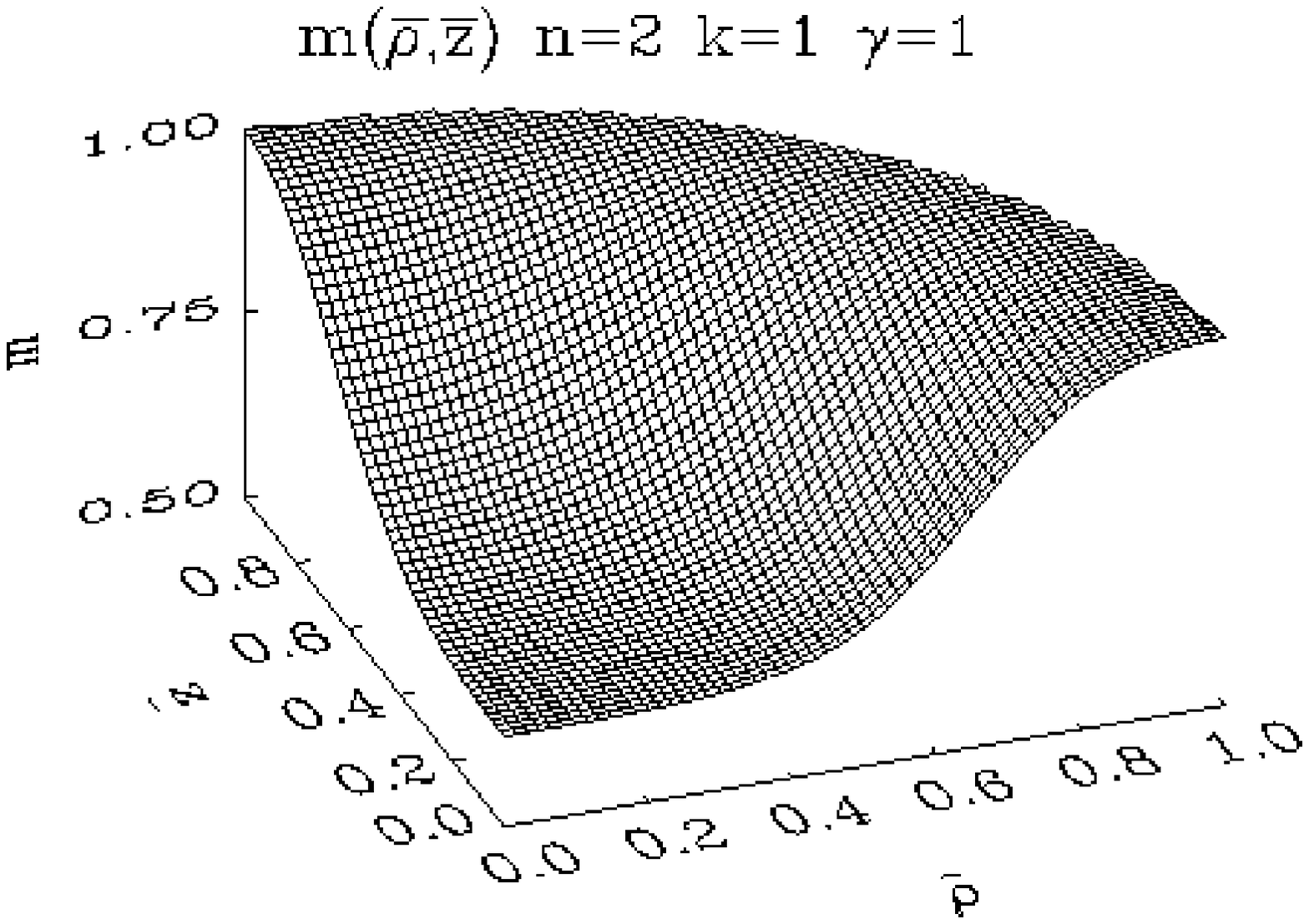
}}\\
Fig.~6b\\
Same as Fig.~6a for the metric function $m$.
\end{figure} 

\newpage
\begin{figure}
\centering
\epsfysize=12cm
\mbox{\epsffile{
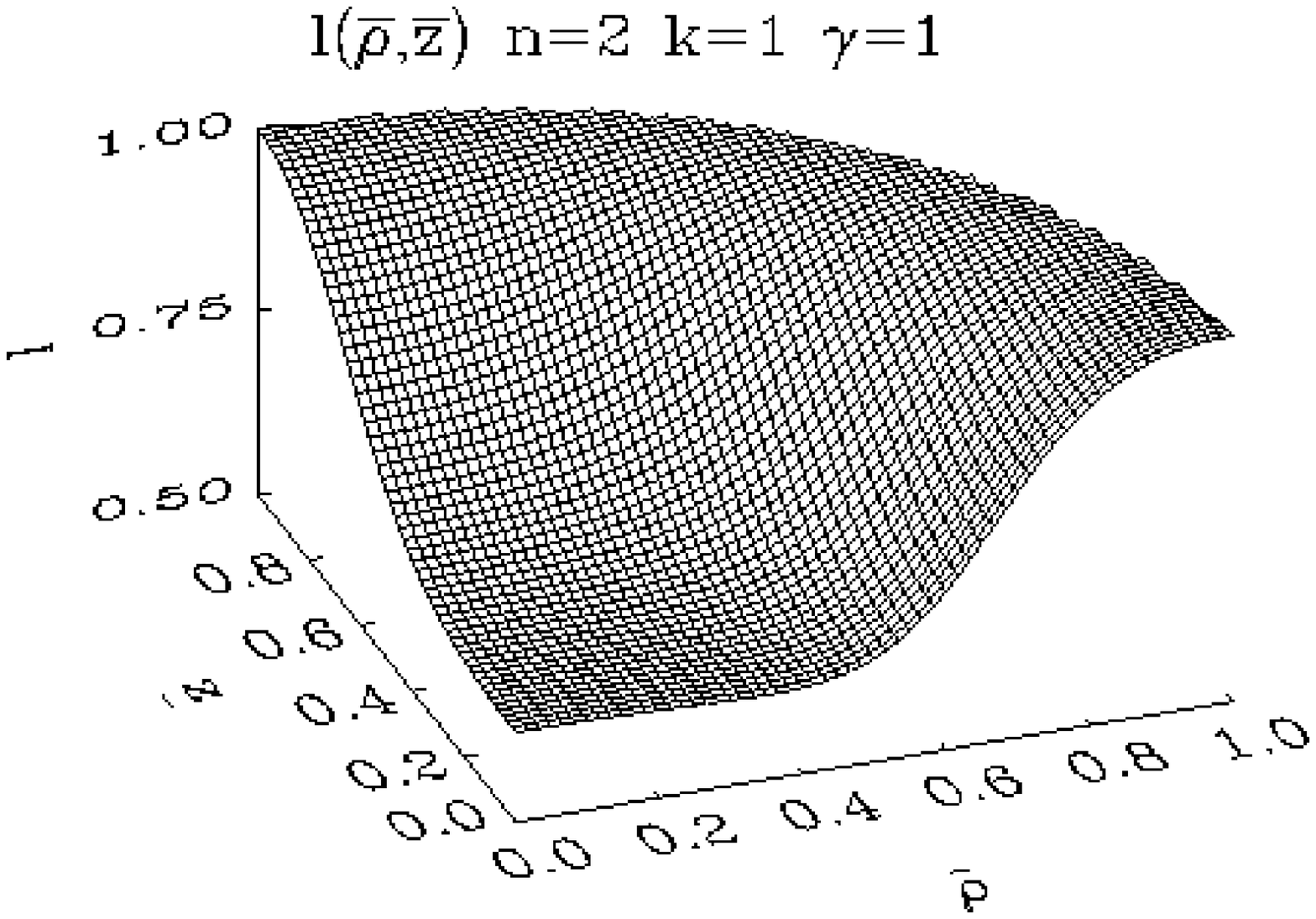
}}\\
Fig.~6c\\
Same as Fig.~6a for the metric function $l$.
\end{figure} 

\newpage
\begin{figure}
\centering
\epsfysize=12cm
\mbox{\epsffile{
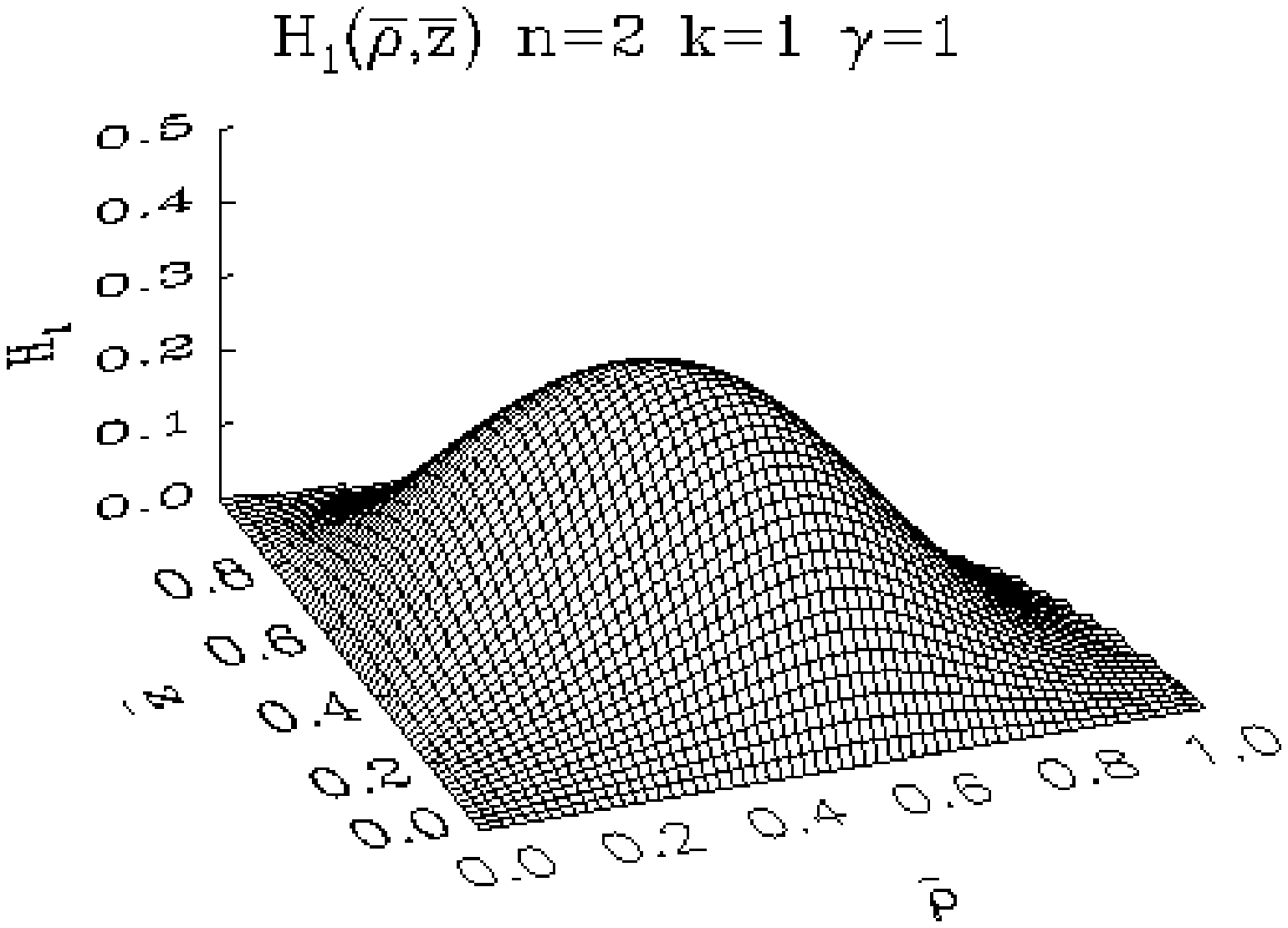
}}\\
Fig.~7a\\
Same as Fig.~6a for the gauge field function $H_1$.
\end{figure} 

\newpage
\begin{figure}
\centering
\epsfysize=12cm
\mbox{\epsffile{
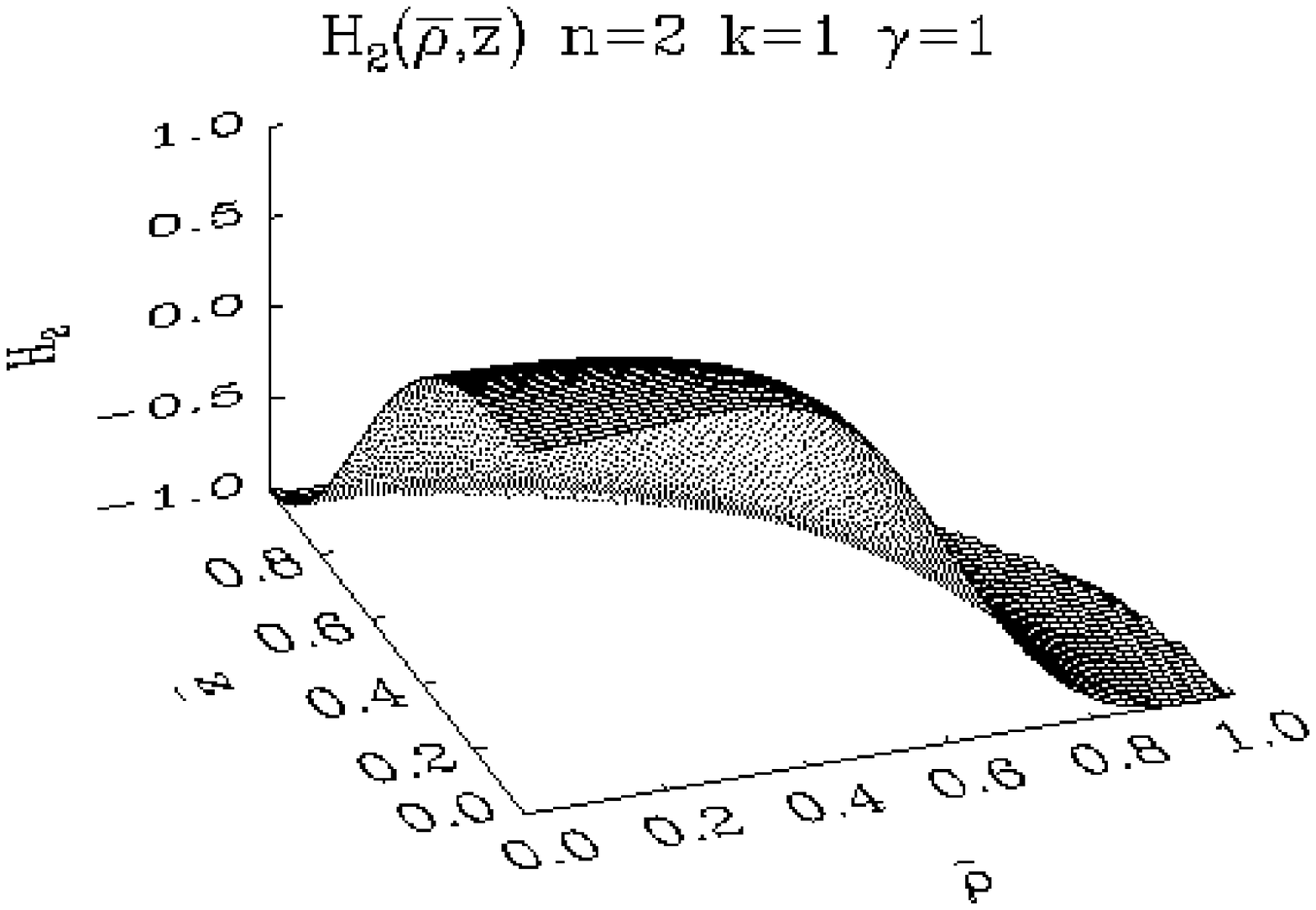
}}\\
Fig.~7b\\
Same as Fig.~6a for the gauge field function $H_2$.
\end{figure}

\newpage
\begin{figure}
\centering
\epsfysize=12cm
\mbox{\epsffile{
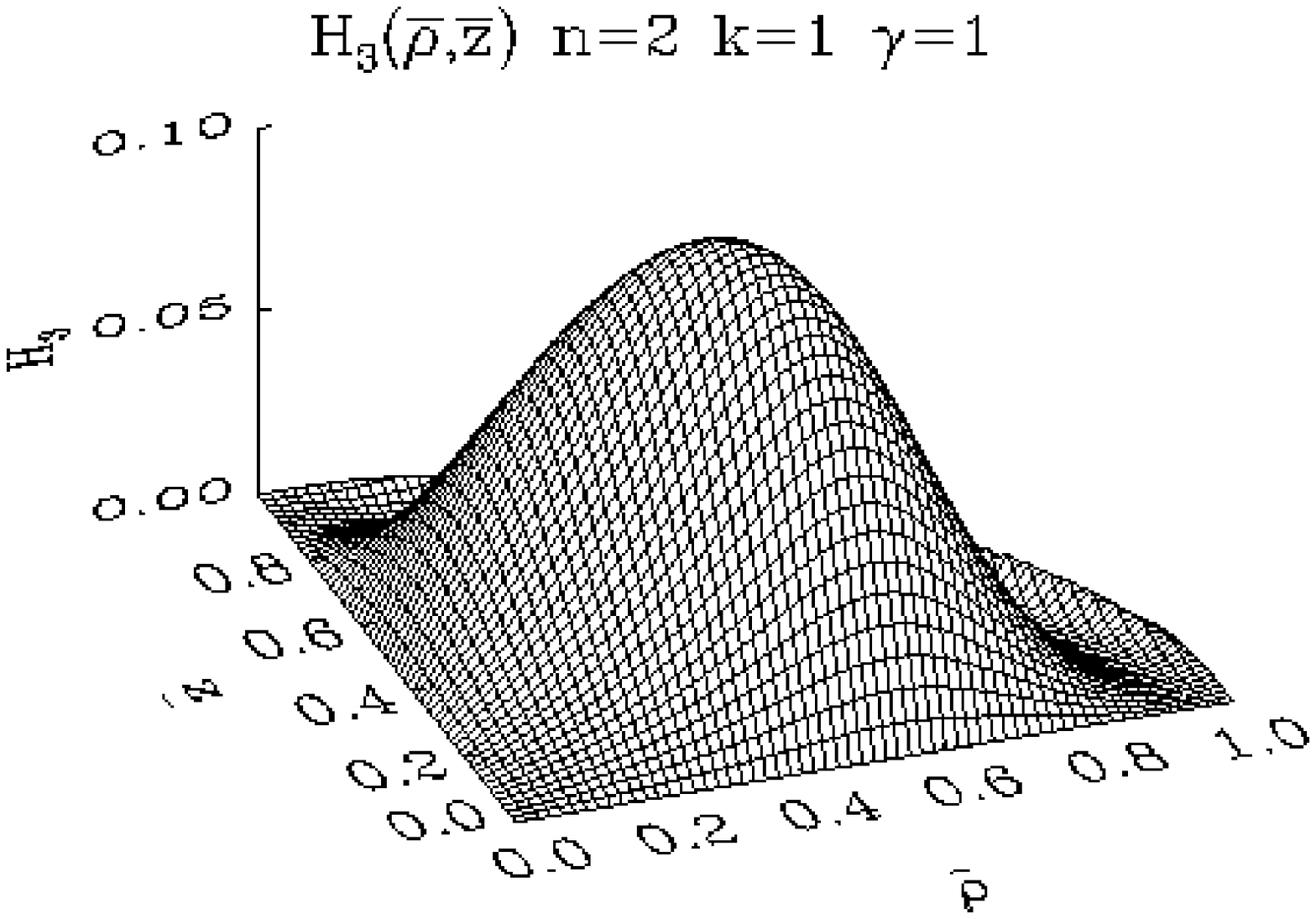
}}\\
Fig.~7c\\
Same as Fig.~6a for the gauge field function $H_3$.
\end{figure}

\newpage
\begin{figure}
\centering
\epsfysize=12cm
\mbox{\epsffile{
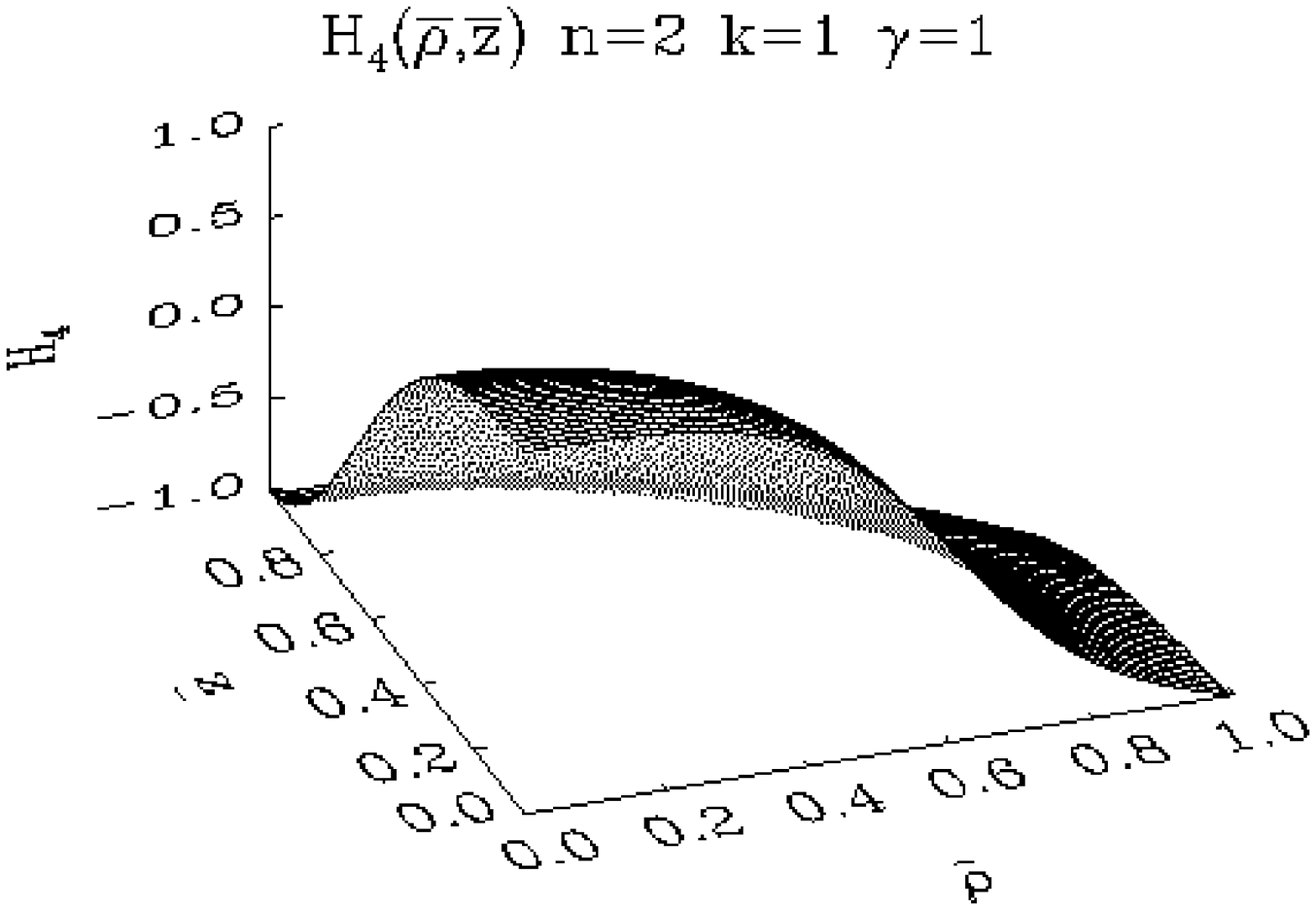
}}\\
Fig.~7d\\
Same as Fig.~6a for the gauge field function $H_4$.
\end{figure}

\newpage
\begin{figure}
\centering
\epsfysize=12cm
\mbox{\epsffile{
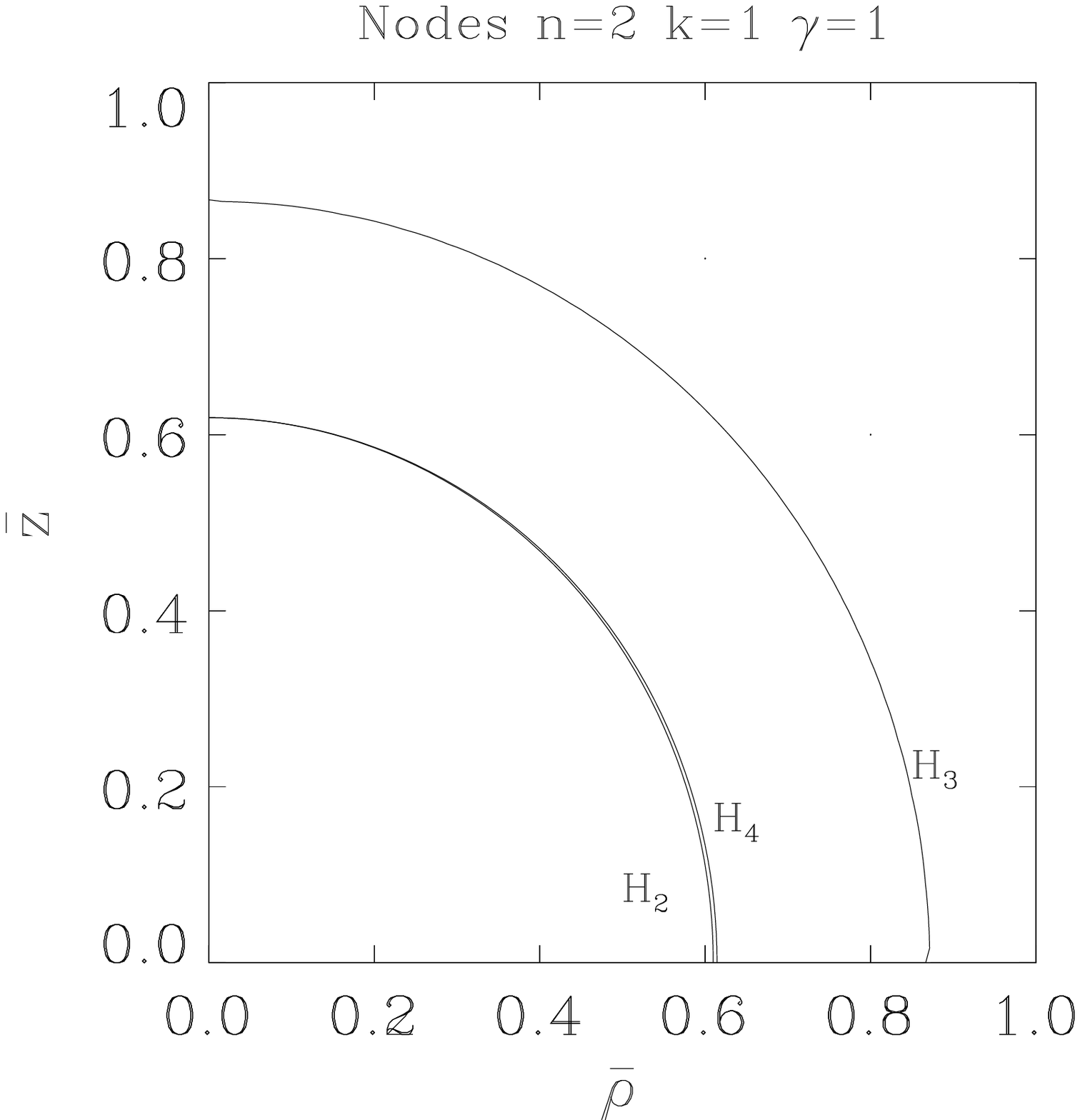
}}\\
Fig.~7e\\
Same as Fig.~6a for the nodal lines of the gauge field functions
$H_2-H_4$.
\end{figure}
\clearpage

\newpage
\begin{figure}
\centering
\epsfysize=12cm
\mbox{\epsffile{
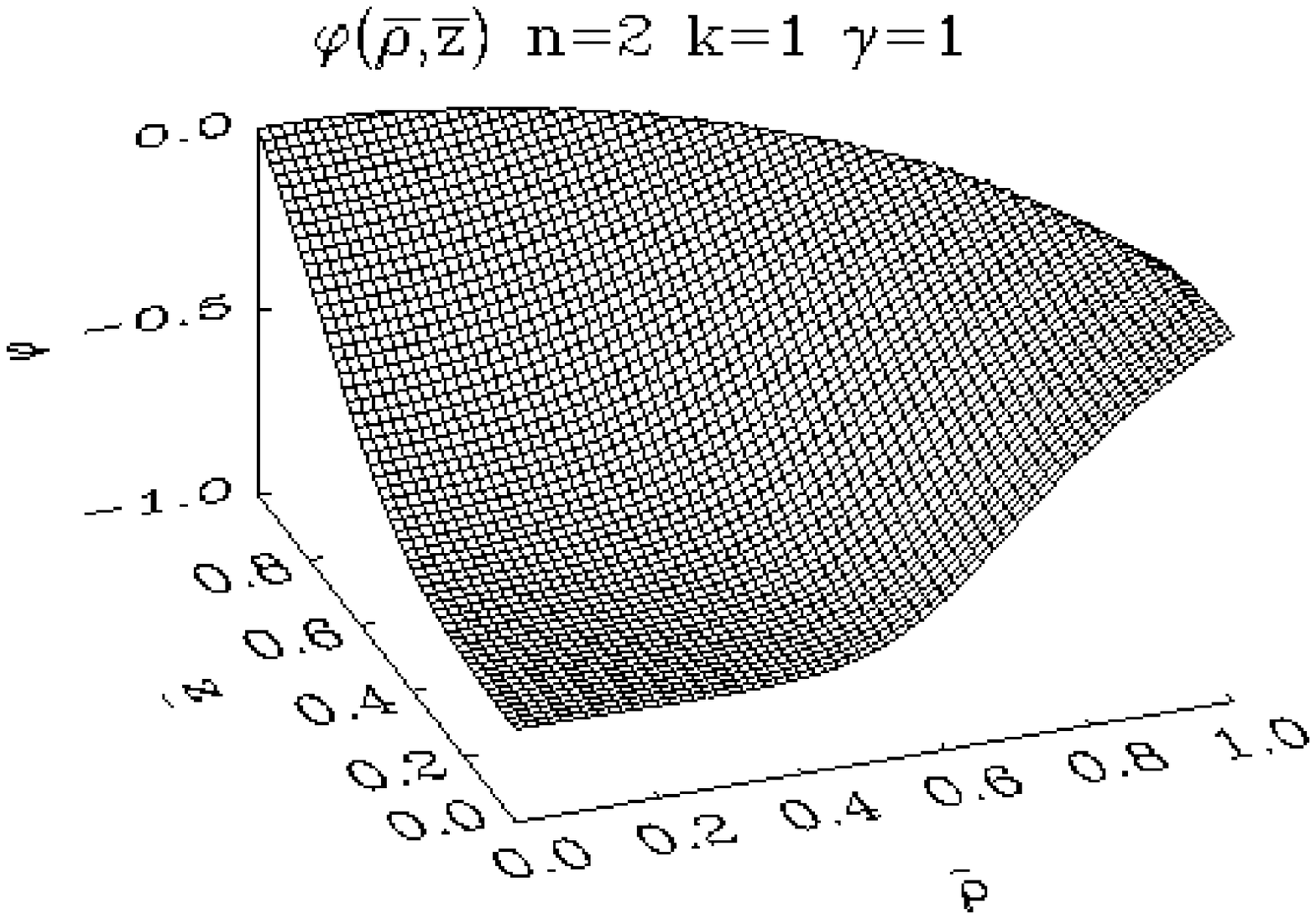
}}\\
Fig.~8\\
Same as Fig.~6a for the dilaton function $\varphi$.
\end{figure}

\newpage
\begin{figure}
\centering
\epsfysize=12cm
\mbox{\epsffile{
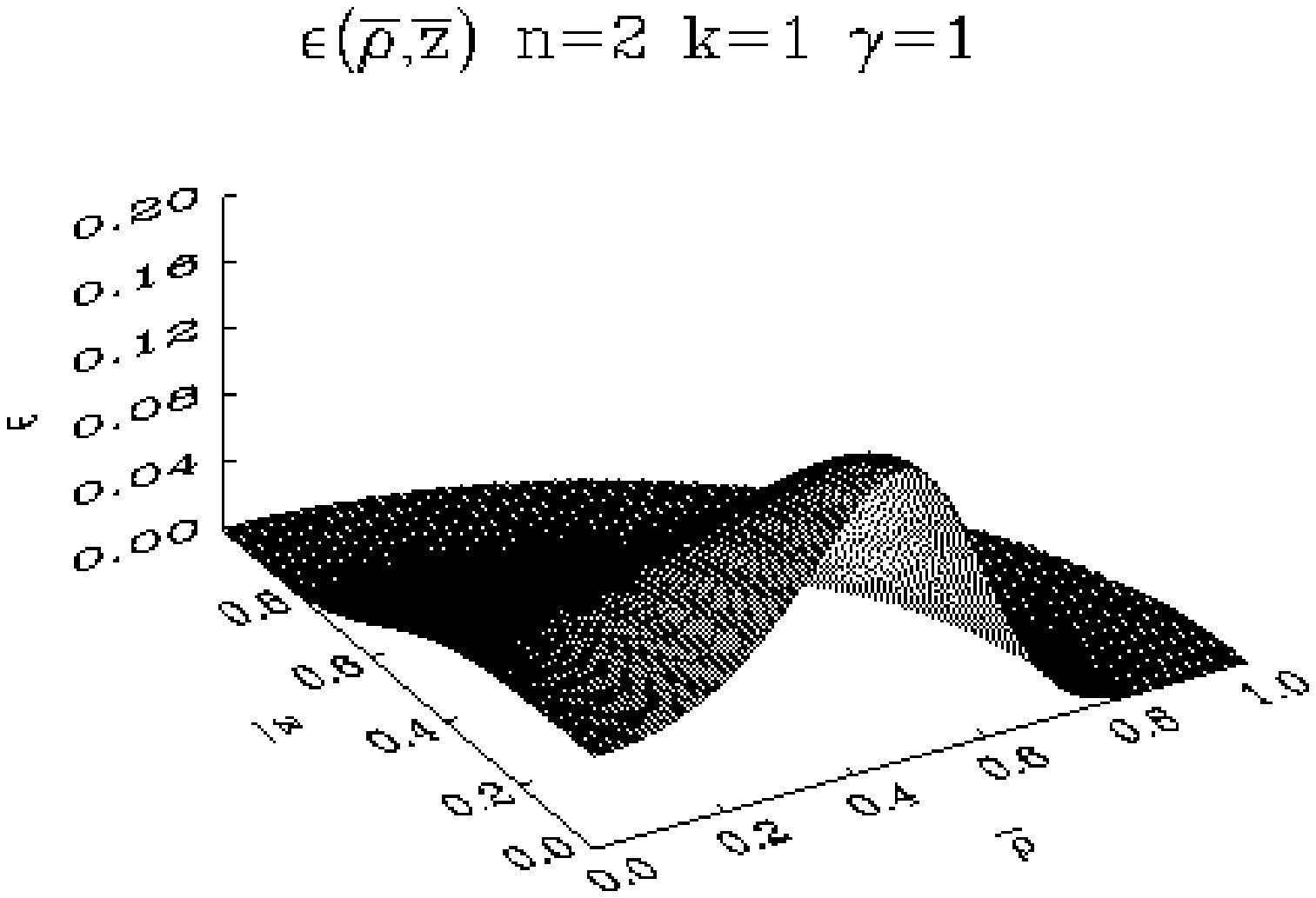
}}\\
Fig.~9\\
Same as Fig.~6a for the energy density of the matter fields $\epsilon$.
\end{figure}

\newpage
\begin{figure}
\centering
\epsfysize=12cm
\mbox{\epsffile{
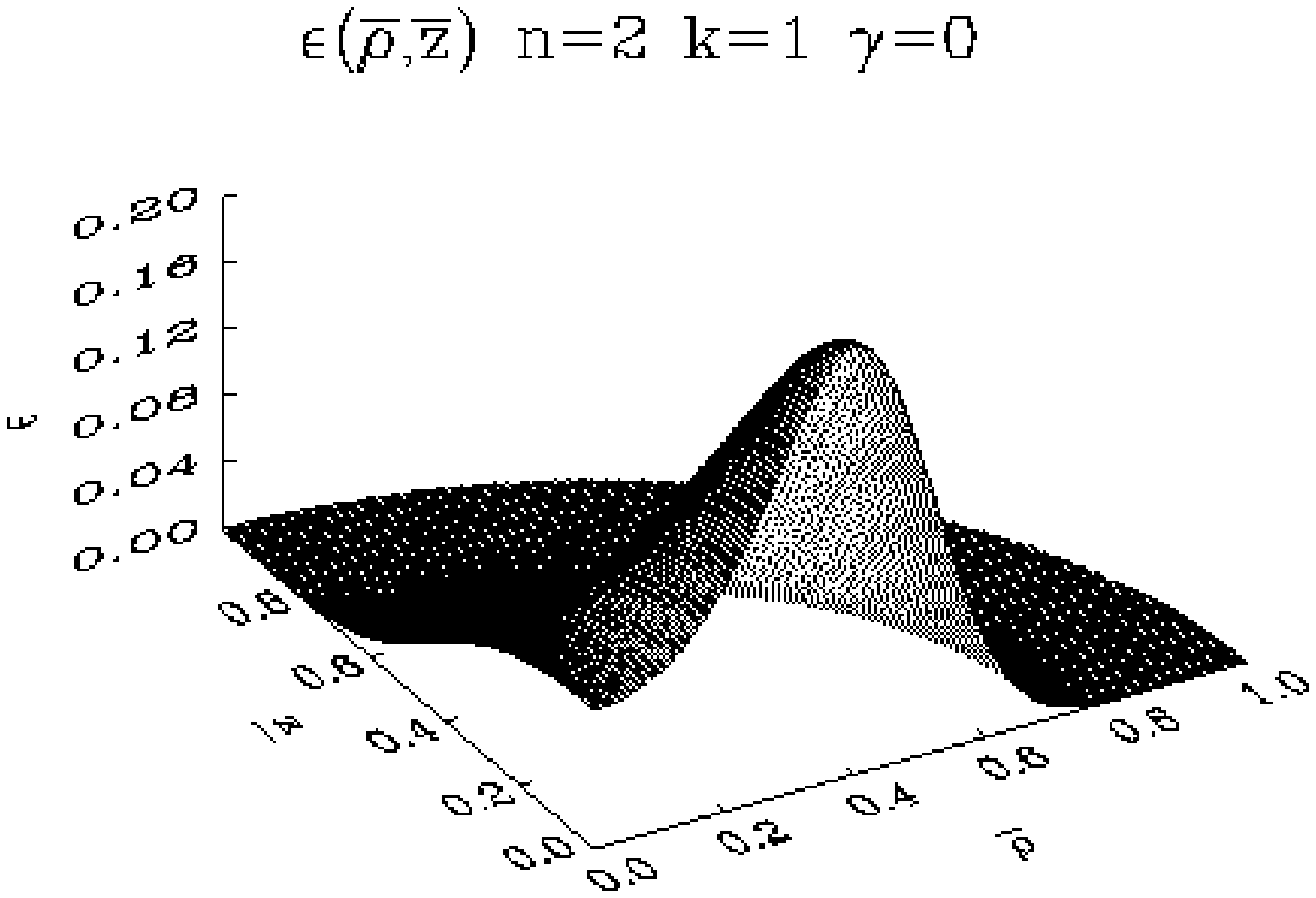
}}\\
Fig.~10\\
The energy density of the matter fields $\epsilon$
of the EYM solution with $n=2$,
$k=1$ is shown as a function of the compactified coordinates
$\bar \rho$ and $\bar z$.
\end{figure}
\clearpage

\newpage
\begin{figure}
\centering
\epsfysize=12cm
\mbox{\epsffile{
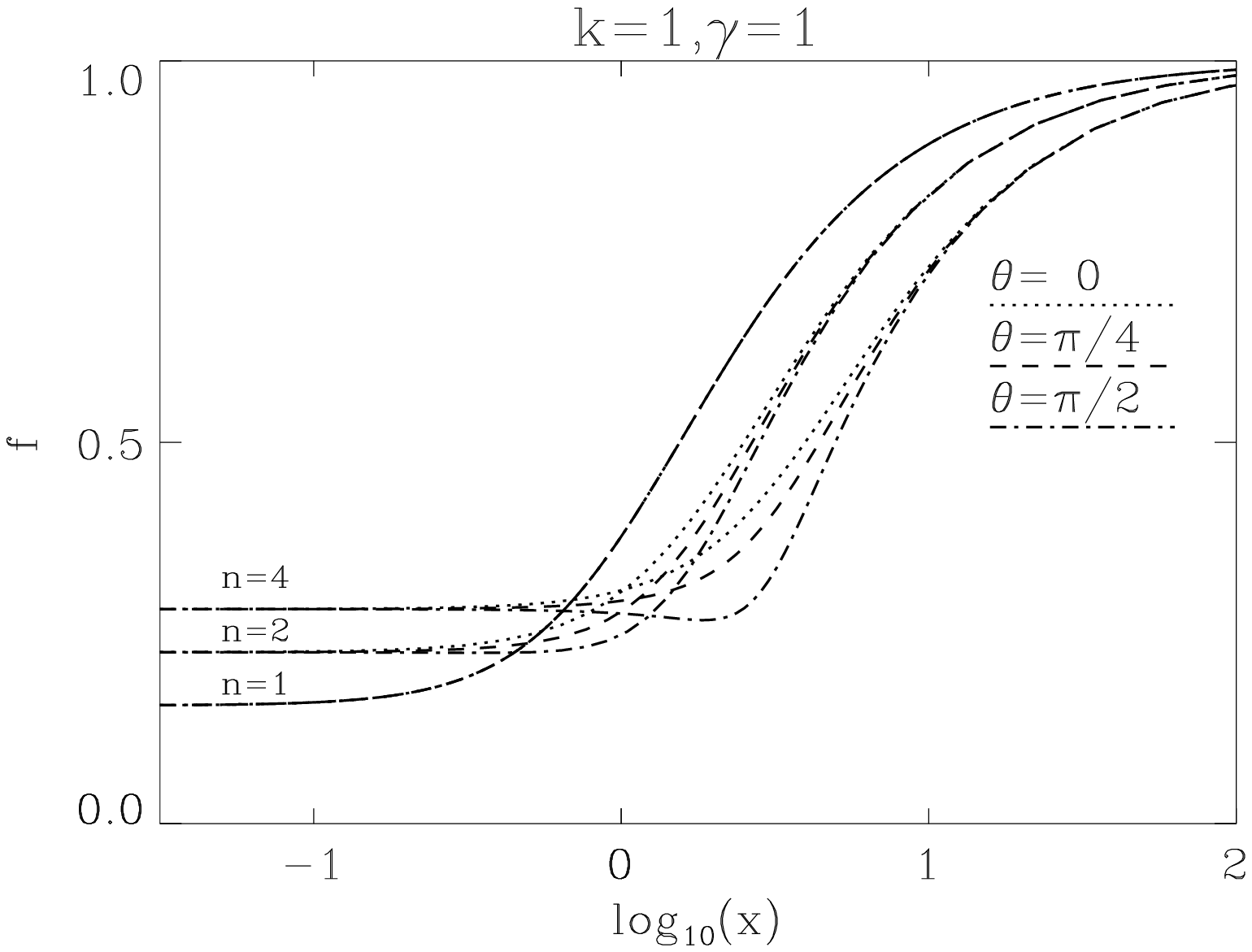
}}\\
Fig.~11a\\
The metric function $f$ of the EYMD solutions with $\gamma=1$, $n=1$,2,
and 4 and $k=1$ is shown as a function of the coordinate $x$
for the angles $\theta=0$, $\pi/4$ and $\pi/2$.
\end{figure}
\clearpage

\newpage
\begin{figure}
\centering
\epsfysize=12cm
\mbox{\epsffile{
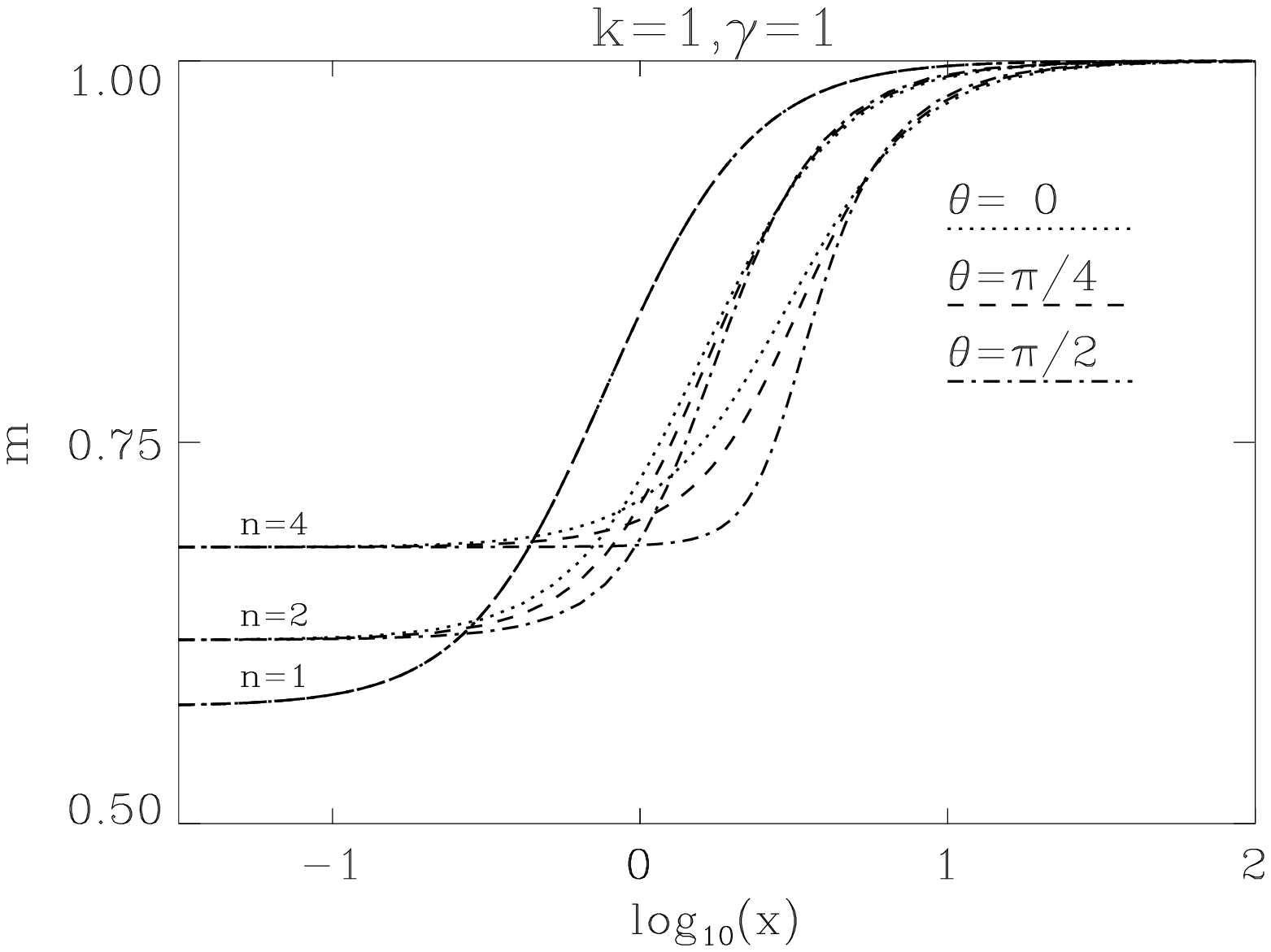
}}\\
Fig.~11b\\
Same as Fig.~11a for the metric function $m$.
\end{figure}
\clearpage

\newpage
\begin{figure}
\centering
\epsfysize=12cm
\mbox{\epsffile{
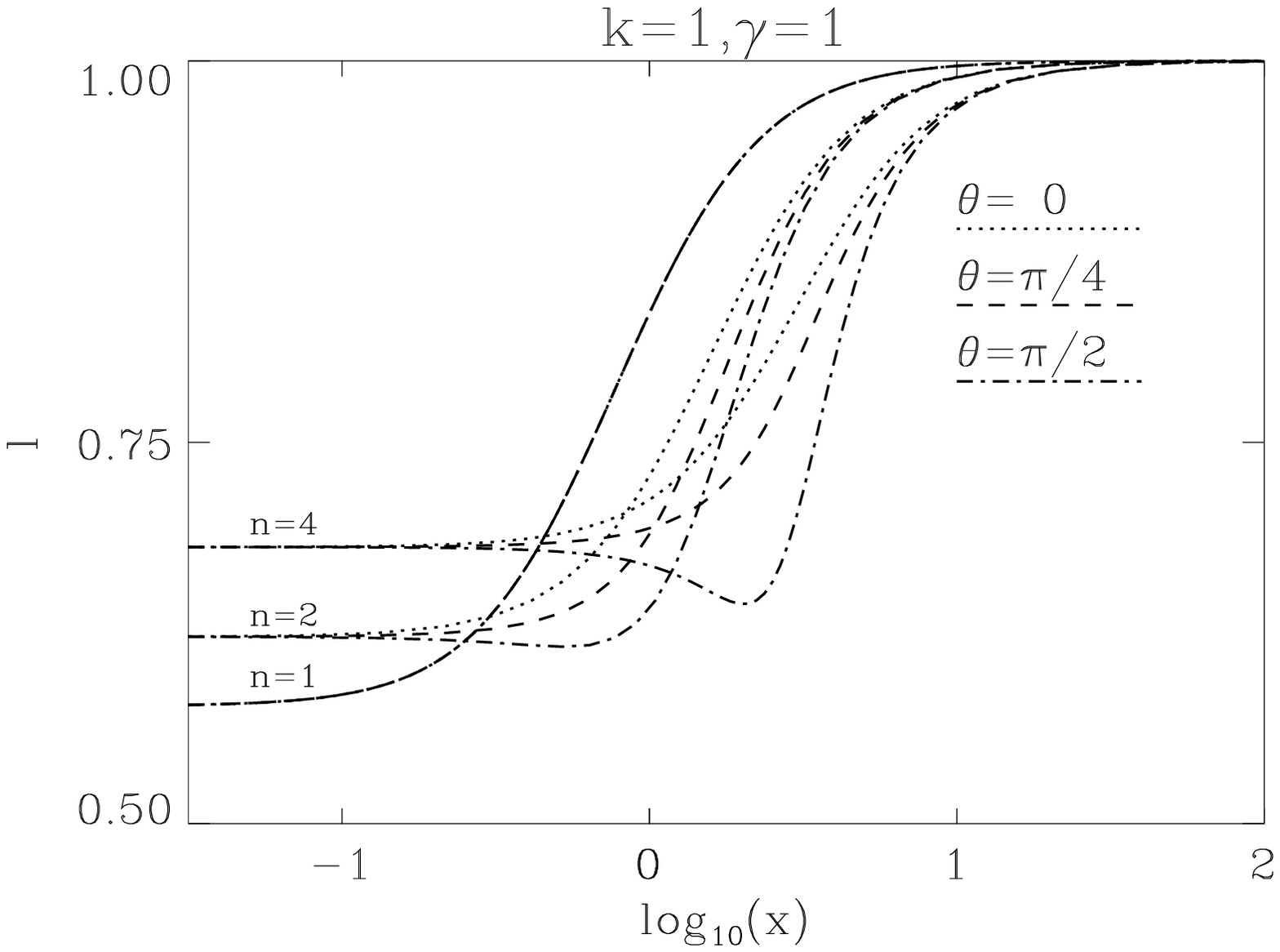
}}\\
Fig.~11c\\
Same as Fig.~11a for the metric function $l$.
\end{figure}
\clearpage

\newpage
\begin{figure}
\centering
\epsfysize=12cm
\mbox{\epsffile{
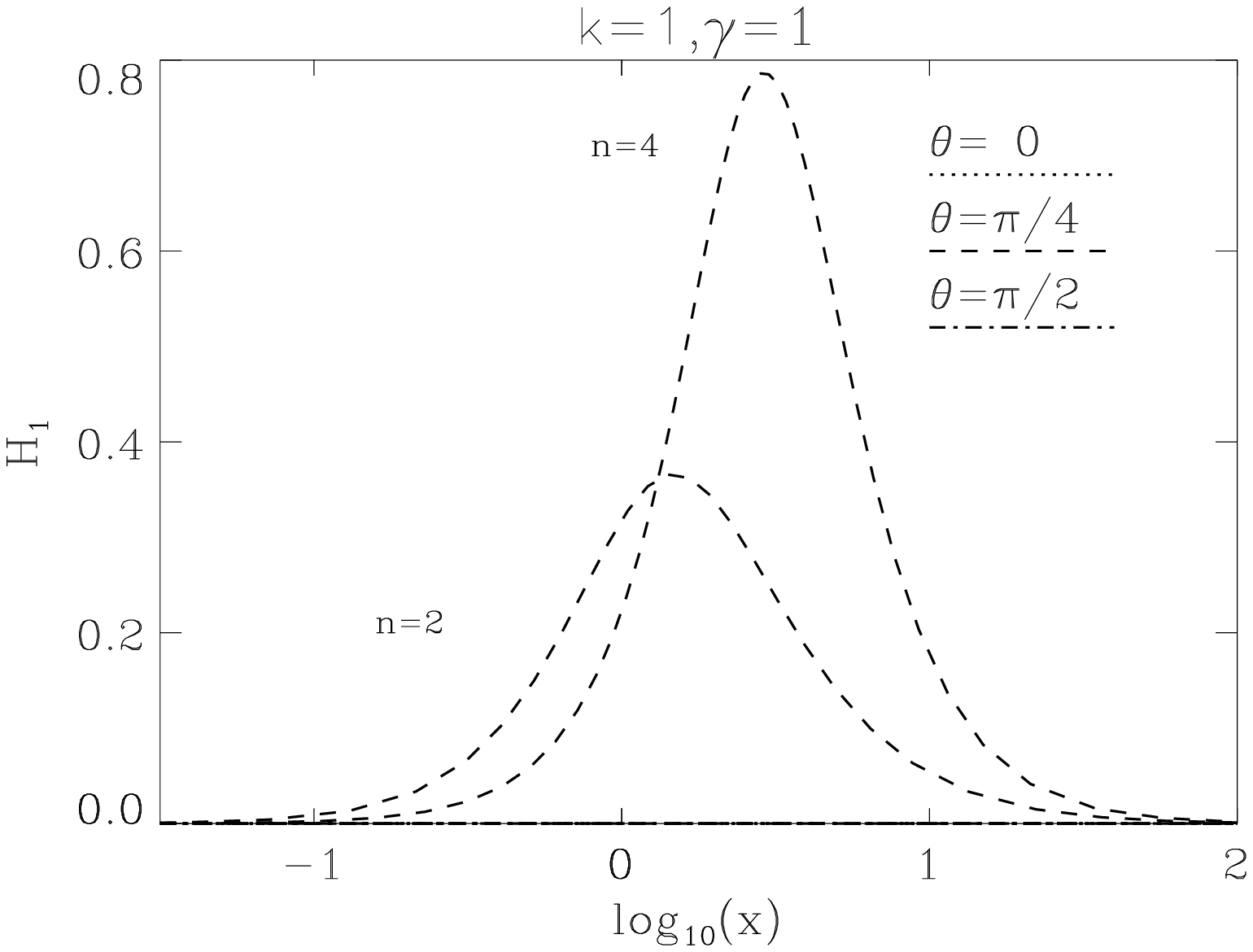
}}\\
Fig.~12a\\
Same as Fig.~11a for the gauge field function $H_1$.
\end{figure}
\clearpage

\newpage
\begin{figure}
\centering
\epsfysize=12cm
\mbox{\epsffile{
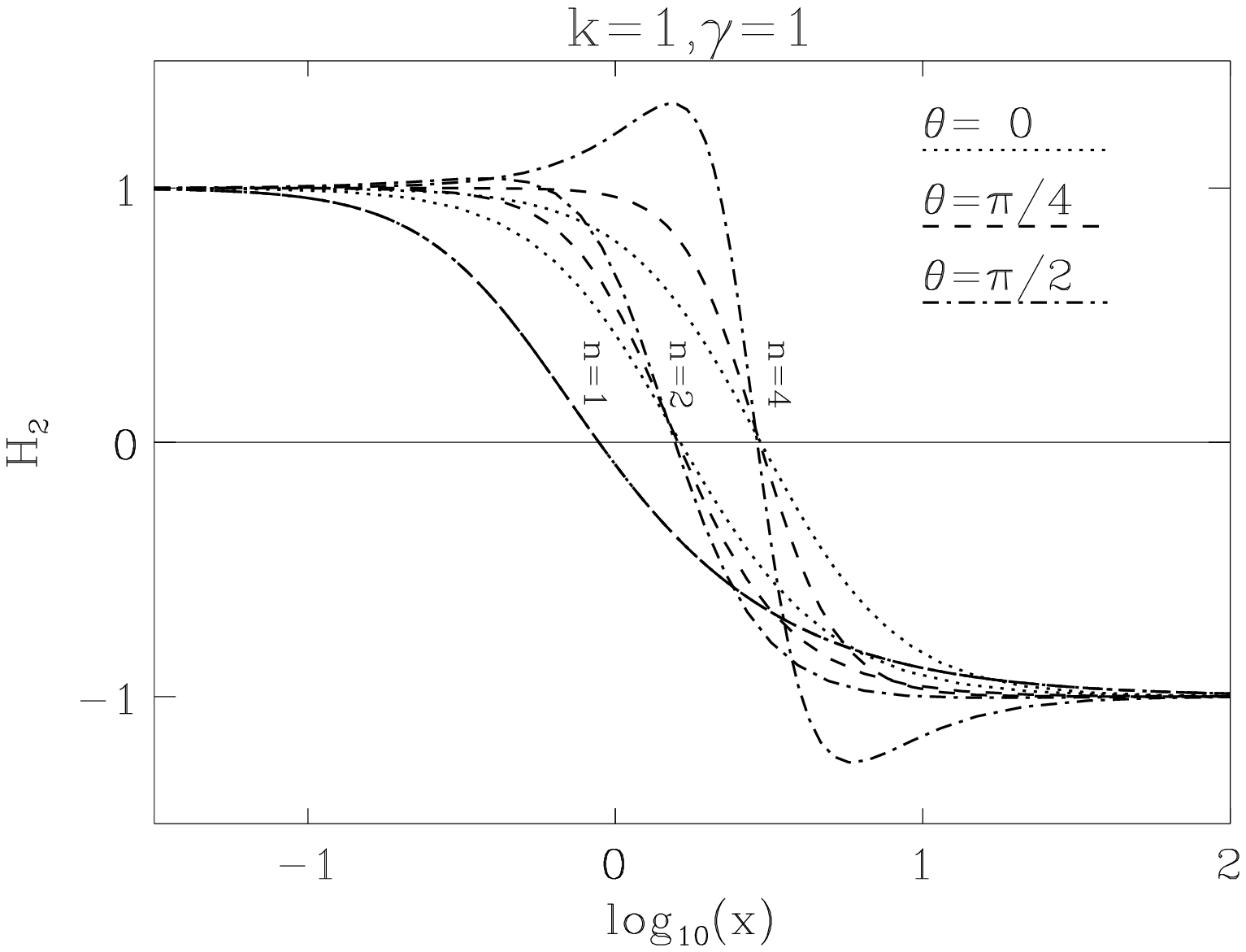
}}\\
Fig.~12b\\
Same as Fig.~11a for the gauge field function $H_2$.
\end{figure}
\clearpage

\newpage
\begin{figure}
\centering
\epsfysize=12cm
\mbox{\epsffile{ 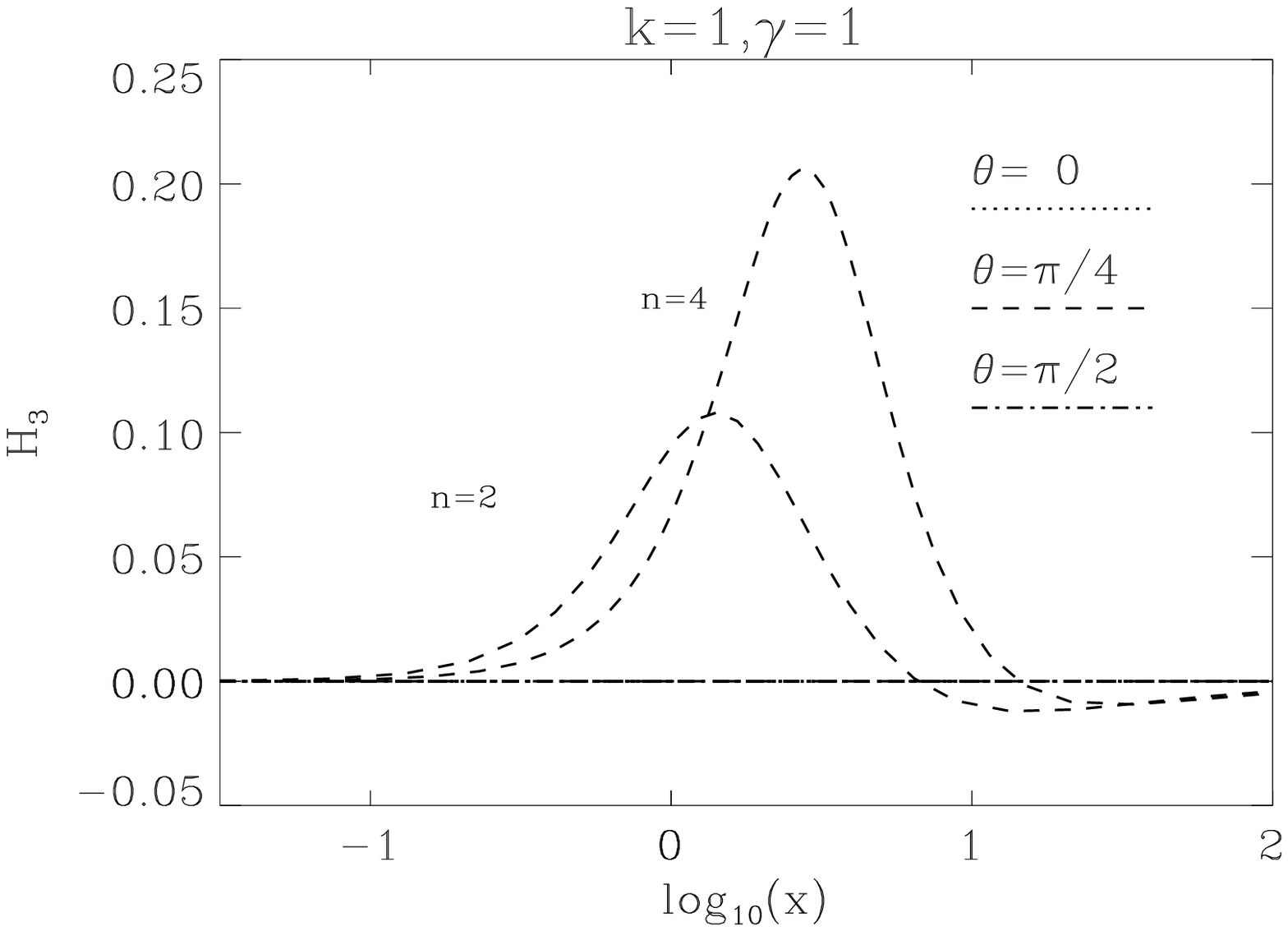 }}\\ 
Fig.~12c\\ 
Same as Fig.~11a for the gauge field function $H_3$.  
\end{figure} 
\clearpage

\newpage
\begin{figure}
\centering
\epsfysize=12cm
\mbox{\epsffile{
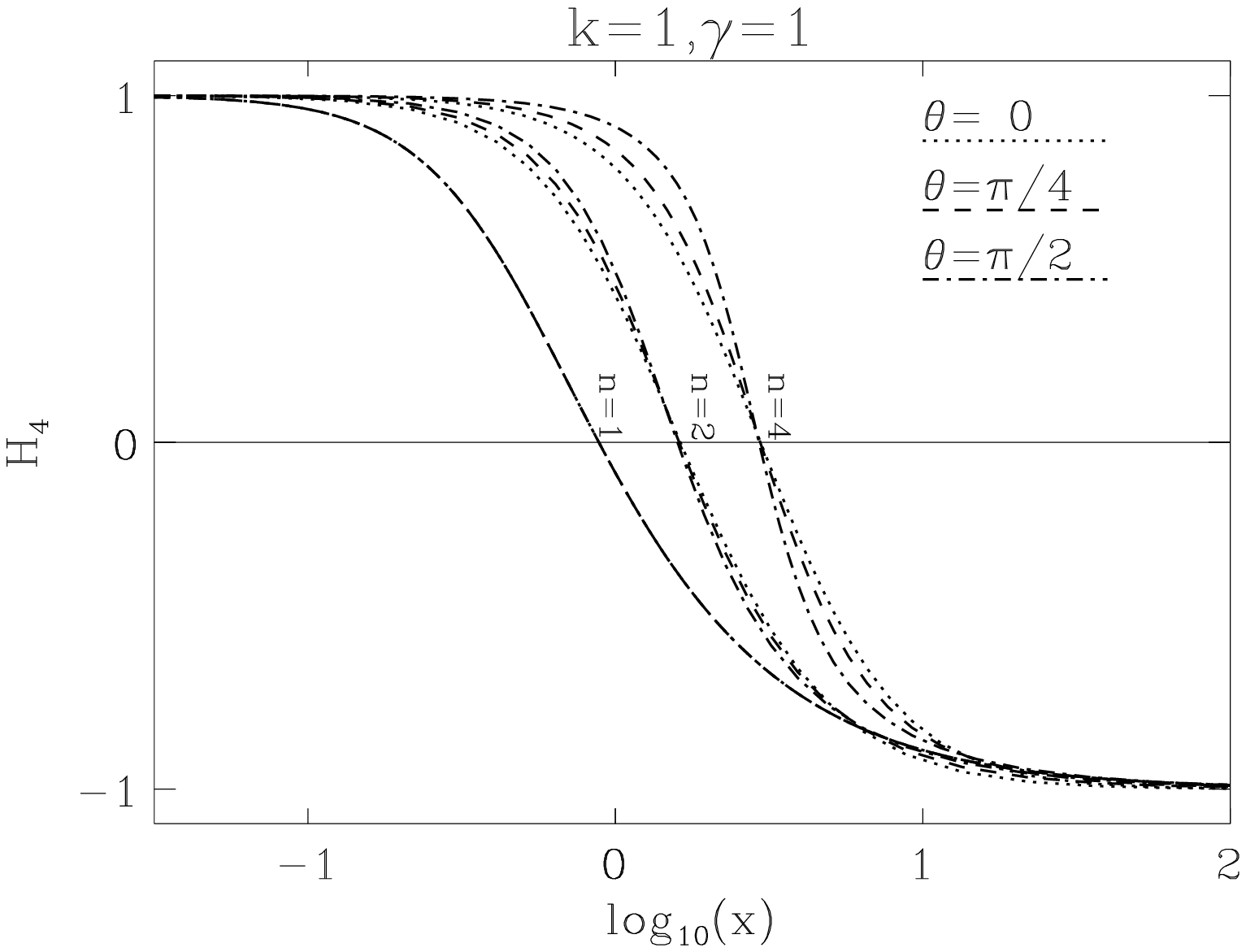
}}\\
Fig.~12d\\
Same as Fig.~11a for the gauge field function $H_4$.
\end{figure}
\clearpage

\newpage
\begin{figure}
\centering
\epsfysize=12cm
\mbox{\epsffile{
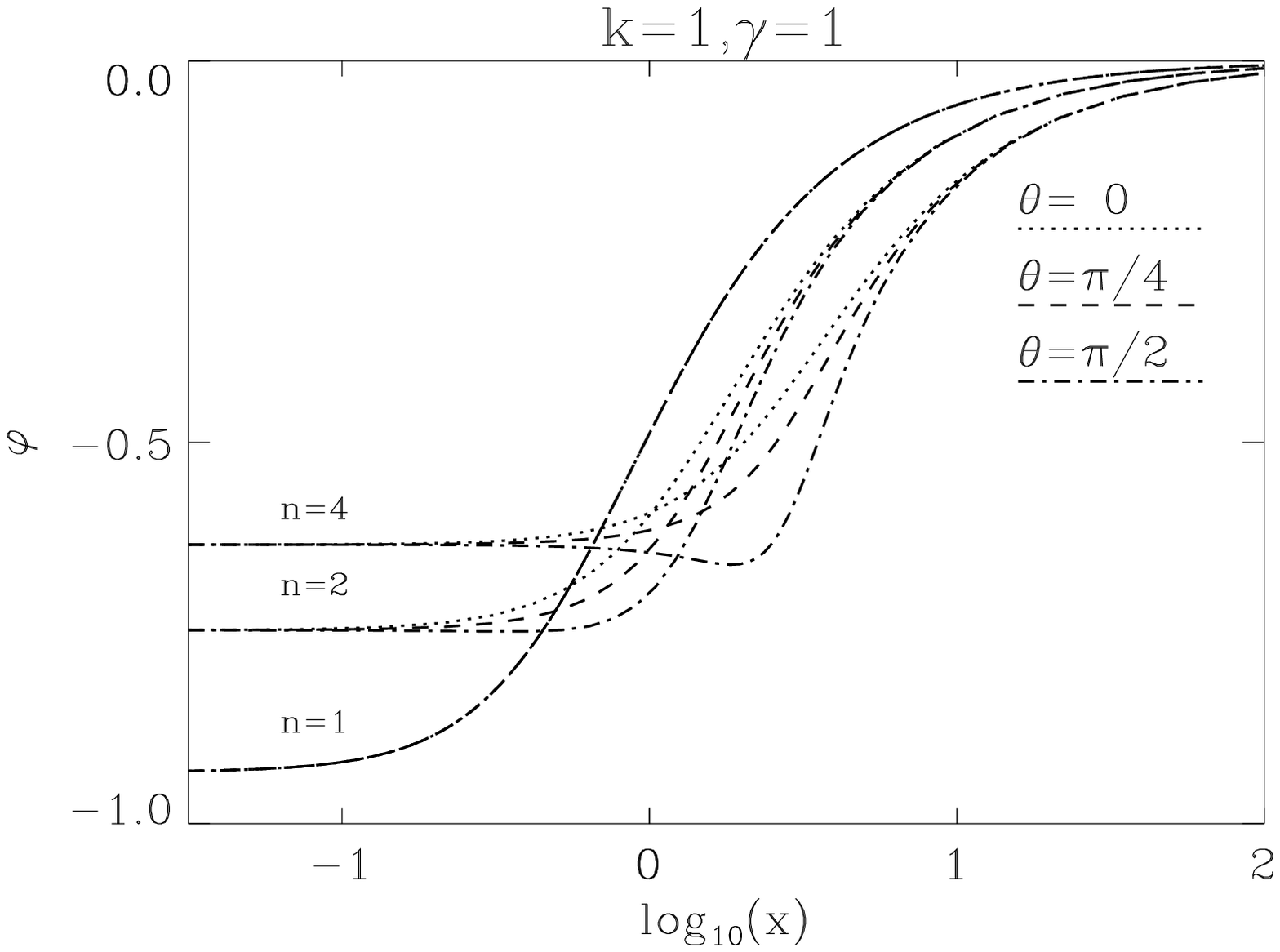
}}\\
Fig.~13\\
Same as Fig.~11a for the dilaton function $\varphi$.
\end{figure}
\clearpage

\newpage
\begin{figure}
\centering
\epsfysize=12cm
\mbox{\epsffile{
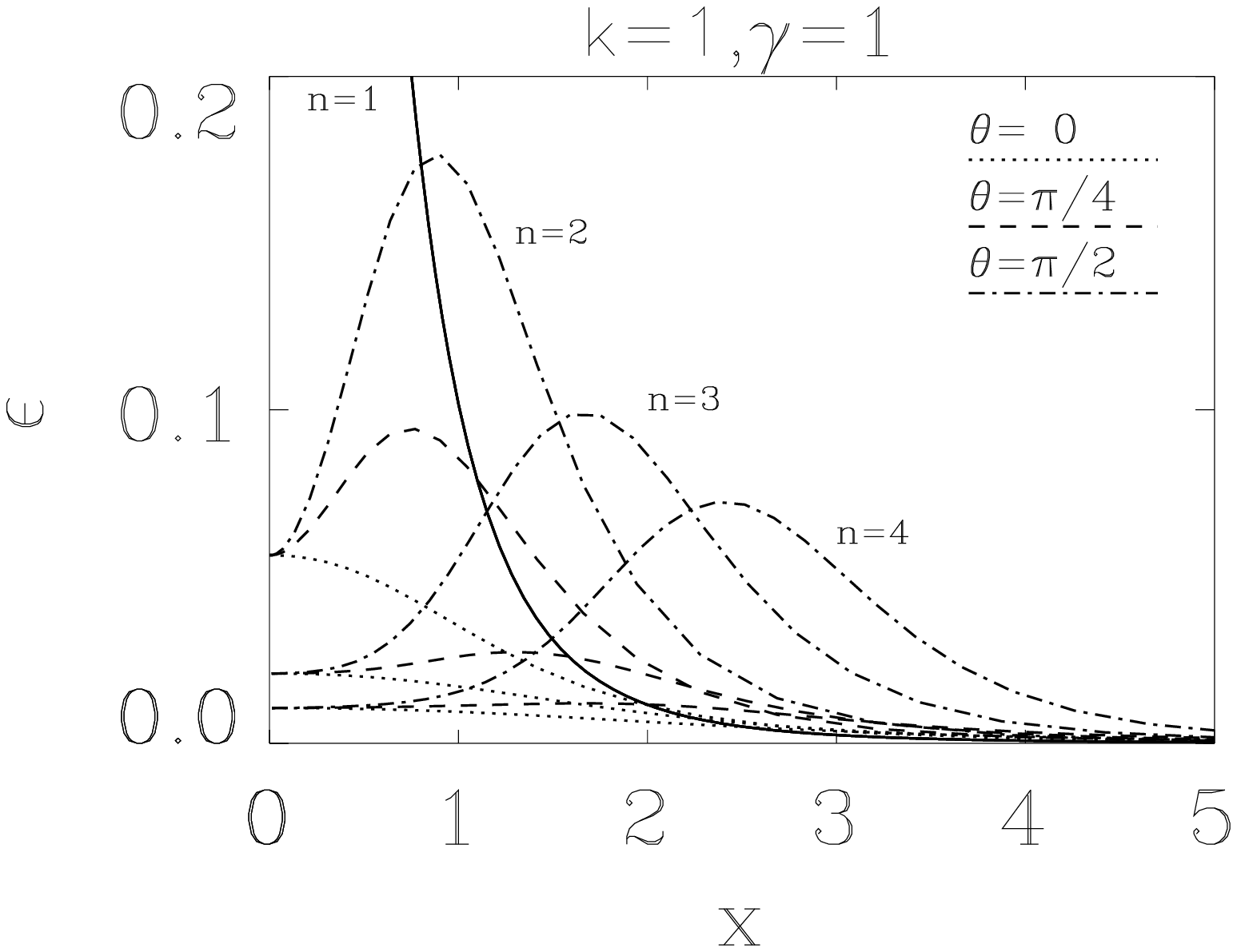
}}\\
Fig.~14\\
Same as Fig.~11a for the energy density of the matter fields
$\epsilon$.
\end{figure}
\clearpage
 
\newpage
\begin{figure}
\centering
\epsfysize=12cm
\mbox{\epsffile{
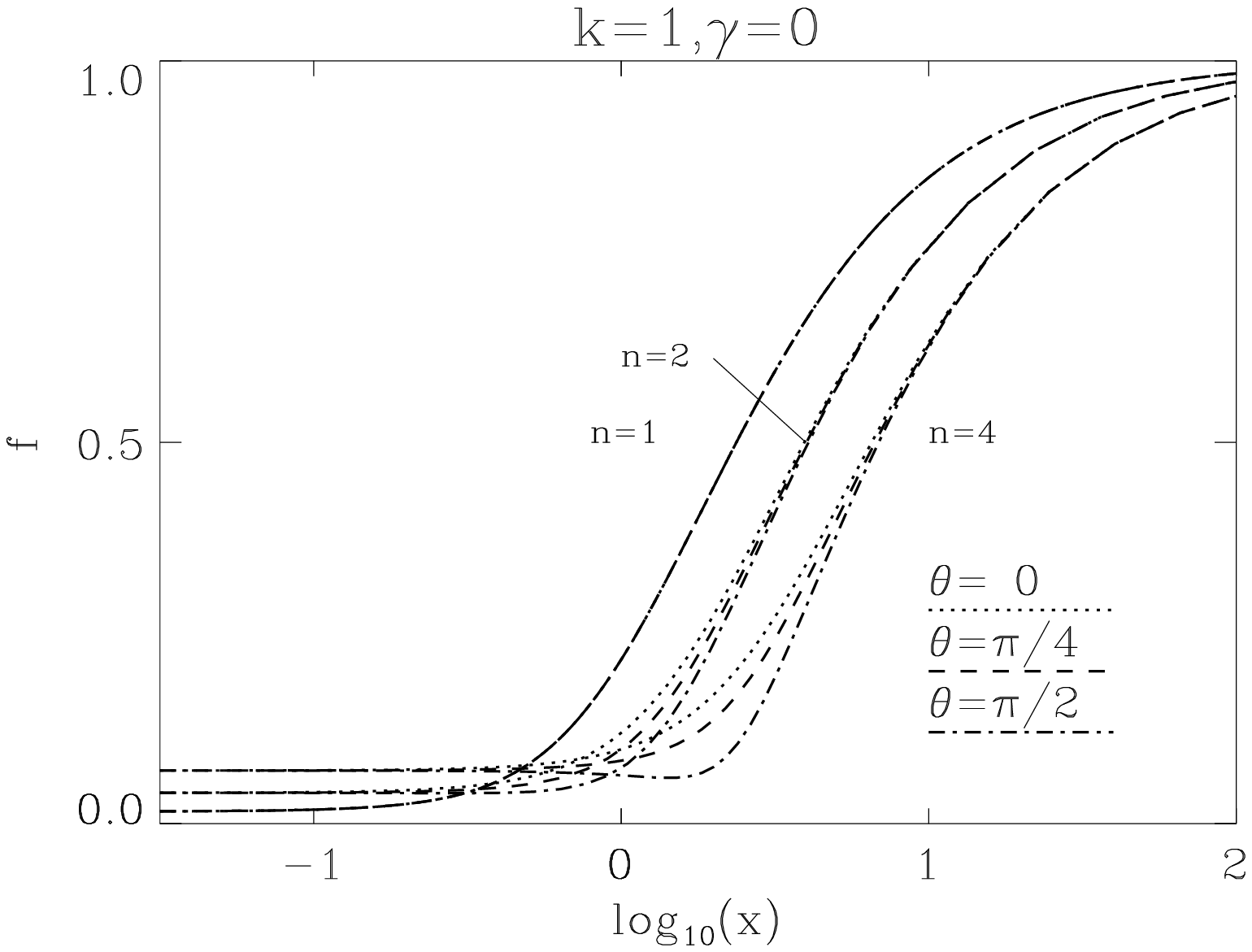
}}\\
Fig.~15a\\
The metric function $f$ of the EYM solutions with $n=1$,2,
and 4 and $k=1$ is shown as a function of the coordinate $x$
for the angles $\theta=0$, $\pi/4$ and $\pi/2$.
\end{figure}
\clearpage

\newpage
\begin{figure}
\centering
\epsfysize=12cm
\mbox{\epsffile{
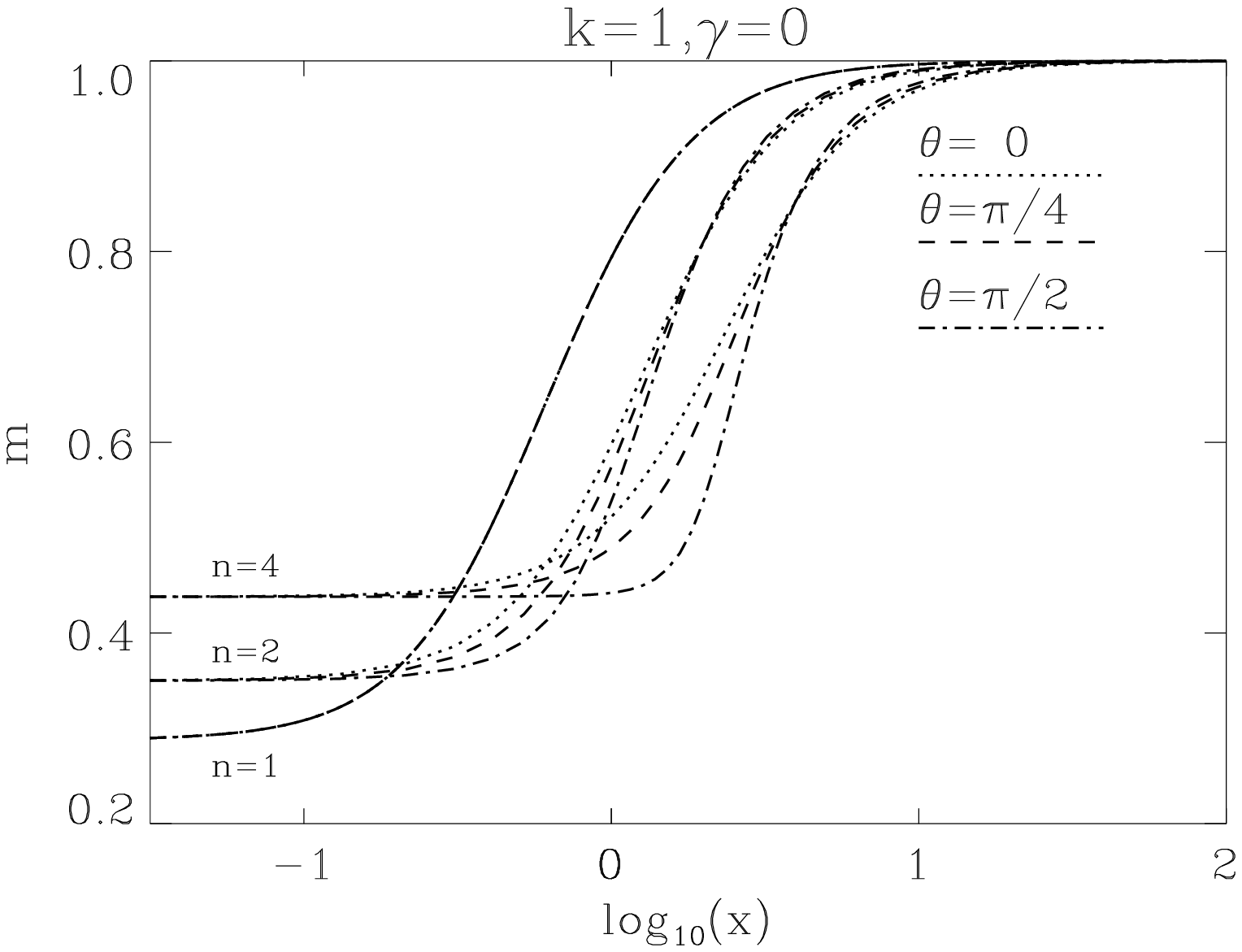
}}\\
Fig.~15b\\
Same as Fig.~15a for the metric function $m$.
\end{figure}
\clearpage

\newpage
\begin{figure}
\centering
\epsfysize=12cm
\mbox{\epsffile{
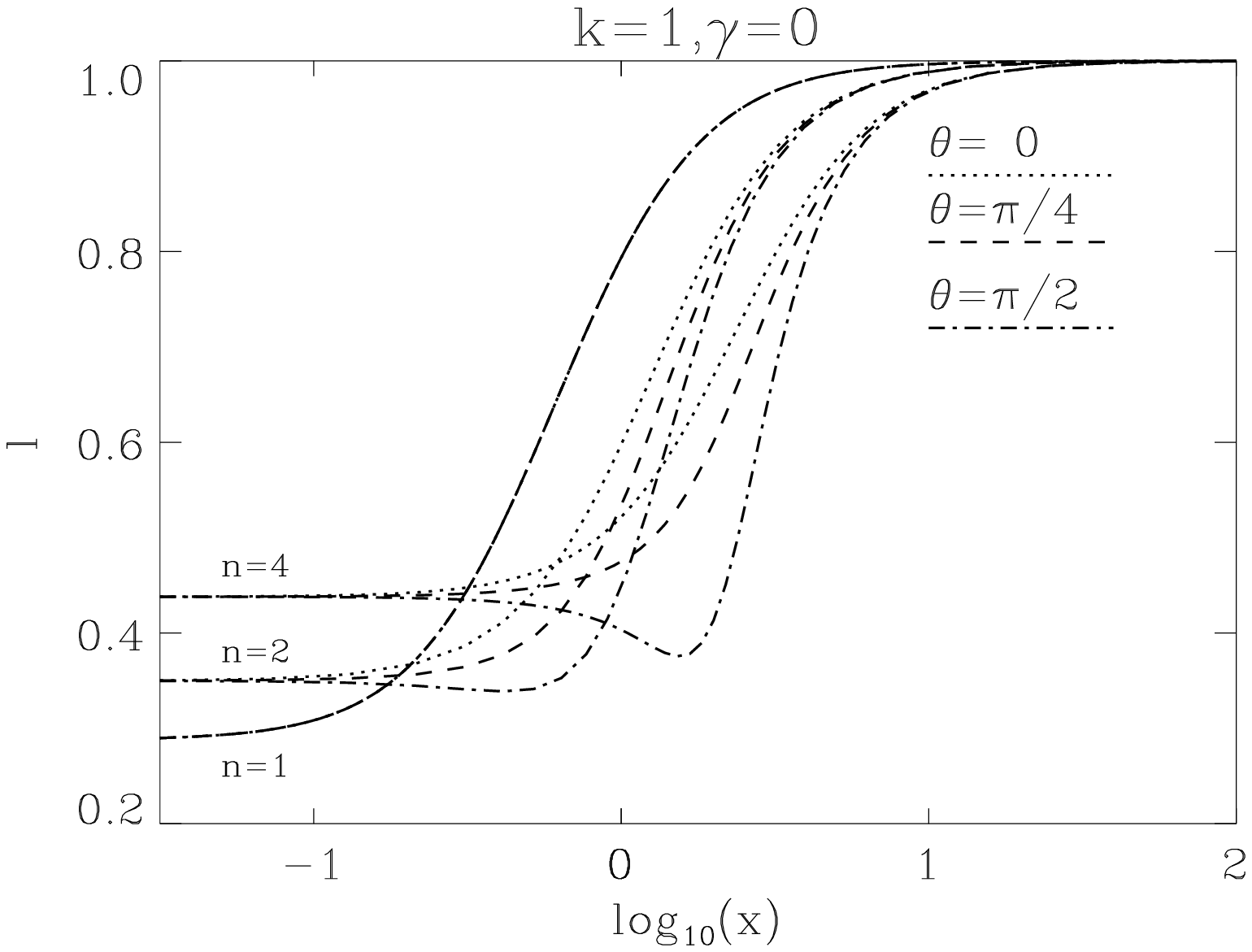
}}\\
Fig.~15c\\
Same as Fig.~15a for the metric function $l$.
\end{figure}
\clearpage

\newpage
\begin{figure}
\centering
\epsfysize=12cm
\mbox{\epsffile{
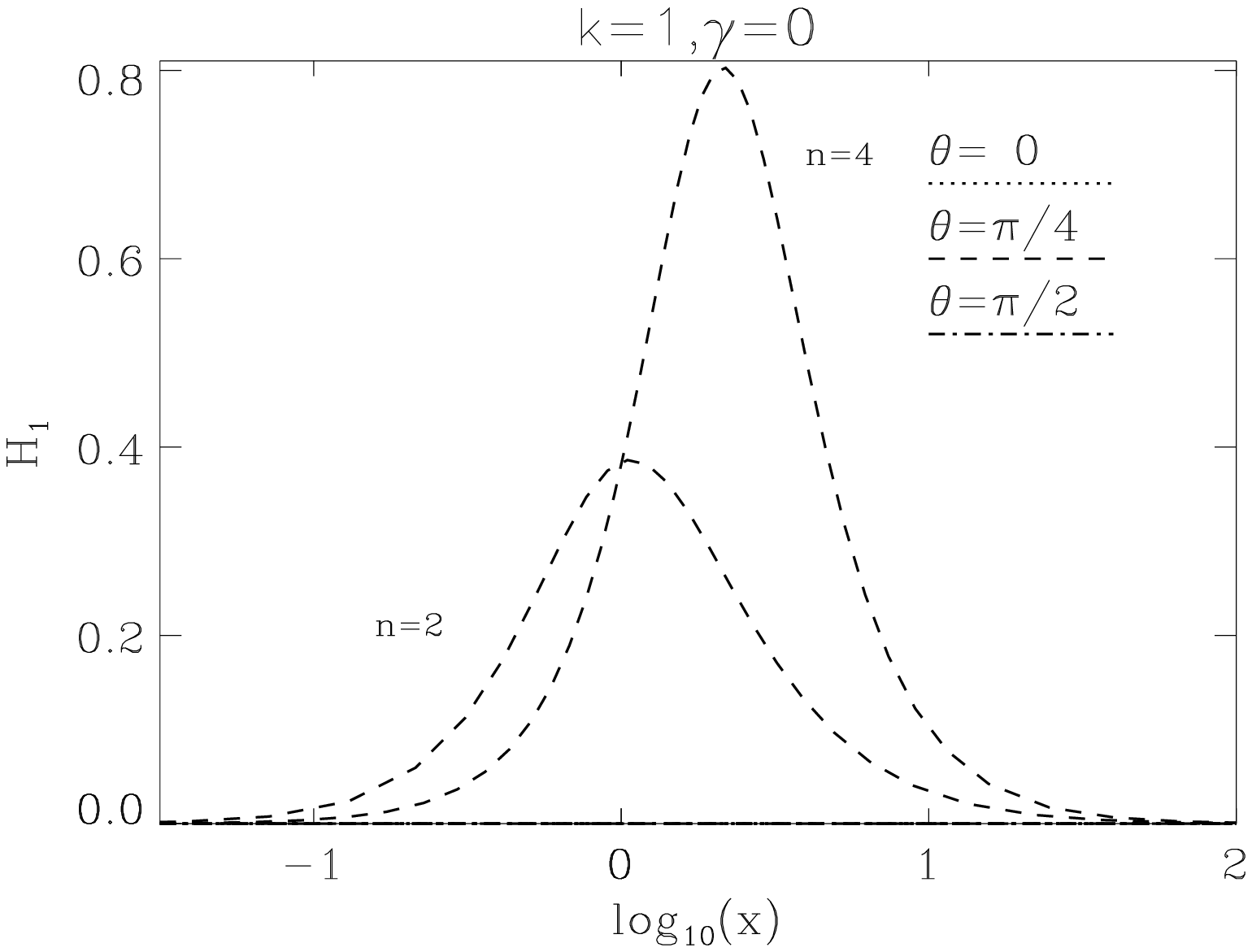
}}\\
Fig.~16a\\
Same as Fig.~15a for the gauge field function $H_1$.
\end{figure}
\clearpage

\newpage
\begin{figure}
\centering
\epsfysize=12cm
\mbox{\epsffile{
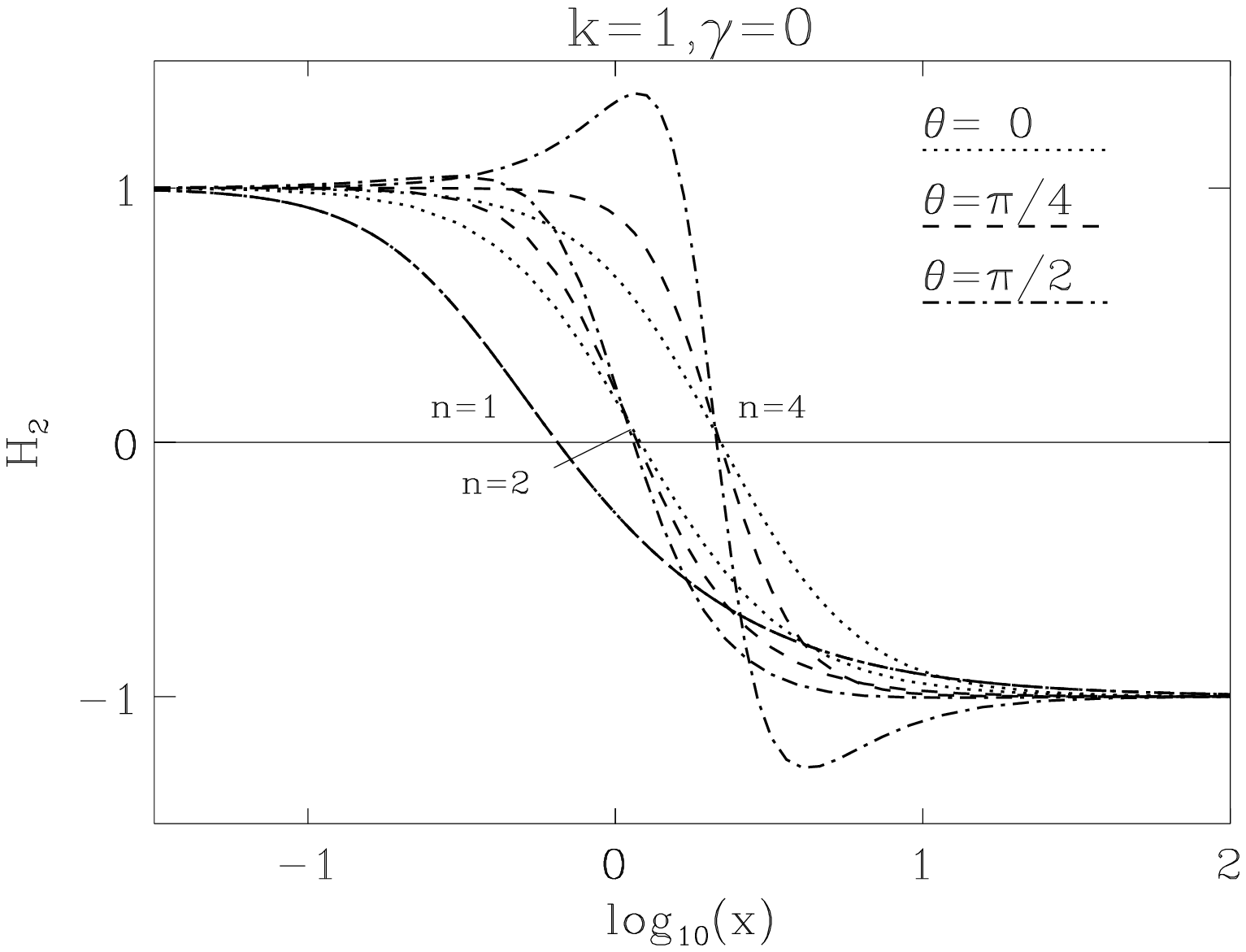
}}\\
Fig.~16b\\
Same as Fig.~15a for the gauge field function $H_2$.
\end{figure}
\clearpage

\newpage
\begin{figure}
\centering
\epsfysize=12cm
\mbox{\epsffile{
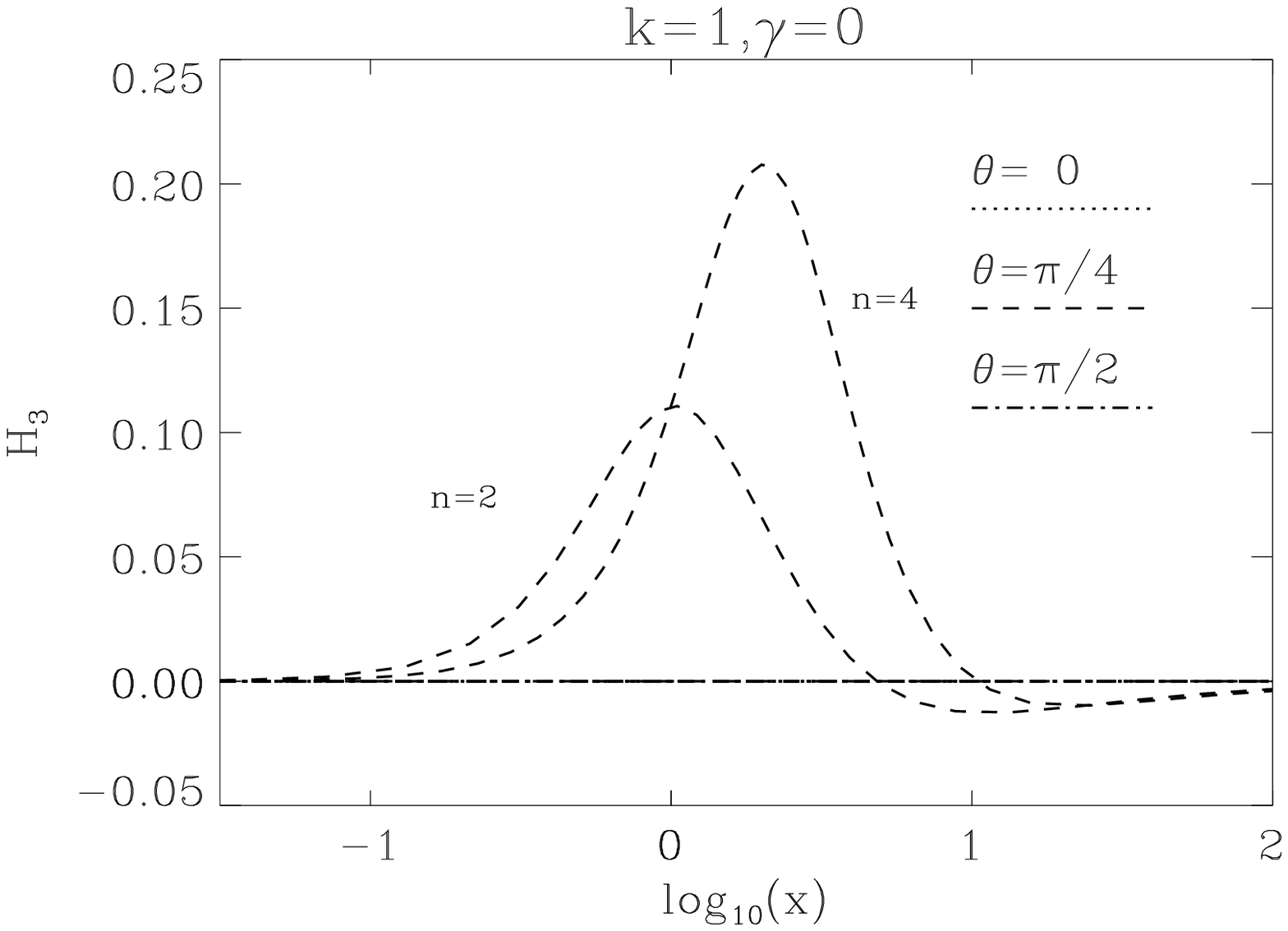
}}\\
Fig.~16c\\
Same as Fig.~15a for the gauge field function $H_3$.
\end{figure}
\clearpage

\newpage
\begin{figure}
\centering
\epsfysize=12cm
\mbox{\epsffile{
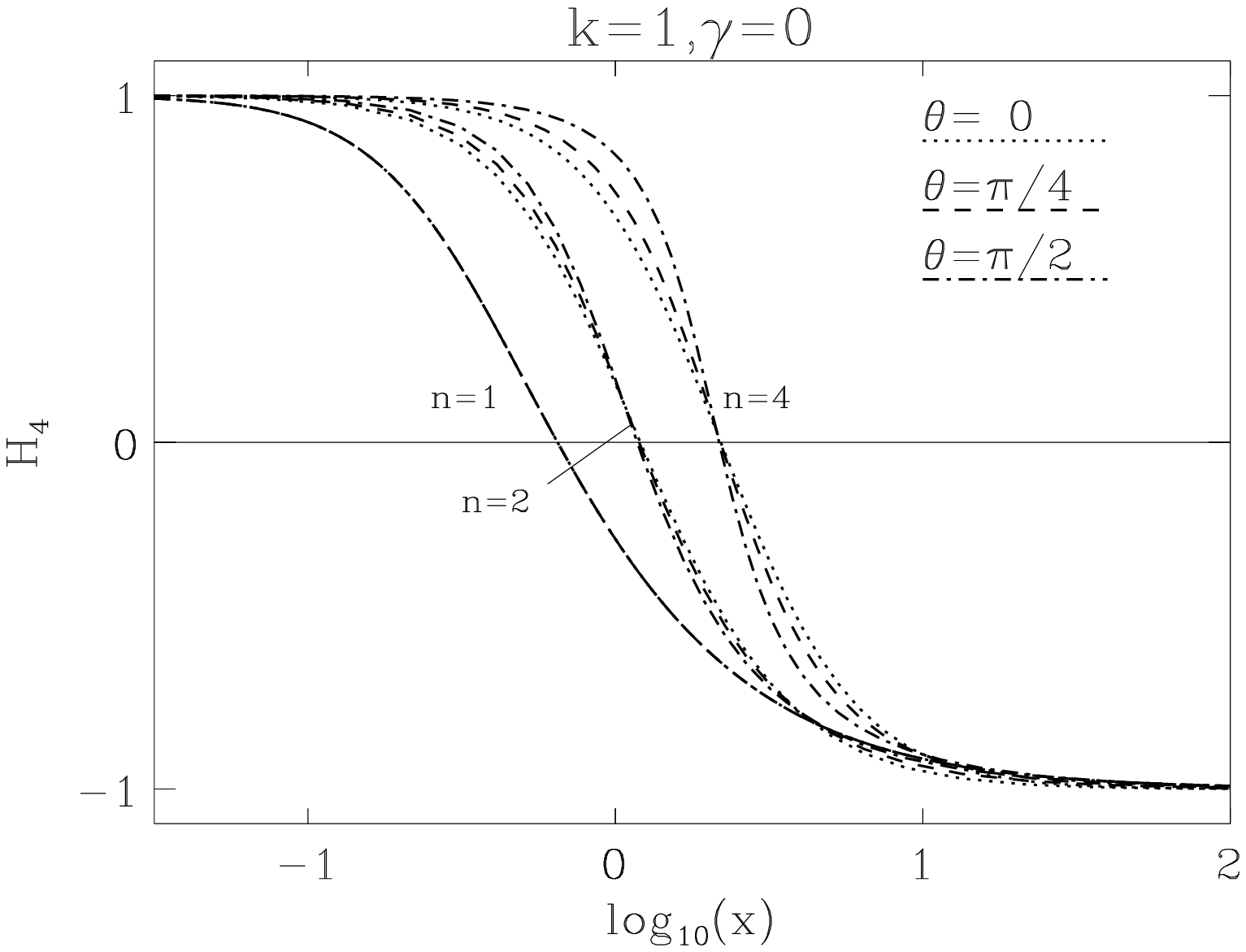
}}\\
Fig.~16d\\
Same as Fig.~15a for the gauge field function $H_4$.
\end{figure}
\clearpage

\newpage
\begin{figure}
\centering
\epsfysize=12cm
\mbox{\epsffile{
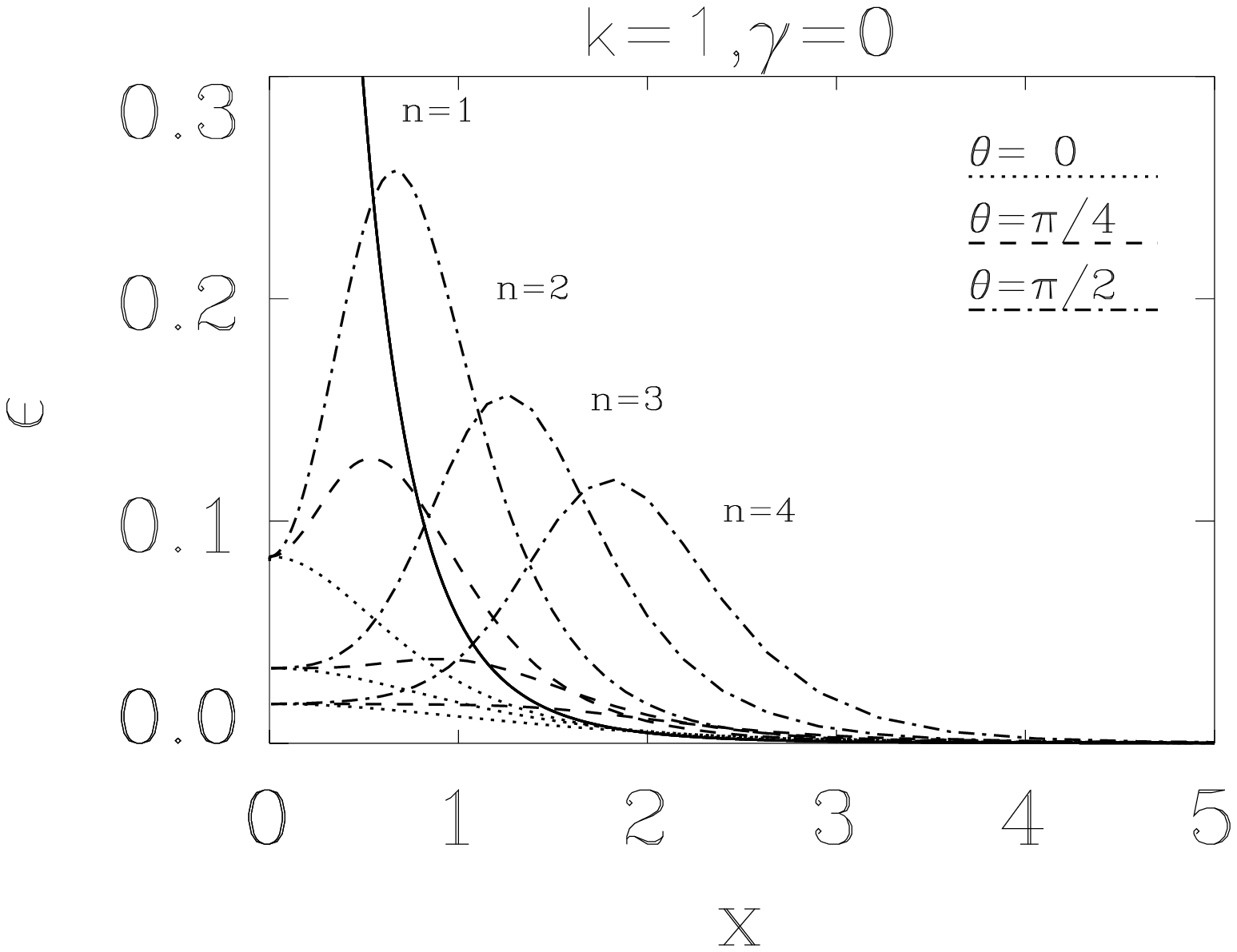
}}\\
Fig.~17\\
Same as Fig.~15a for the energy density of the matter fields
$\epsilon$.
\end{figure}
\clearpage

\newpage
\begin{figure}
\centering
\epsfysize=12cm
\mbox{\epsffile{
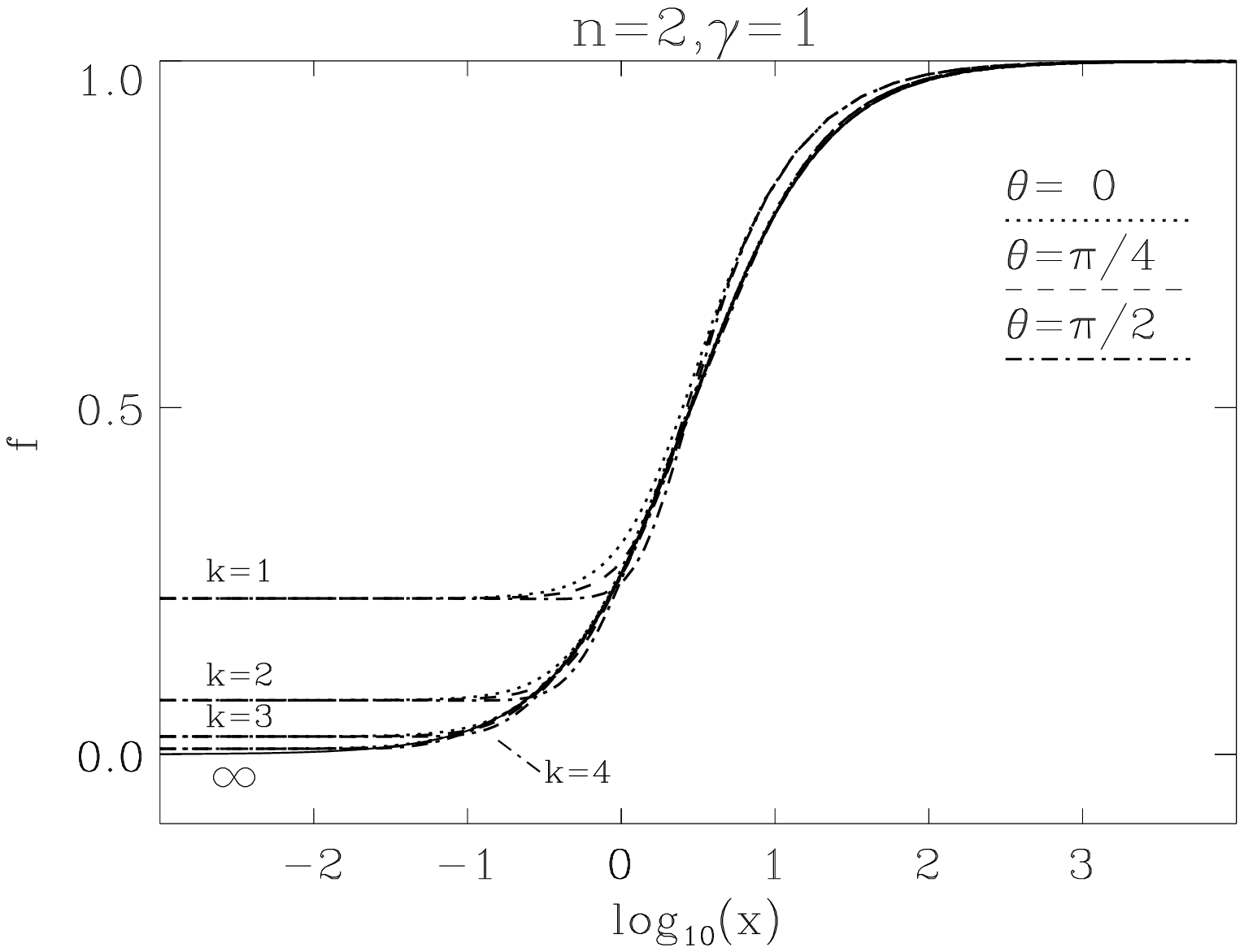
}}\\
Fig.~18a\\
The metric function $f$ of the EYMD solutions with $\gamma=1$, $n=2$,
and $k=1-4$ is shown as a function of the coordinate $x$
for the angles $\theta=0$, $\pi/4$ and $\pi/2$.
Also shown is the metric function of the limiting EMD solution.
\end{figure}
\clearpage

\newpage
\begin{figure}
\centering
\epsfysize=12cm
\mbox{\epsffile{
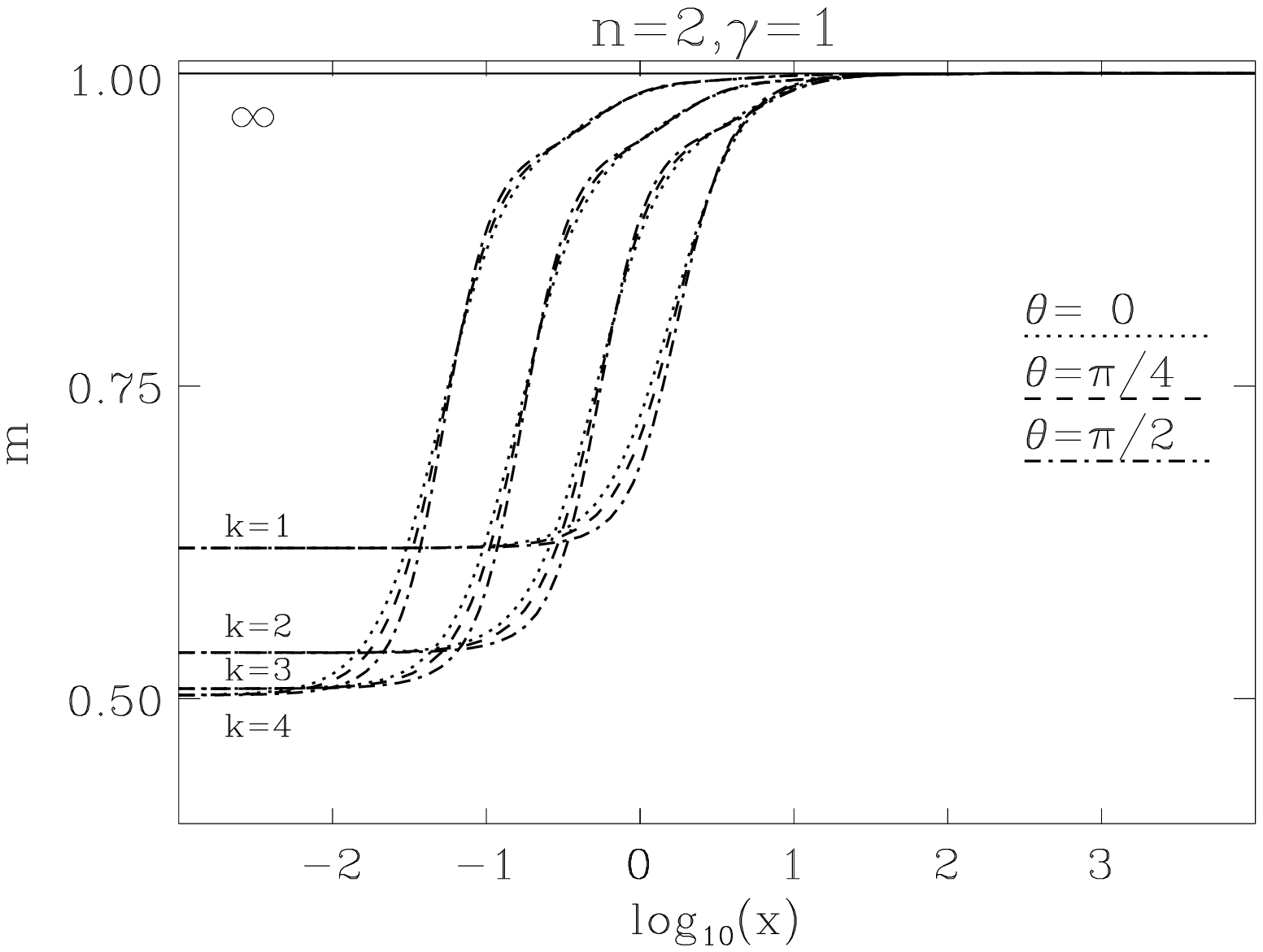
}}\\
Fig.~18b\\
Same as Fig.~18a for the metric function $m$.
\end{figure}
\clearpage

\newpage
\begin{figure}
\centering
\epsfysize=12cm
\mbox{\epsffile{
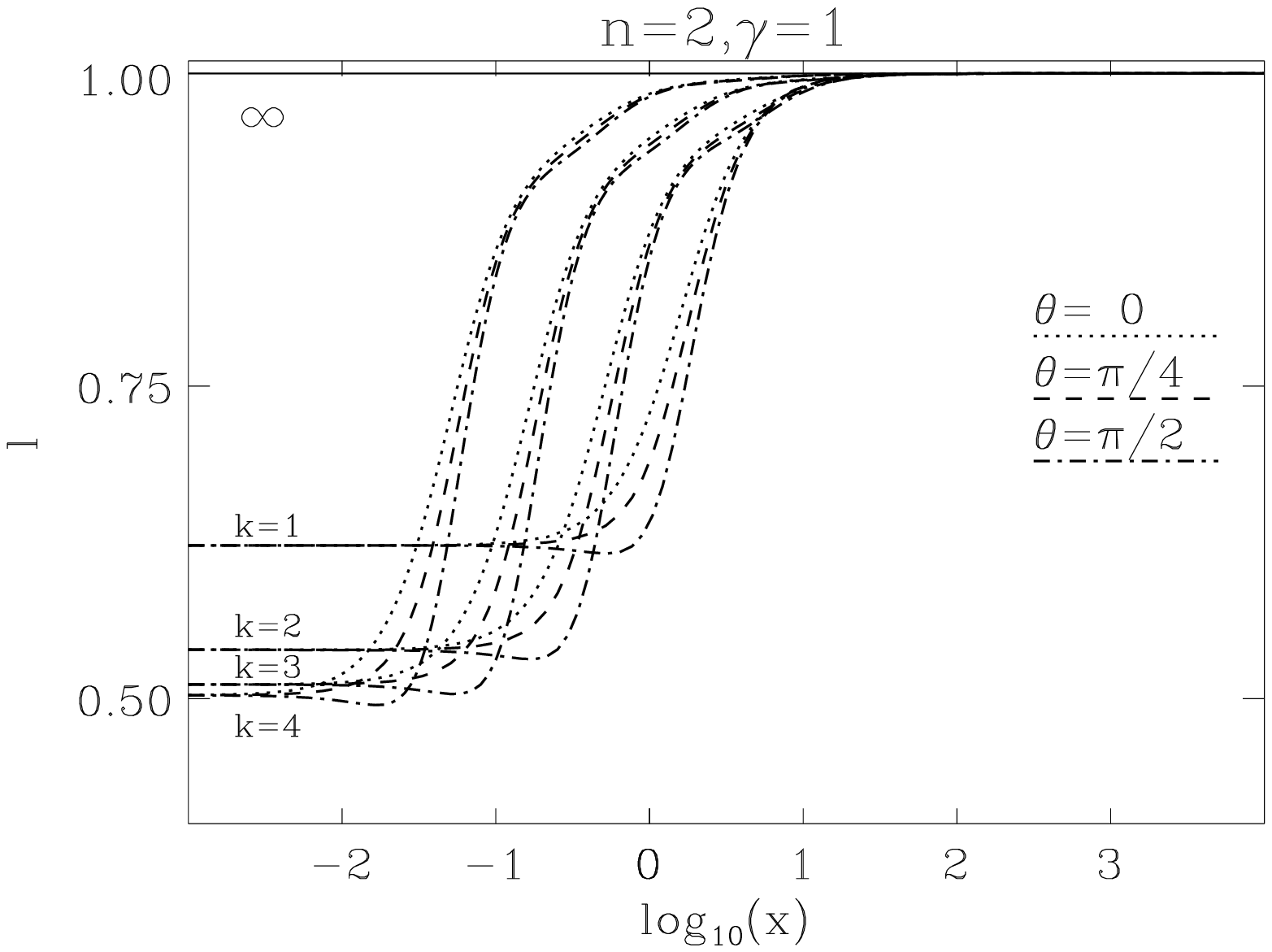
}}\\
Fig.~18c\\
Same as Fig.~18a for the metric function $l$.
\end{figure}
\clearpage

\newpage
\begin{figure}
\centering
\epsfysize=12cm
\mbox{\epsffile{
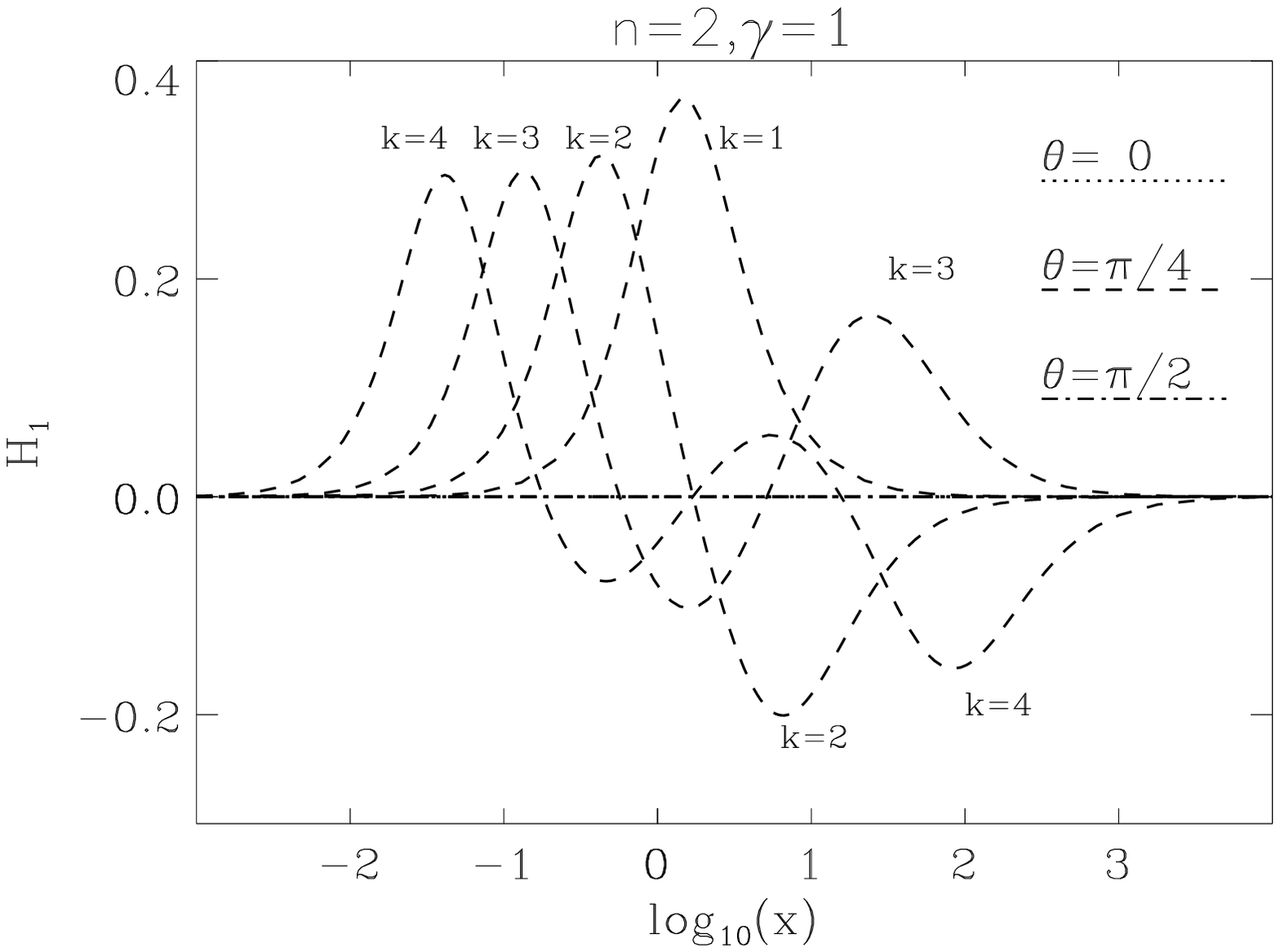
}}\\
Fig.~19a\\
Same as Fig.~18a for the gauge field function $H_1$.
\end{figure}
\clearpage

\newpage
\begin{figure}
\centering
\epsfysize=12cm
\mbox{\epsffile{
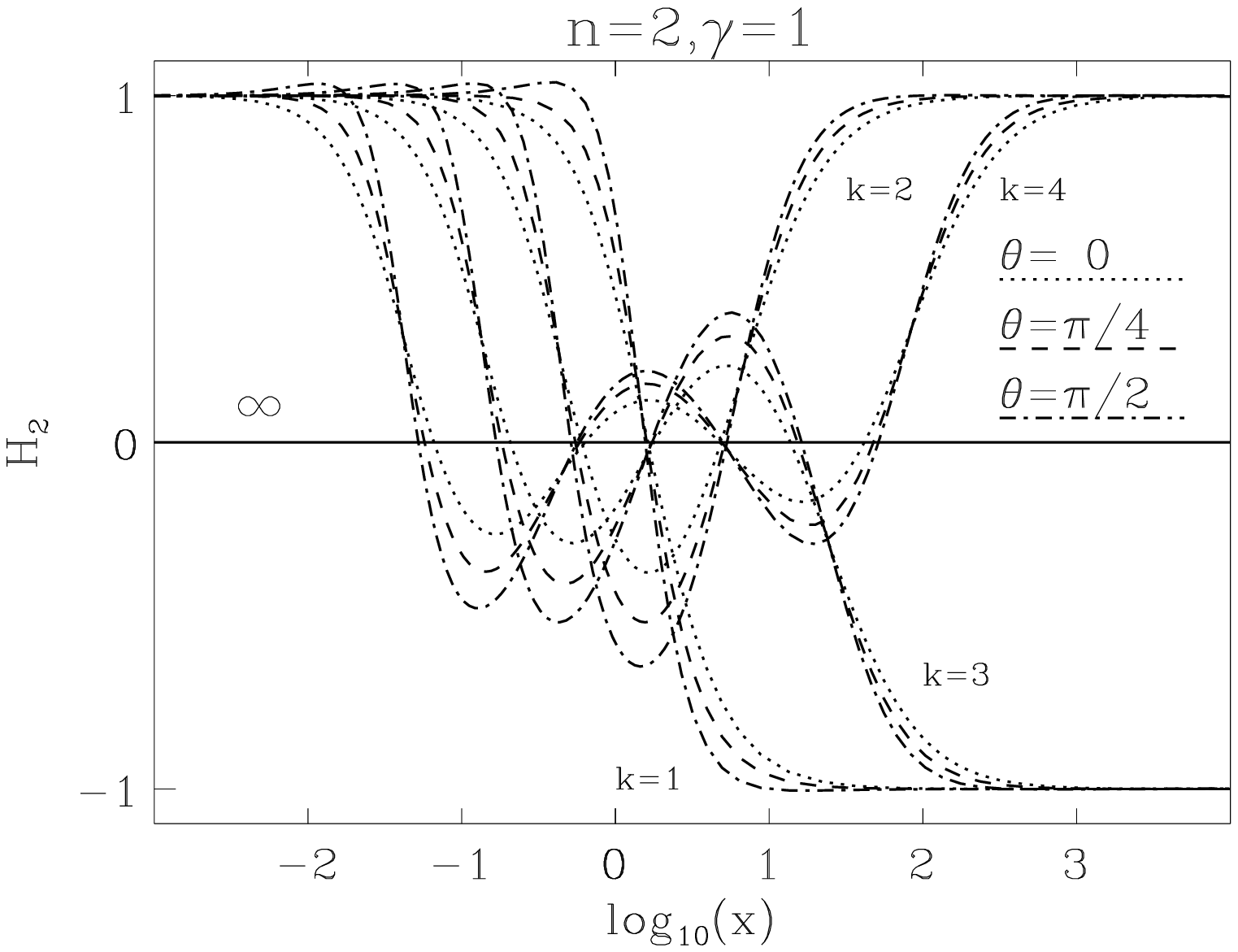
}}\\
Fig.~19b\\
Same as Fig.~18a for the gauge field function $H_2$.
\end{figure}
\clearpage

\newpage
\begin{figure}
\centering
\epsfysize=12cm
\mbox{\epsffile{
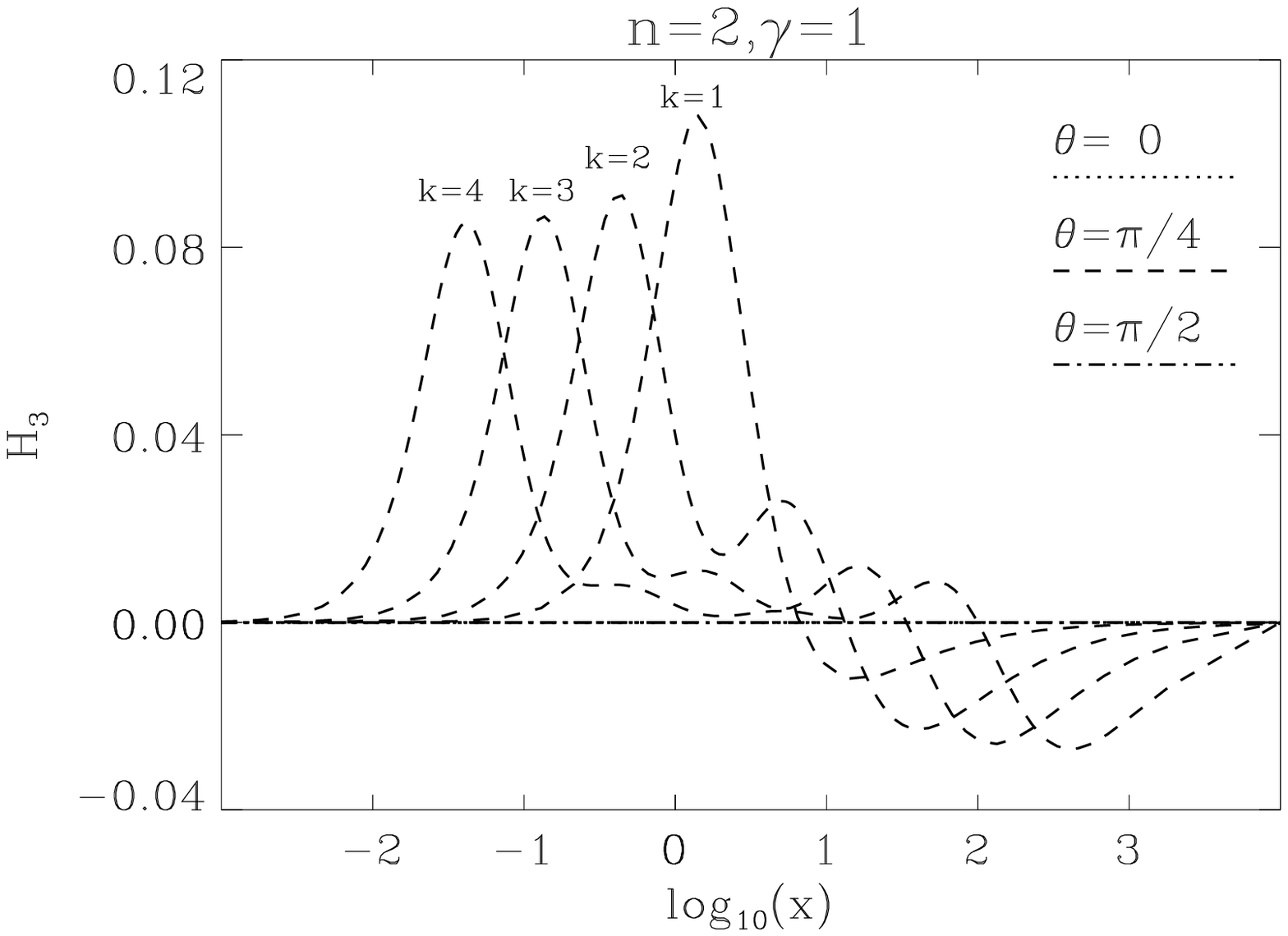
}}\\
Fig.~19c\\
Same as Fig.~18a for the gauge field function $H_3$.
\end{figure}
\clearpage

\newpage
\begin{figure}
\centering
\epsfysize=12cm
\mbox{\epsffile{
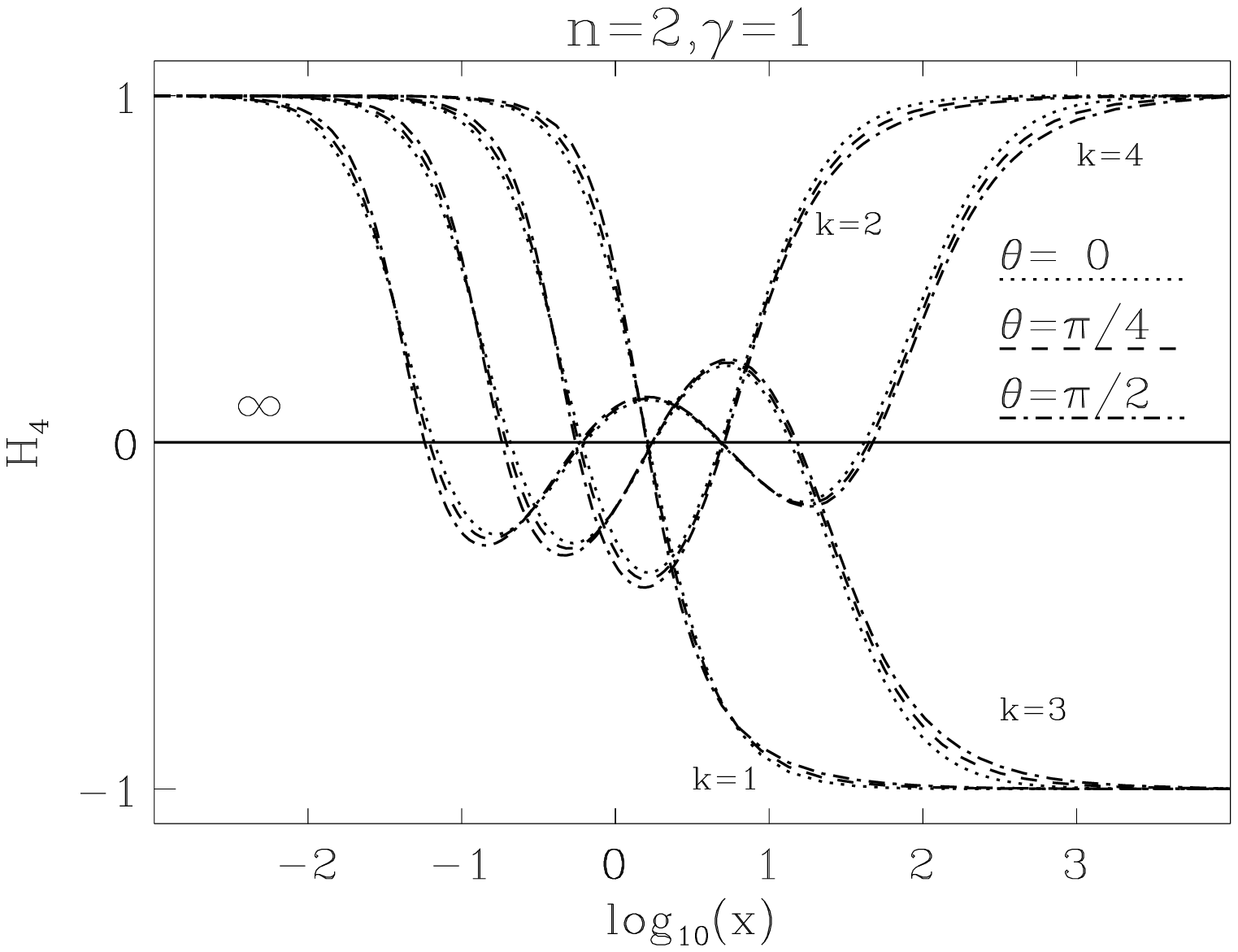
}}\\
Fig.~19d\\
Same as Fig.~18a for the gauge field function $H_4$.
\end{figure}
\clearpage

\newpage
\begin{figure}
\centering
\epsfysize=12cm
\mbox{\epsffile{
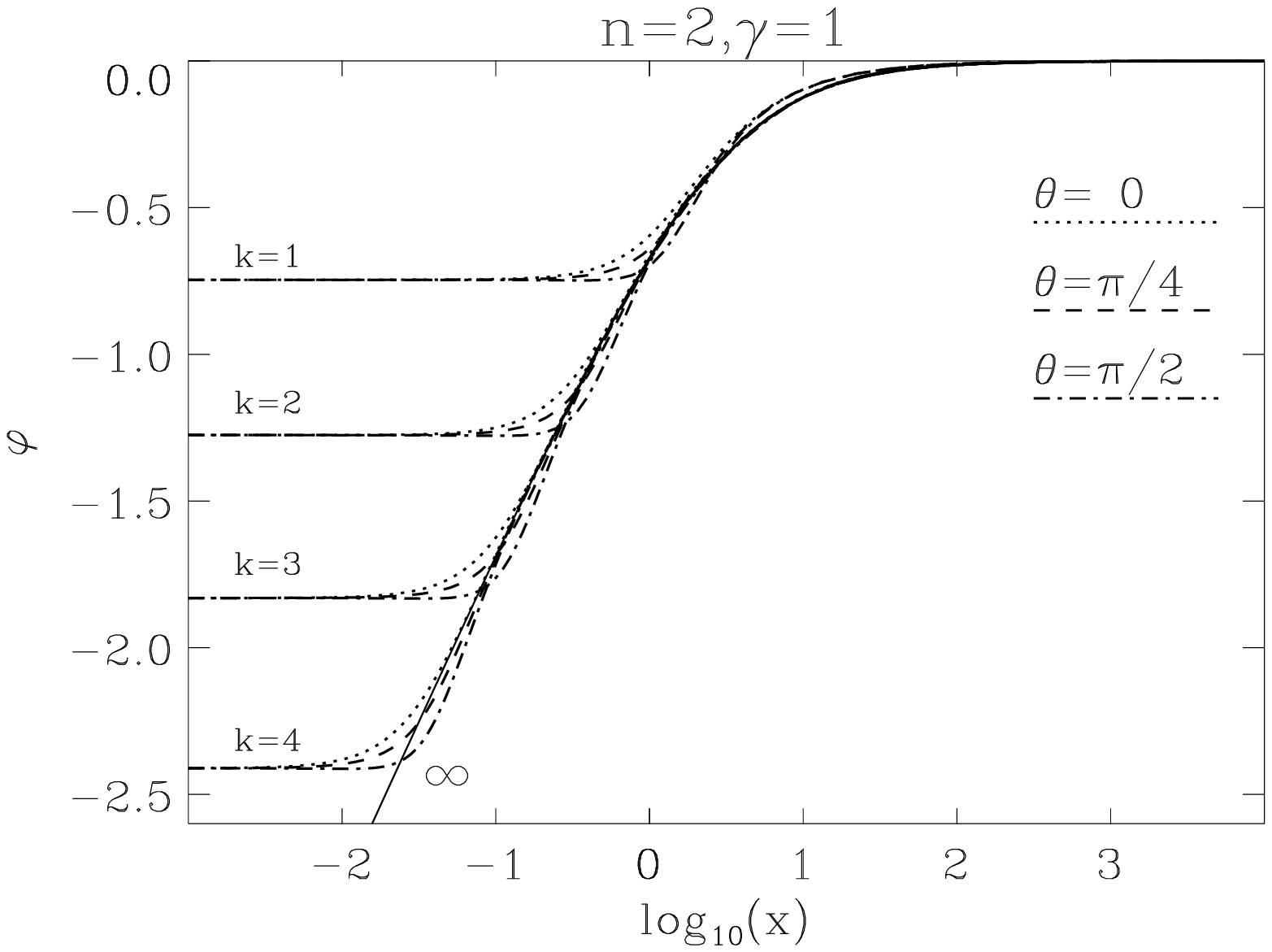
}}\\
Fig.~20\\
Same as Fig.~18a for the dilaton function $\varphi$.
\end{figure}
\clearpage

\newpage
\begin{figure}
\centering
\epsfysize=12cm
\mbox{\epsffile{
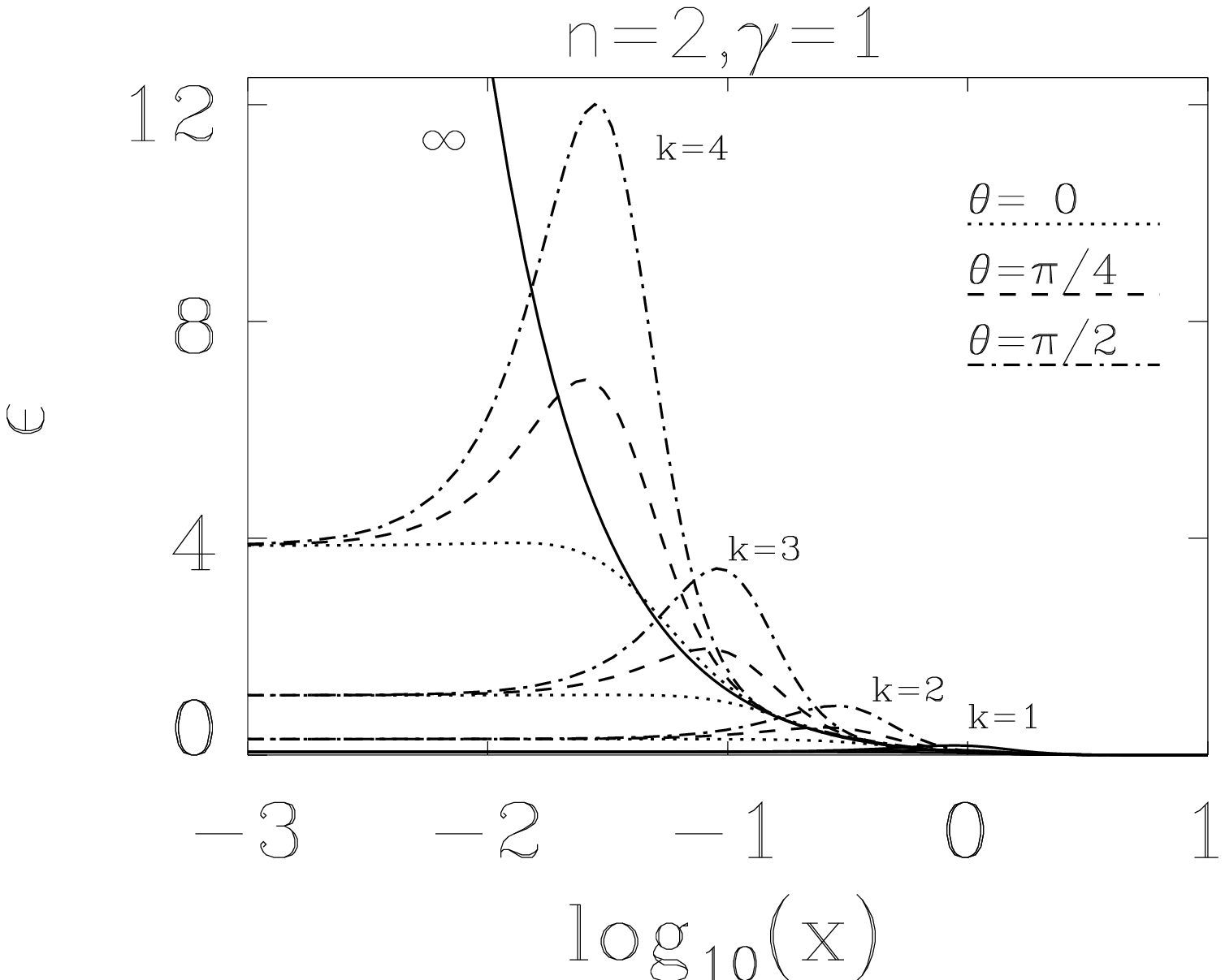
}}\\
Fig.~21\\
Same as Fig.~18a for the energy density of the matter fields
$\epsilon$.
\end{figure}
\clearpage

\newpage
\begin{figure}
\centering
\epsfysize=12cm
\mbox{\epsffile{
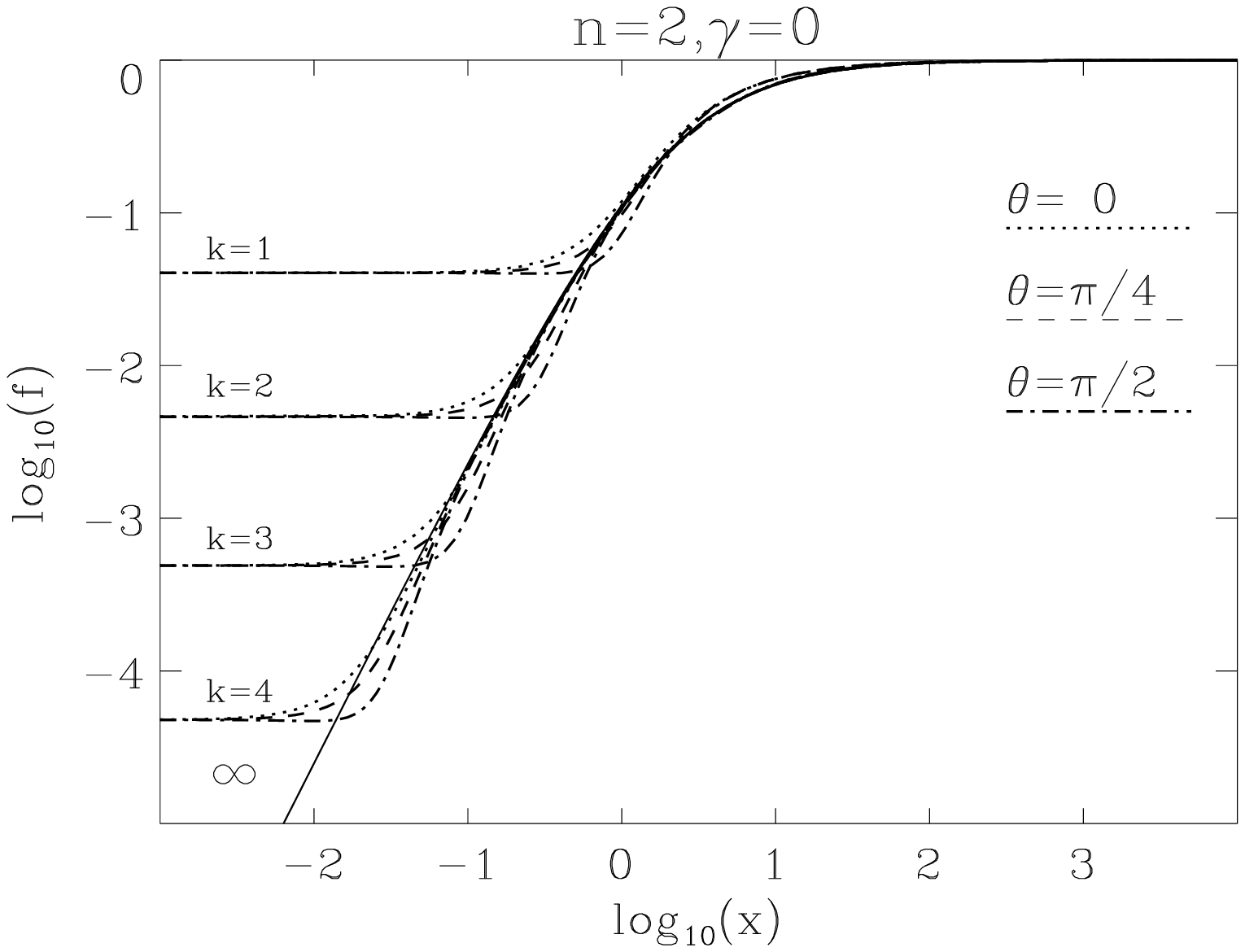
}}\\
Fig.~22a\\
The metric function $f$ of the EYM solutions with $n=2$,
and $k=1-4$ is shown as a function of the coordinate $x$
for the angles $\theta=0$, $\pi/4$ and $\pi/2$.
Also shown is the metric function of the limiting RN solution.
\end{figure}
\clearpage

\newpage
\begin{figure}
\centering
\epsfysize=12cm
\mbox{\epsffile{
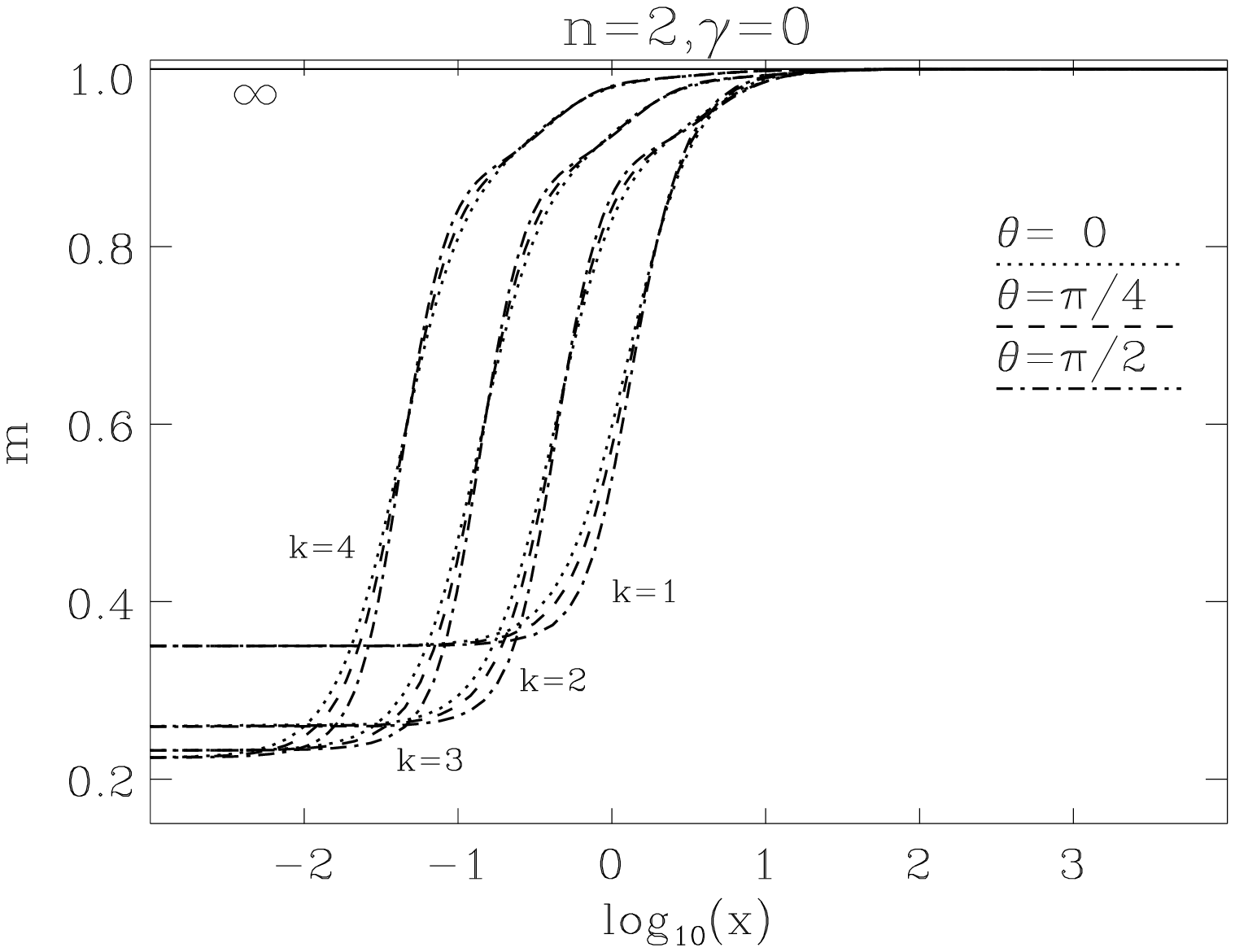
}}\\
Fig.~22b\\
Same as Fig.~22a for the metric function $m$.
\end{figure}
\clearpage

\newpage
\begin{figure}
\centering
\epsfysize=12cm
\mbox{\epsffile{
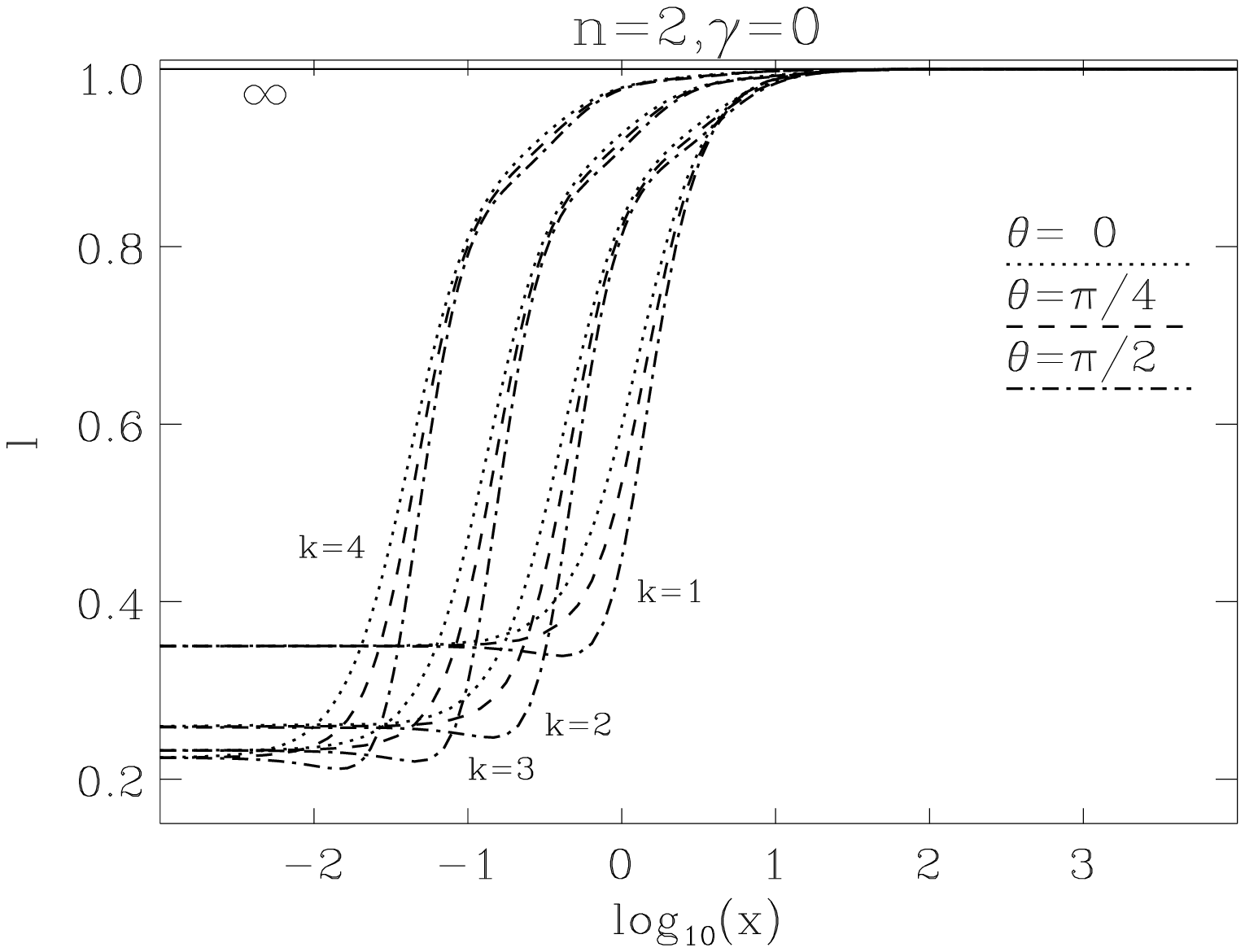
}}\\
Fig.~22c\\
Same as Fig.~22a for the metric function $l$.
\end{figure}
\clearpage

\newpage
\begin{figure}
\centering
\epsfysize=12cm
\mbox{\epsffile{
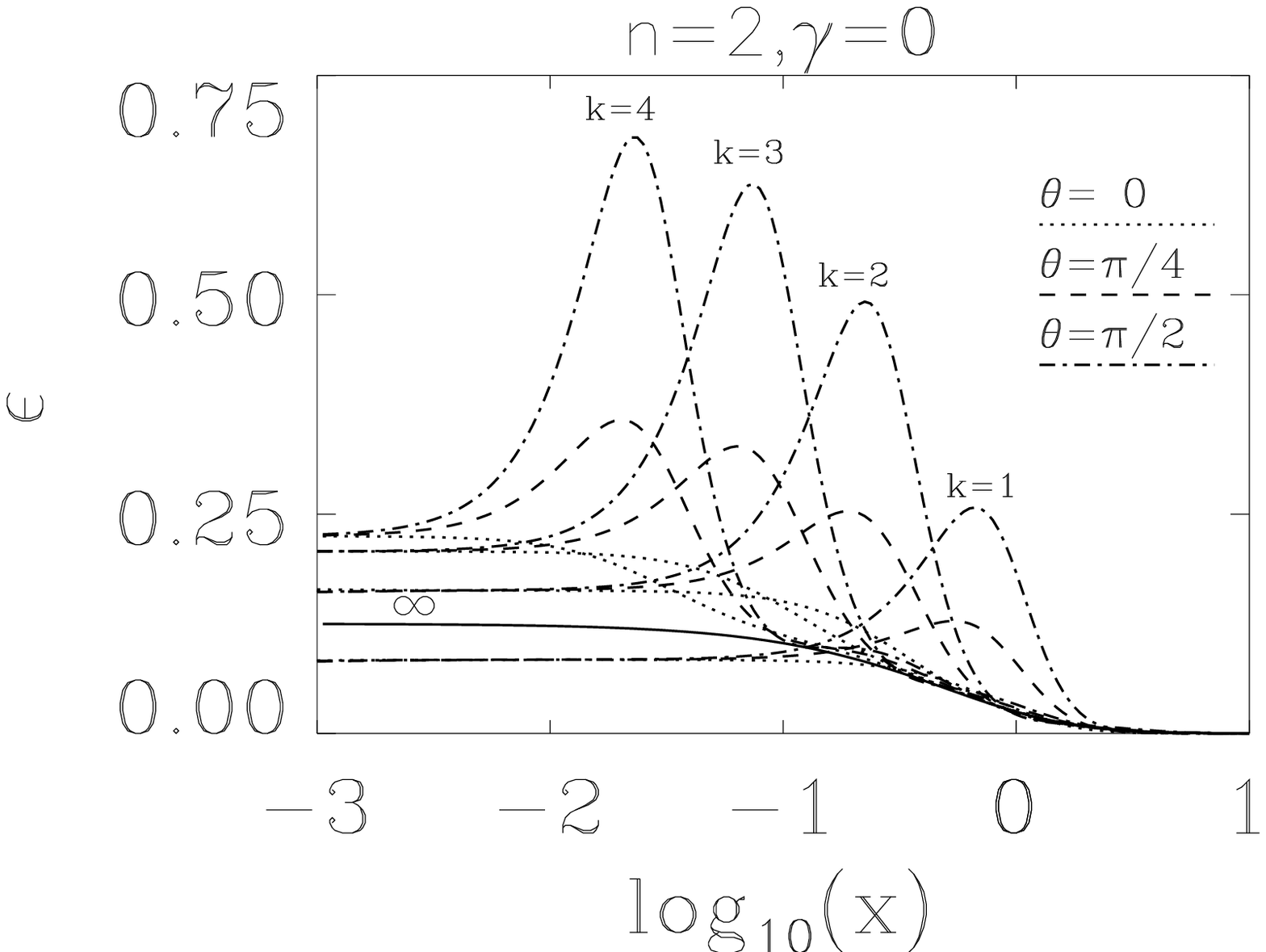
}}\\
Fig.~23\\
Same as Fig.~22a for the energy density of the matter fields
$\epsilon$.
\end{figure}
\clearpage

\newpage
\begin{figure}
\centering
\epsfysize=12cm
\mbox{\epsffile{
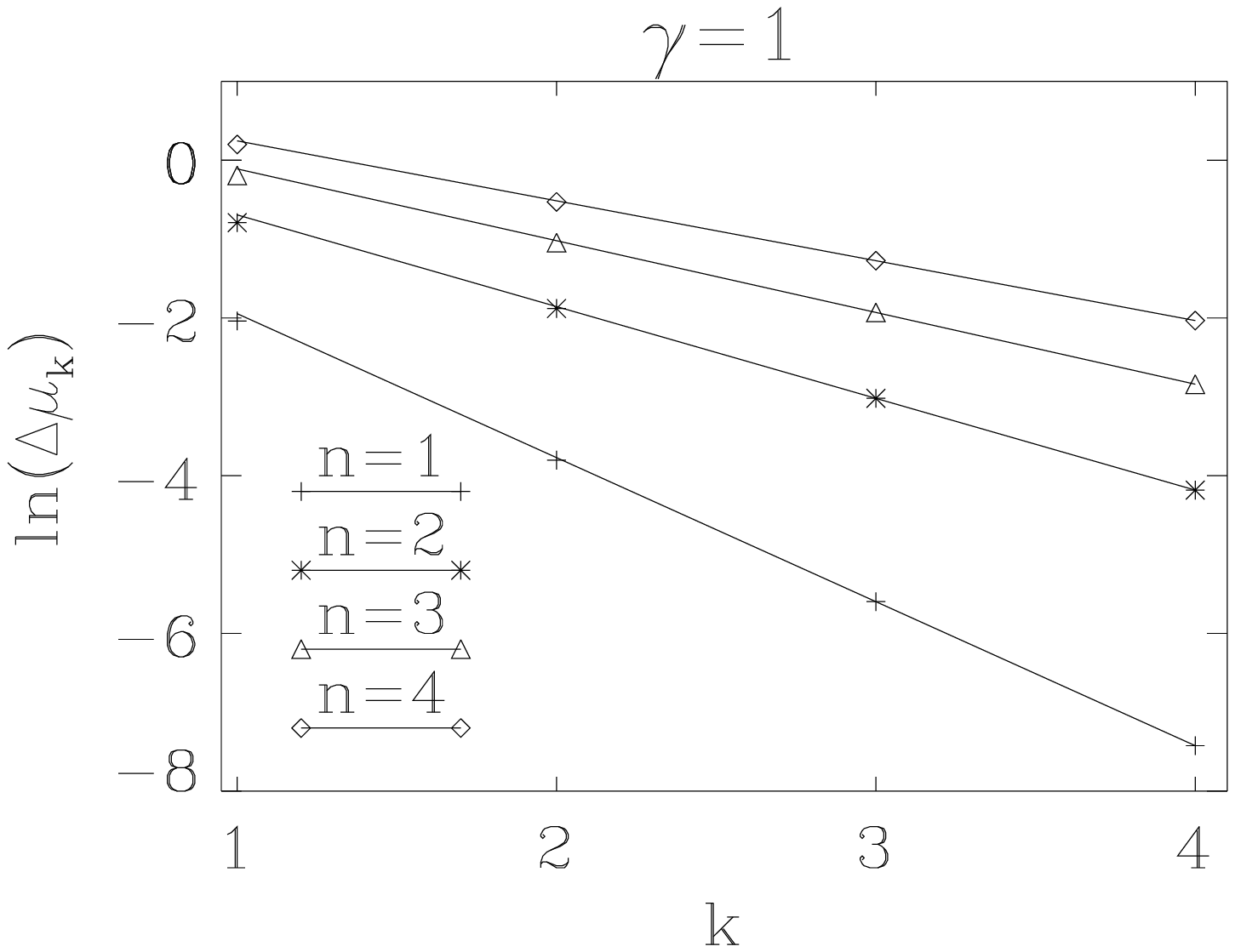
}}\\
Fig.~24a\\
The logarithm of the absolute deviation from the limiting solution
$\Delta \mu_k = \mu_\infty (\infty)-\mu_k (\infty)$
for the EYMD masses as a function of the node number $k$ with dilaton
coupling constant $\gamma=1$ for winding number $n=1-4$. The straight lines
are fitted to the data with $k=3$ and $k=4$.
\end{figure}
\clearpage

\newpage
\begin{figure}
\centering
\epsfysize=12cm
\mbox{\epsffile{
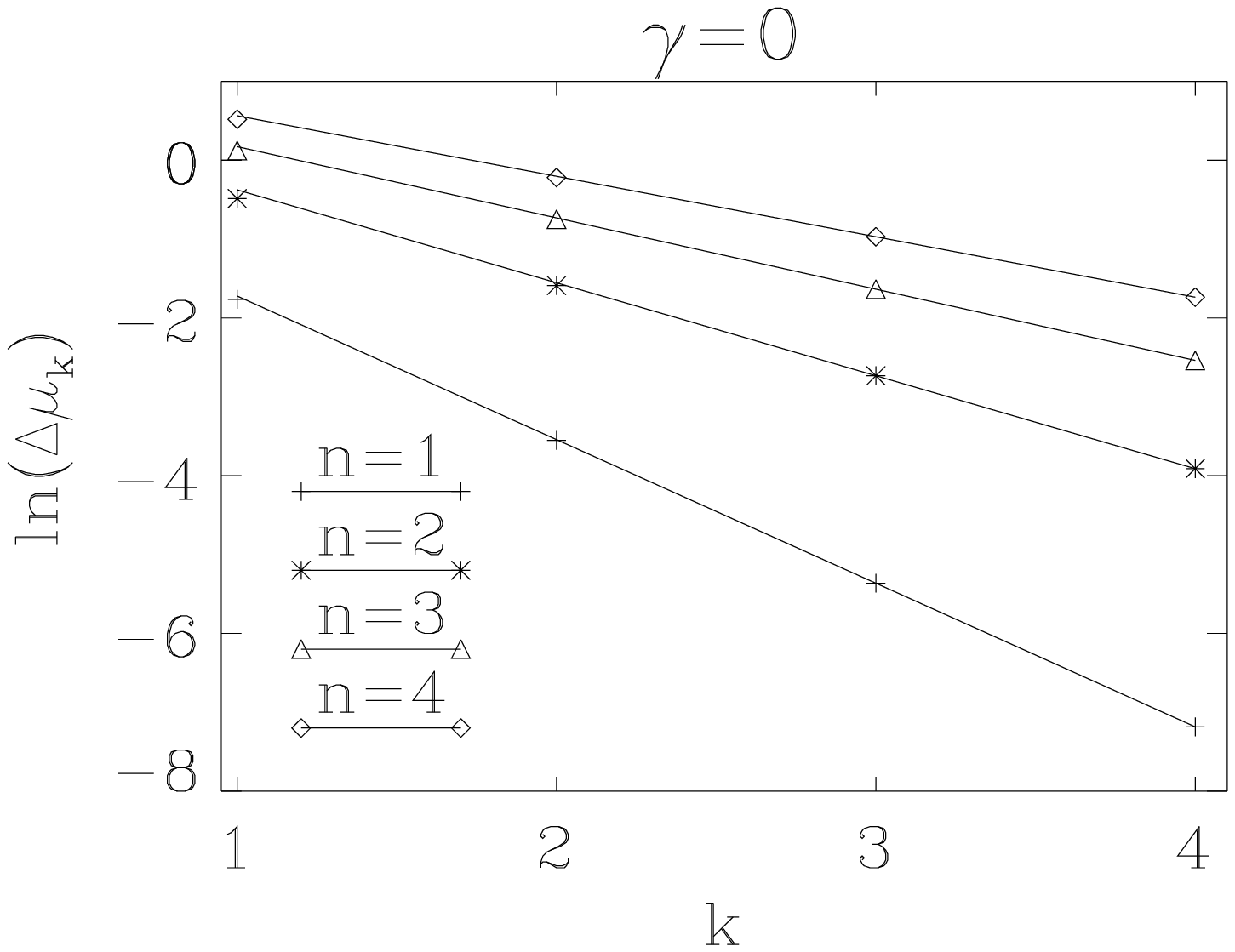
}}\\
Fig.~24b\\
Same as Fig.~24a for the EYM solutions, i.~e. $\gamma=0$.
\end{figure}
\clearpage


\begin{thebibliography}{000}
\bibitem{foot2}
 Another parametrization of the isotropic metric is given
 in Appendix D.
\end{thebibliography}
\end{document}